\documentclass[fleqn,usenatbib]{mnras}

\usepackage{newtxtext,newtxmath}
\usepackage{url}
\usepackage{color}
\usepackage{tabulary}
\usepackage{multirow}
\usepackage{amsfonts}
\usepackage{array}
\usepackage{ulem}
\usepackage[T1]{fontenc}
\usepackage{ae,aecompl}

\usepackage{graphicx}	
\usepackage{amsmath}	
\usepackage{amssymb}	
\usepackage{cases}

\graphicspath{ {pictures/} }
\newcommand{\beq}{\begin{equation}}
\newcommand{\eeq}{\end{equation}}
\newcommand{\simlt}{\mathrel{\hbox{\rlap{\hbox{\lower4pt\hbox{$\sim$}}}\hbox{$<$}}}}
\newcommand{\simgt}{\mathrel{\hbox{\rlap{\hbox{\lower4pt\hbox{$\sim$}}}\hbox{$>$}}}}

\newcommand{\s}{\;\mathrm{s}}

\newcommand{\Msol}{\;\mathrm{M}_{\odot}}

\newcommand{\km}{\;\mathrm{km}}

\newcommand{\pc}{\;\mathrm{pc}}

\newcommand{\Myr}{\;\mathrm{Myr}}

\def\apj{ApJ}
\def\mnras{M.N.R.A.S.}
\def\aap{A\&A}

\def\araa{Ann. Rev. A\&A}

\def\apjs{ApJ Supp.}
\def\aj{AJ}

\defcitealias{Gualandris+2004}{GPZE04}

\title[runaway stars]{An analytic method for identifying dynamically-formed runaway stars}

\author[T. Ryu et al.]{
Taeho Ryu$^{1}$\thanks{email: taeho.ryu@stonybrook.edu}, Nathan W. C. Leigh$^{2}$, Rosalba Perna$^{1}$
\\
$^{1}$Department of Physics and Astronomy, Stony Brook University, Stony Brook, NY 11794-3800, USA\\
$^{2}$Department of Astrophysics, American Museum of Natural History, Central Park West and 79th Street, New York, NY 10024}

\date{Accepted XXX. Received YYY; in original form ZZZ}

\pubyear{2016}

\begin{document}
\label{firstpage}
\pagerange{\pageref{firstpage}--\pageref{lastpage}}
\maketitle

\begin{abstract}
In this paper, we study the 3-body products (two single
stars and a binary) of binary-binary (2+2) scattering interactions.
This is done using a combination of analytic methods and numerical
simulations of 2+2 scattering interactions, both in isolation and in a
homogeneous background potential. We derive analytically a simple
formula relating the angle between the velocity vectors of the two
ejected single stars and the orbital separation of the remaining
binary.  We compare our analytic formulation to numerical scattering
simulations, and illustrate that the agreement is excellent, both in
isolation and in a homogeneous background potential.  Our results are
ideally suited for application to the GAIA database, which is expected
to identify many hundred runaway stars.  The analytic relation
presented here has the potential to identify runaway stars formed
dynamically with high confidence.  Finally, by applying our method to
the runaways AE Aur and $\mu$ Col, we illustrate that it
can be used to constrain the history of the background potential,
which was denser than the presently observed density in the case of
the Trapezium cluster.
\end{abstract}




\section{introduction}

\label{sec: introduction}

In the general gravitational $N$-body problem, all masses are non-zero
and their initial positions and velocities are not arranged in any
specific way.  The simplest subset of this more general problem,
called the three-body problem, involves only three particles.  And
yet, the problem is analytically intractable without simplifying
assumptions.  This is in contrast to the two-body problem, in which
the positions and velocities of both particles are known exactly at
any time given any initial configuration.  This complication stems
from the fact that, for $N >$ 2, there are no coordinate
transformations that simplify the problem sufficiently (e.g., the
centre of mass coordinate system).  Consequently, it is not possible
to express the positions and velocities of all particles at any future
time given a set of initial conditions using analytic theory alone.
What's more, the long-term evolution of these systems are very
sensitive to the initial conditions.  If one waits long enough,
however, a final stable state will always emerge.  This usually occurs
after the system has disrupted and all particles can be regarded as
having escaped to spatial infinity.  But the final state cannot be
calculated directly from the initial conditions without the need for
computer simulations.  The problem is inherently chaotic.

Few-body systems have been extensively investigated over the last few decades \citep[e.g][]{Heggie1975,Mikkola1983,Mikkola1984}. 
	These studies typically adopted numerical approaches. However, analytic theory remains a powerful tool
in the study of gravitational dynamics.  For example, if the outcome
of the interaction is known a priori, aspects of the problem can
become deterministic.  That is, energy and momentum conservation can
be invoked to directly relate the initial conditions to the final
outcome properties.  

Here, we apply this logic to the four-body problem and binary-binary
scattering.  In particular, for (nearly) identical initial binaries
and large viral ratios, most four-body interactions decay to produce two
single stars and a binary \citep{Leigh+2016}.  As we will show,
knowledge of the initial interaction energy and the momentum are
sufficient to directly relate the properties of the ejected single
stars to the remaining binary.  This offers a clear prediction for
observations of dynamically-formed runaway stars; young O/B stars that
have been ejected from their host star cluster at high velocity
($\gtrsim$ 30~km s$^{-1}$) \citep[e.g][]{Fujii2011,Gualandris+2004}.
Consequently, the method presented here can be used to unambiguously
distinguish runaways formed during binary-binary interactions from
those formed from supernova explosions in a binary system
\citep[e.g][]{Stone1979,Stone1982}, by making specific predictions for
the observable properties of the left-over binary.  Our results are
particularly useful in light of the expected hundreds of detections of
runaways stars with the GAIA satellite (i.e. \citealt{Kenyon+2014}).

Our paper is organized as follows: Section~2 presents the derivation
of the analytical formulation of the 2+1+1 problem, first in the
purely stellar dynamical case (Sec.2.1), and then in the presence of
a homogeneous background potential (Sec.2.2).  A comparison of the
predictions of the analytical formulae and the results of numerical
scattering experiments is presented in Section~3. Astrophysical
implications of our findings are discussed in Section~4, with a
particular emphasis on the application to runaway stars. We summarize
in Section~5.

\section{Analytic formulation of 2+1+1 case}
\subsection{Analytic formula : Purely stellar dynamics}
\label{sec:analyticformula:puerlystellardynamics}

\begin{figure}
	\centering
	\includegraphics[width=8.7cm]{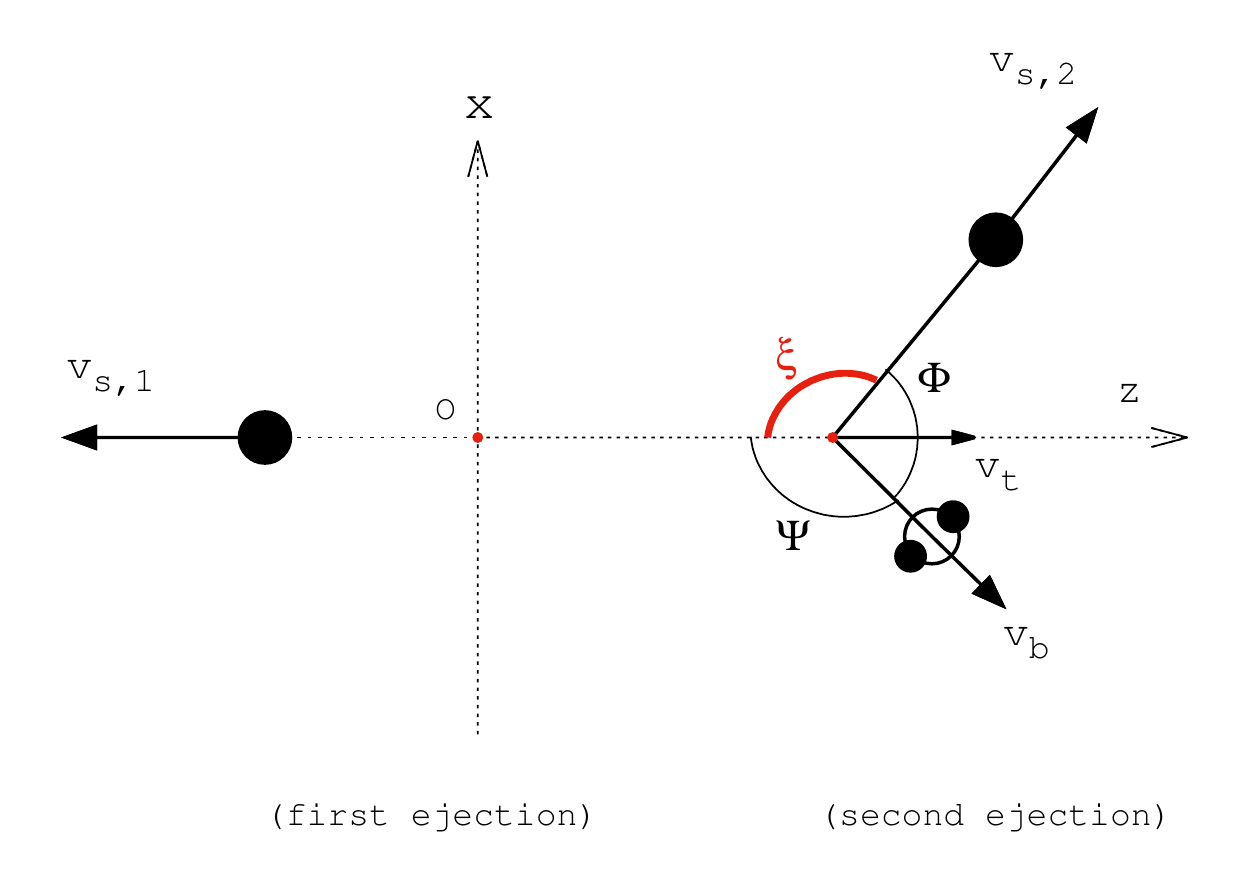}
	\caption{A schematic diagram on $x-z$ plane showing the two ejection events. At the first ejection event, a star of mass $m=m_{s,1}$ is ejected in -z direction with velocity $\textbf{\textit{v}}_{\rm s,1}$, leaving behind a system of mass $m=m_{\rm t}$ moving at $\textbf{\textit{v}}_{\rm t}=(m_{\rm s,1}/m_{\rm t})\textbf{\textit{v}}_{\rm s,1}$. At the second (last) ejection event, another star of mass $m=m_{\rm s,2}$ is ejected with velocity $\textbf{\textit{v}}_{\rm s,2}$ and the final binary of mass $m=m_{\rm b}$ is recoiled with velocity $(m_{\rm s,2}/m_{\rm b})\textbf{\textit{v}}_{\rm s,2}$. $\xi$ is the relative angle between the two single stars and $\Psi$ ($\Phi$) between the binary and the first (second) ejected star.}
	\label{fig:coordinate}
\end{figure}
\label{appendix:analyticformula_1}
When two binaries collide, a chaotic interaction ensues until two single stars 
are ejected sequentially, leaving behind a binary pair. 
After the first ejection 
event, a single star of mass $m_{\rm s,1}$ and the remaining system of 
mass $m_{\rm t}$ recede in opposite directions (in the CM frame of all four stars).  We 
assume a velocity of $\textbf{\textit{v}}_{\rm s,1}$ (so the momentum $\textbf{\textit{p}}_{\rm s,1}=m_{\rm s,1}\textbf{\textit{v}}_{\rm s,1}$) for the first ejected single star.
The remaining system (with three stars) recedes with velocity $\textbf{\textit{v}}_{\rm t}=(m_{\rm s,1}/m_{\rm t})\textbf{\textit{v}}_{\rm s,1}$.
At the time of the second ejection event, another single star is ejected with velocity 
$\textbf{\textit{v}}_{\rm s,2}$ (the momentum $\textbf{\textit{p}}_{\rm s,2}$) and leaves behind the binary with a recoil velocity of $\textbf{\textit{v}}_{\rm b}$ (the momentum $\textbf{\textit{p}}_{\rm b}$). We present a schematic diagram in 
Figure \ref{fig:coordinate} showing the two ejection 
events. 

Now, in the CM of the four stars, the three momentum vectors satisfy the following relation,
\begin{align}
\label{eq:conservationofp}
\textbf{\textit{p}}_{\rm s,1}+\textbf{\textit{p}}_{\rm s,2}+\textbf{\textit{p}}_{\rm b}=\textbf{0}.
\end{align}
Using Equation \ref{eq:conservationofp}, we find an expression for $v_{\rm b}$ as follows,
\begin{align}
\label{eq:solution_vbrest_overall}
v_{\rm b}&=\frac{1}{m_{\rm b}}\sqrt{p_{\rm s,1}^{2}+p_{\rm s,2}^{2}+2p_{\rm s,1}^{}p_{\rm s,2}^{}\cos\xi}
\end{align}
where $p_{\rm s,1}$, $p_{\rm s,2}$ and  $p_{\rm b}$ are the momenta of the first, second ejected stars and the binary, respectively, i.e., $p_{i}=m_{\rm i}v_{i}$. And $\xi$ is the 
relative angle between the two ejected single stars, defined as, 
\begin{align}
\label{eq:relativeangle}
\cos\xi
&=\frac{\textbf{\textit{v}}_{\rm s,1}\cdot\textbf{\textit{v}}_{\rm s,2}}{v_{\rm s,1}v_{\rm s,2}}.
\end{align}
We also provide the derivations for Equation~\ref{eq:solution_vbrest_overall} using their 
vector components in Appendix \ref{appendix:derivation1}. In this derivation, we consider the two ejection events in their respective center of mass frames separately. This allows us to find explicit relations for the relative angles between the two ejected stars in the different reference frames.  This approach could be useful for identifying the first ejected single star, when applying our method to observations of runaways where only one single star and an associated binary have been identified.

Using Equation \ref{eq:solution_vbrest_overall}, we can find an
expression for the relative angle between the first ejected star and
the binary as a function of $p_{\rm s,1}$, $p_{\rm s,2}$ and $\xi$. We
denote this angle as $\Psi$ (see Figure~\ref{fig:coordinate}). Then
the angle between the second ejected single star and the binary
(denoted by $\Phi$) is automatically determined since
$\xi+\Psi+\Phi=2\pi$.  Equation~\ref{eq:conservationofp} also gives
the following equation,
\begin{align}
\label{eq:Psi_1}
p_{\rm s,2}=\sqrt{p_{\rm s,1}^{2}+p_{\rm b}^{2}+2p_{\rm s,1}^{}p_{\rm b}^{}\cos\Psi}\,.
\end{align}
Plugging Equation~\ref{eq:solution_vbrest_overall} into Equation~\ref{eq:Psi_1} yields 
\begin{align}
\label{eq:Psi_2}
\cos\Psi=-\frac{p_{\rm s,1}+p_{\rm s,2}\cos\xi}{\sqrt{p_{\rm s,1}^{2}+p_{\rm s,2}^{2}+2p_{\rm s,1}^{}p_{\rm s,2}^{}\cos\xi}}\,.
\end{align}
Therefore, given the velocities and the masses of the two ejected stars,
with Equations \ref{eq:solution_vbrest_overall} and \ref{eq:Psi_2} we
can determine the momenta of the binary (and hence the binary speed,
given the binary mass) and the relative angles between the binary and
each of the two ejected stars ($\Psi$ and $\Phi$).

There are several points worth emphasizing here.
\begin{enumerate}
	\item We assume only conservation of momentum to derive Equations \ref{eq:solution_vbrest_overall}, \ref{eq:relativeangle} and \ref{eq:Psi_2} 
	without additional conditions. 
	
	\item These formulas can be applied independently of the details of the initial 
	binary-binary encounter (i.e., head-on, or with non-zero impact 
	parameter) since we consider only the velocities at the time of the last ejection event.
	The only condition is that two binaries interact to produce a 2+1+1 outcome. As an example, 
	in Section \ref{results_backgroundpotential}, we apply our analytic formula to 
	the 2+1+1 outcome from scatterings of two binaries with equal/unequal energies.

\item In Equation \ref{eq:solution_vbrest_overall}, the two ejected
  stars are distinguished in terms of which star is ejected
  first. However, Equation \ref{eq:solution_vbrest_overall} is
  symmetric upon exchanging the two single star momenta.  Clearly,
  $v_{\rm b}$ at $\xi=0$ depends only on the total sum of the momenta,
  whereas $v_{\rm b}$ at $\xi=\pi$ depends on the difference between
  the two momenta.  Hence the order of ejection events is irrelevant in
  terms of estimating the final velocity of the binary (see the green
  line in the {left} panel of Figure \ref{fig:analyticformula}).  In
  other words, the subscripts 1 and 2 do not necessarily correspond to
  the order of ejection events.  This is of critical value for
  applying our method to observational data, since in practice it is
  very hard to distinguish the order of ejection events.

	\item We can generalize Equation
          \ref{eq:solution_vbrest_overall} to a stellar system going
          through any number of ejection events. Consider a stellar
          system consisting of $N_{\rm tot}$ stars. In the CM of the
          whole system, the speed of a substellar group left behind
          after the $N$th ejection event can be expressed as
	\begin{align}
	\label{eq:solution_vbrest_anyN}
	v&=\frac{1}{m}\sqrt{\sum_{i}^{N}p_{i}^{2}+2\sum_{i>j}^{N}\textbf{\textit{p}}_{i}^{}\cdot 
\textbf{\textit{p}}_{j}^{}}\,,
	\end{align}
	where $p_{i}$ represents the momentum of the $i_{\rm th}$ ejected object. 

	\item If scatterings between two binaries occur in a background 
	potential (eg., a star cluster), an escape condition can be derived  
	by comparing the escape 
	velocity of the potential with $v_{\rm b}$, as estimated from our formula.  
	We will return to this in the next section.

\end{enumerate}

\subsection{Analytic formula: Homogeneous background potential}
\label{sec:analyticformula_bg}
In this section we generalize our analytic formulation to include a homogeneous 
background potential.  

When we observe at time 
$t=t_{\rm obs}$ the velocities of two field stars (outside the potential) suspected of being a 
causally-related pair of runaway stars, 
we can trace their velocities backward in time. In other words, we can estimate the velocities at any 
time $t~ (<t_{\rm obs})$ post-interaction (i.e., after the second ejection event). 
Starting with the velocities of the ejected single stars, 
$\textbf{\textit{v}}_{\rm s,1}(t=t_{\rm obs})$ 
and $\textbf{\textit{v}}_{\rm s,2}(t=t_{\rm obs})$ as observed outside of the potential 
($r>r_{\rm bg}$) at $t=t_{\rm obs}$, we integrate 
the equations of motion backward in time in order to estimate the stars' initial ejection 
velocities inside 
the potential. Then, we can use equation \ref{eq:solution_vbrest_overall} 
to calculate the velocity of the final binary at the time of the last ejection event. 
Integrating the equation of motion for the binary forward 
in time gives us the speed of the binary $v_{\rm b}(t=t_{\rm obs})$ at $t=t_{\rm obs}$.

We consider a spherically symmetric potential with constant density $\rho$ and 
outer boundary $r_{\rm bg}$. For a given total background mass $M_{\rm bg}$, 
the outer boundary is automatically set. The mass enclosed within a spherical volume of radius $r$ can be written, 
\begin{align}
M_{\rm en,bg}(r)&=
\begin{cases}
\frac{4\uppi}{3}\rho r^{3} \hfill \hspace{0.8in} r \leq r_{\rm bg}\,;\\
\frac{4\uppi}{3}\rho r_{\rm bg}^{3}=M_{\rm bg} \hfill r>r_{\rm bg}\,.
\end{cases}
\label{eq:engasmass}
\end{align}

The gravitational force imparted by the background mass on a given star particle 
at $r$ is given by: 
\begin{align}
\label{eq:backforce_1}
\textbf{\textit{f}}_{\rm bg}(r)&=-\frac{Gm M_{\rm en}(r)}{r^{3}}\textbf{\textit{r}}\nonumber\\ 
&=
\begin{cases}
-\frac{4}{3}\uppi G m\rho \;\textbf{\textit{r}} \hfill \hspace{0.3in} r \leq r_{\rm bg}\,;\\
-\frac{4}{3}\uppi G m\rho \Big(\frac{r_{\rm bg}}{r}\Big)^{3}\textbf{\textit{r}} \hfill \hspace{0.3in} r > r_{\rm bg}\,,\\
\end{cases}
\end{align}
where $m$ is the mass of the star and $\textbf{\textit{r}}$ is the
vector pointing from the system CM to the star. The term
$\sqrt{(4\uppi G\rho)/3}$ corresponds to the frequency of the
resulting harmonic motion, denoted by $w$. Accordingly, the background
potential has the following form:
\begin{align}
\label{eq:backpotential_1}
V_{\rm bg}(r)&=
\begin{cases}
\frac{2}{3}\uppi G m\rho (r^{2}-3r_{\rm bg}^{2})\hfill \hspace{0.3in} r \leq r_{\rm bg}\,;\\
-\frac{GmM_{\rm bg}}{r}=-\frac{4}{3}\uppi G m\rho \frac{r_{\rm bg}^{3}}{r} \hfill \hspace{0.3in} r > r_{\rm bg}\,,\\
\end{cases}
\end{align}
and the escape velocity of the potential $v_{\rm esc}$ at $r=r_{\rm bg}$ is written as,
\begin{equation}
v_{\rm esc}=\sqrt{\frac{2GM_{\rm bg}}{r_{\rm bg}}}=\sqrt{2}r_{\rm bg}w\label{eq:escapevelocity}.
\end{equation}
This is the same background potential as the one adopted by \citet{Ryu+2017}. See \citet{Ryu+2017} for more details.

Given the adopted background potential, we solve for the equations of
motion of the stars inside and outside the potential individually.  To
do this, we make the following assumptions: At each ejection event,
the ejected star reaches a sufficiently high velocity to escape the
underlying potential well (without becoming trapped and undergoing
subsequent oscillations within the background potential). Therefore the
angles between stars do not significantly change while they escape,
which we will show in Figure \ref{fig:angle}.

\subsubsection{Inside the potential, $r\leq r_{\rm bg}$}
Given the radial components of the velocities of stars, we can then solve the equations of motion in 1-D.
For $r\leq r_{\rm bg}$, the solution for Equation \ref{eq:backforce_1} is,
\begin{align}
	r(t)&=\frac{v_{\rm ej}}{w}\sin w(t-t_{\rm ej})+r_{\rm ej}.\label{eq:r_harmonic}
\end{align}
The speed is
\begin{align}
	v(t)&=v_{\rm ej}\cos w(t-t_{\rm ej})\,,\label{eq:v_harmonic}
\end{align}
where $v_{\rm ej}$ and $r_{\rm ej}$ denote the speed and radial distance from the system CM (or the center of the background potential) at the time of an ejection event $t=t_{\rm ej}$.  

When a star crosses the outer boundary $r_{\rm bg}$ with speed $v_{\rm bg}$ 
at time $t=t_{\rm bg}$, Equations \ref{eq:r_harmonic} and \ref{eq:v_harmonic} yield the following 
relations,
\begin{align}
	[(r_{\rm bg}-r_{\rm ej}) w]^{2}+v_{\rm bg}^{2}=v_{\rm ej}^{2},
\end{align}
or, 
\begin{align}
	\label{eq:v_ej_v_bg}
	v_{\rm ej}=\sqrt{	[(r_{\rm bg}-r_{\rm ej}) w]^{2}+v_{\rm bg}^{2}}=\sqrt{\frac{1}{2}v_{\rm esc}^{2}\Big[1-\frac{\sqrt{2}r_{\rm ej}w}{v_{\rm esc}}\Big]^{2}+v_{\rm bg}^{2}}.
\end{align}

It is useful to consider how much the speeds of the stars change inside the potential 
from the moment of ejection until the time of escape. For this, we introduce a fractional 
difference between $v_{\rm ej}$ and $v_{\rm bg}$,
\begin{align}
	\label{eq:v_error}
	\frac{\Delta v}{v_{\rm ej}}&=\Big|\frac{v_{\rm ej}-v_{\rm bg}}{v_{\rm ej}}\Big|=\Big|\frac{v_{\rm ej}-\sqrt{v_{\rm ej}^{2}-\frac{1}{2}v_{\rm esc}^{2}\Big[1-\frac{\sqrt{2}r_{\rm ej}w}{v_{\rm esc}}\Big]^{2}}}{v_{\rm ej}}\Big|.
	\end{align}
Especially for stars ejected near centre ($r_{\rm ej}w\ll v_{\rm esc}$) at 
sufficiently high velocities to escape,
\begin{align}
\label{eq:v_error_fast}
\frac{\Delta v}{v_{\rm ej}}\sim\Big|1-\sqrt{1-\frac{1}{2}\Big(\frac{v_{\rm esc}}{v_{\rm ej}}\Big)^{2}}\Big|\sim\Big(\frac{v_{\rm esc}}{v_{\rm ej}}\Big)^{2}+O(\frac{v_{\rm esc}^{4}}{v_{\rm ej}^{4}}).
\end{align}
This implies that: 1) for a given $v_{\rm ej}$ and $r_{\rm ej}$, $v_{\rm esc}$ is a good diagnostic 
(rather than either $M_{\rm bg}$ or $\rho$ individually) to gauge the overall 
decrease in the speeds of the escaping stars;  2) the fractional difference 
($\Delta v/v_{\rm ej}$) or absolute difference ($\Delta v$) in $v_{\rm ej}$ depends on $v_{\rm ej}$ itself.
\subsubsection{Outside the potential, $r>r_{\rm bg}$}
For $r>r_{\rm bg}$, we apply conservation of energy instead of directly solving 
the second-order differential equation in Equation \ref{eq:backforce_1}. At 
$r=r_{\rm bg}$, the total specific energy $e_{\rm bg}$ is defined as,

\begin{align}
	\label{eq:e_bg}
	e_{\rm bg}=\frac{1}{2}v_{\rm bg}^{2}-\frac{GM_{\rm bg}}{r_{\rm bg}},
\end{align}
Similarly, at $r>r_{\rm bg}$ and $t>t_{\rm bg}$, 
\begin{align}
	\label{eq:e_t}
	e(t)=\frac{1}{2}v(t)^{2}-\frac{GM_{\rm bg}}{r(t)}\,.
\end{align}
Equating Equations \ref{eq:e_bg} and \ref{eq:e_t} and rearranging terms,
\begin{align}
	\label{eq:e_Dequation}
	v(t)=\frac{dr(t)}{dt}=\sqrt{2e_{\rm bg}}\sqrt{1+\frac{GM_{\rm bg}}{e_{\rm bg}}\frac{1}{r}}.
\end{align}
Defining $GM_{\rm bg}/e_{\rm bg}\equiv{R}$ and integrating both sides of the above gives, 
\begin{align}
	\label{eq:solution}
	\int_{t_{\rm bg}}^{t}dt \sqrt{2e_{\rm bg}}&=\int_{r_{\rm bg}}^{r}\frac{1}{\sqrt{1+{R}\frac{1}{r}}}dr\nonumber\\
	\sqrt{2e_{\rm bg}}(t-t_{\rm bg})&=\sqrt{{R}~r+r^{2}}-\frac{1}{2}{R} \log\Big[1+2\Big(1+\sqrt{1+\frac{{R}}{r}}\Big)\frac{r}{{R}}\Big]\Bigg|^{r}_{r_{\rm bg}}\,.
\end{align}
We are unable to solve further for a simple expression for $r(t)$. However, 
we note that the second term (with $\log$) on the right hand side can be 
neglected when $r/{R}$ is sufficiently large.

It useful to further consider the range in ${R}/r_{\rm bg}$ and its physical 
meaning. $r_{\rm bg}/{R}$ measures the relative difference between the 
absolute magnitude of the kinetic energy and that of the potential energy 
at the outer boundary of the potential. Assuming $v_{\rm bg}>v_{\rm esc}$, 
we can express ${R}$ in terms of $r_{\rm bg}$ and $v_{\rm esc}$,
\begin{align}
	\frac{r_{\rm bg}}{{R}}&=\frac{\frac{1}{2}v_{\rm bg}^{2}-\frac{GM_{\rm bg}}{r_{\rm bg}}}{\frac{GM_{\rm bg}}{r_{\rm bg}}}=\Big(\frac{r_{\rm bg}}{2GM_{\rm bg}}\Big)v_{\rm bg}^{2}-1\nonumber\\
	&=\Big(\frac{v_{\rm bg}}{v_{\rm esc}}\Big)^{2}-1\simeq\Big(\frac{v_{\rm bg}}{v_{\rm esc}}\Big)^{2}\simeq10^{2}\Big(\frac{v_{\rm bg}}{70\km/\s}\Big)^{2}\Big(\frac{v_{\rm esc}}{10\km/\s}\Big)^{-2}\,,
\end{align}
where $v_{\rm bg}\sim 60-70\km\s^{-1}$ are typical speeds of runaway stars
formed during binary-binary numerical scattering experiments 
(\citealt{RLP2017}, ``simulation paper'' in the following).
Typically, $r_{\rm bg},~r(t)> {R}$. Such large values for
$r_{\rm bg}/{R}$ imply that the kinetic energy is dominant over the
background potential energy when stars are escaping from the
potential. Moreover, this may mean that the decrease in speed outside
of the potential is insignificant as long as $v_{\rm esc}$ is not too
high (see also the $v_{\rm ej}$ dependence in Equation
\ref{eq:v_error}).

		\begin{figure}
			\centering
			\includegraphics[width=8.7cm]{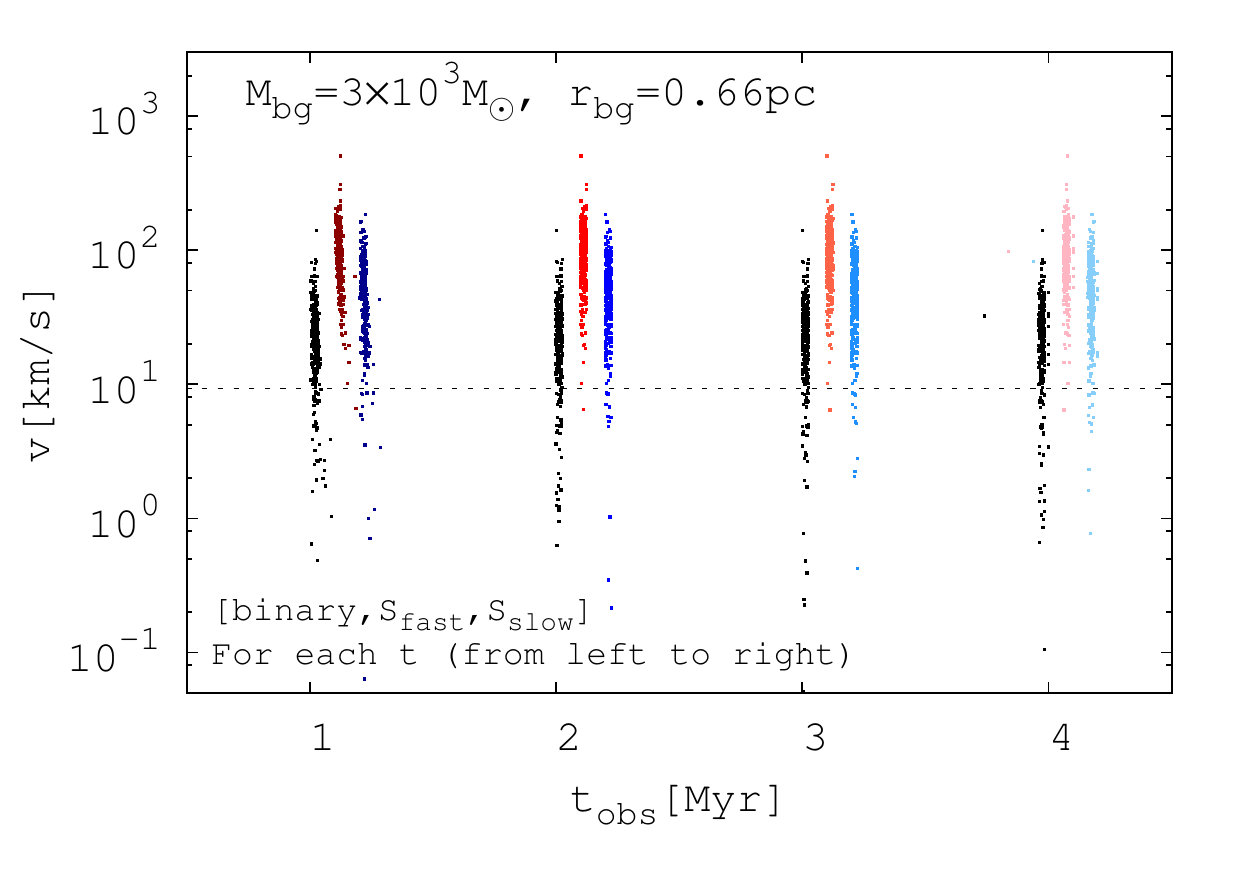}
			\includegraphics[width=8.7cm]{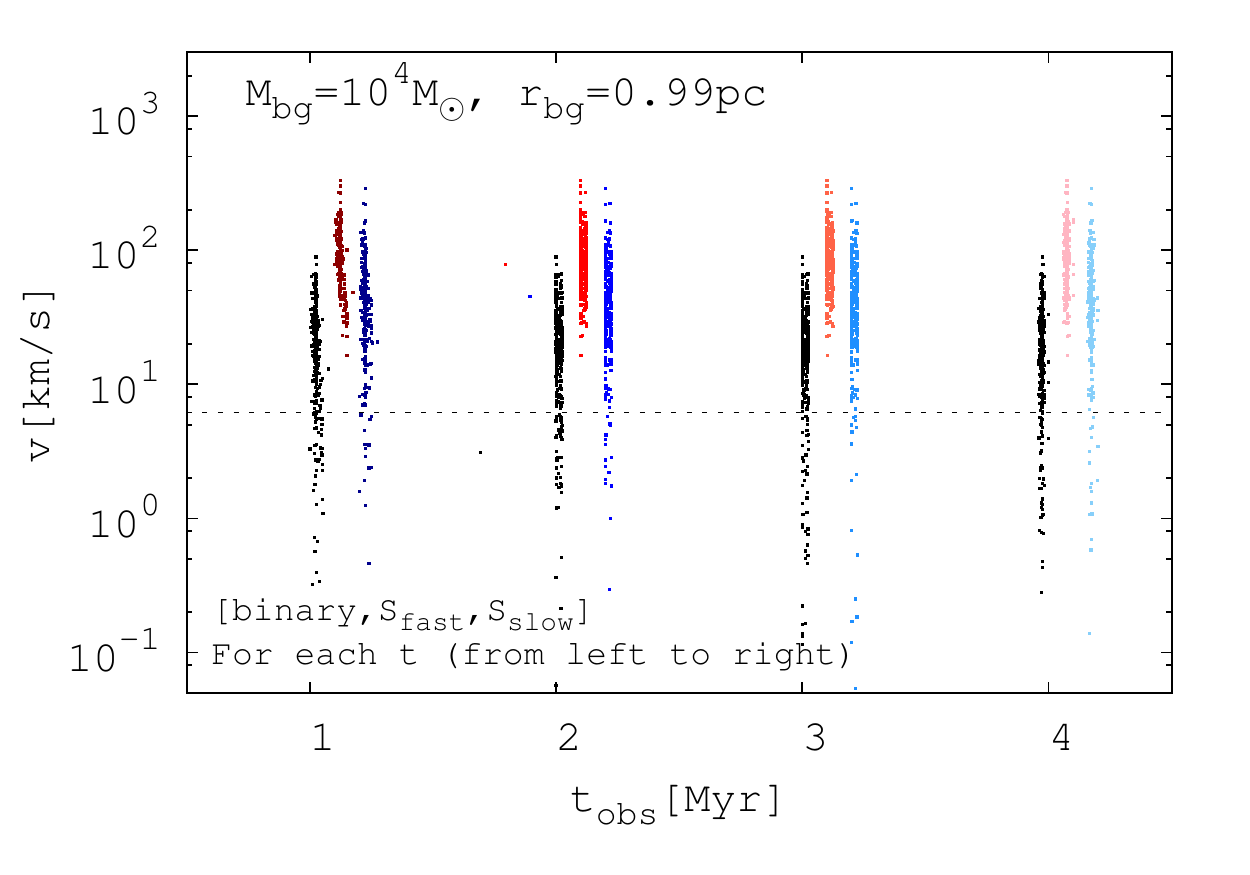}
			\caption{The time evolution of the velocities of the binaries, $S_{\rm fast}$ 
				(the more rapidly moving single star for a given run) and $S_{\rm slow}$ 
				(the more slowly moving single star for a given run) formed during binary-binary 
				encounters in scattering experiments, at $t=$1, 2, 3 and 4$\Myr$ 
				for two background potential models. $M_{\rm bg}$ and $r_{\rm bg}$ 
				are given in the plots. The horizontal dotted line indicates the escape velocity. 
				See the simulation paper for more details regarding the scattering experiments. 
				At each $t$, three vertically distributed groups of dots correspond to 
				(from left to right) the binary (black dots), $S_{\rm fast}$ (red dots) and 
				$S_{\rm slow}$ (blue dots), respectively. To avoid overcrowding, 
				we offset the distributions for $S_{\rm fast}$ and $S_{\rm slow}$ 
				by $+0.1\Myr$ and $+0.2\Myr$ for a given $t$ (i.e., shift to the right). 
				This shows that the velocities of rapidly moving stars with 
				$v/v_{\rm esc}\geq 5-7$ remain roughly the same after escaping the potential.}
			\label{fig:velocity_propagation}
		\end{figure}

Given a high value for $r/{R}$ (i.e., for rapidly moving stars, hence a
suitable value for runaway stars), we can ignore the log-term in
Equation \eqref{eq:solution}. Then we can find a relatively simple
expression for $r(t)$ at $t\geq t_{\rm bg}$, which is,
\begin{align}
	\label{eq:r_out}
	r(t)=\frac{1}{2}{R}\Bigg[-1\pm \sqrt{
		1+4\Big[\sqrt{\frac{r_{\rm bg}}{{R}}+\Big(\frac{r_{\rm bg}}{{R}}\Big)^{2}}+\sqrt{\frac{2GM_{\rm bg}}{{R}^{3}}}(t-t_{\rm bg})    \Big]^{2}
	}\Bigg]
\end{align}
where we must choose the plus sign since $r(t=t_{\rm bg})=r_{\rm bg}$. 
Differentiating with respect to $t$ gives,
\begin{align}
	\label{eq:v_out}
	v(t)=\frac{dr(t)}{dt}=\frac{2\sqrt{\frac{2GM_{\rm bg}}{{R}}}\Big[\sqrt{\frac{r_{\rm bg}}{{R}}+\Big(\frac{r_{\rm bg}}{{R}}\Big)^{2}}+\sqrt{\frac{2GM_{\rm bg}}{{R}^{3}}}(t-t_{\rm bg})  \Big]}
	{\sqrt{
			1+4\Big[\sqrt{\frac{r_{\rm bg}}{{R}}+\Big(\frac{r_{\rm bg}}{{R}}\Big)^{2}}+\sqrt{\frac{2GM_{\rm bg}}{{R}^{3}}}(t-t_{\rm bg})    \Big]^{2}
		}}
	\end{align}
	Since we made an approximation in Equation \ref{eq:solution} in order to derive the above, 
	we must also ensure that $v(t)$ is continuous at $t=t_{\rm bg}$ and well-defined at 
	$t\rightarrow\infty$.

	\begin{enumerate}
		\item $v(t)$ at $t=t_{\rm bg}$.
		
		\begin{align}
			v(t=t_{\rm bg})&=\frac{2\sqrt{\frac{2GM_{\rm bg}}{{R}}}\sqrt{\frac{r_{\rm bg}}{{R}}+\Big(\frac{r_{\rm bg}}{{R}}\Big)^{2}}}
			{\sqrt{
					1+4\Big[\sqrt{\frac{r_{\rm bg}}{{R}}+\Big(\frac{r_{\rm bg}}{{R}}\Big)^{2}}\Big]^{2}
				}}=\frac{2v_{\rm esc}\frac{r_{\rm bg}}{{R}}\sqrt{1+\frac{r_{\rm bg}}{{R}}}}{1+2\frac{r_{\rm bg}}{{R}}}\nonumber\\
				&=2v_{\rm bg}\frac{\Big(\frac{v_{\rm bg}}{v_{\rm esc}}\Big)^{2}-1}{2\Big(\frac{v_{\rm bg}}{v_{\rm esc}}\Big)^{2}-1}\simeq v_{\rm bg}
			\end{align}
			for $v_{\rm bg}/v_{\rm esc}\gg1$, which is the same assumption that we made to 
			solve Equation \ref{eq:solution}.
			
			\item $v(t)$ at $t\rightarrow \infty$.
			
			\begin{align}
				v(t\rightarrow \infty)&\simeq\frac{\frac{4GM_{\rm bg}}{{R}^{2}}t}{\sqrt{1+8\frac{GM_{\rm bg}}{{R}^{3}}t^{2}}}\simeq\sqrt{\frac{2GM_{\rm bg}}{{R}}}=\sqrt{2e_{\rm bg}}
			\end{align}
			which is the velocity at $t=\infty$ (see equation \ref{eq:e_t}).
			
		\end{enumerate}
		
		We can simplify the solution further for stars moving
                at high speeds (i.e., large $v_{\rm bg}$).  We can
                approximate their speeds as being constant since
                $v(t)$ is a monotonically decreasing function with the
                same two boundary values ($v_{\rm bg}$ at $t=t_{\rm
                  bg}$ and $t\rightarrow\infty$). Note that for such
                high speeds, $\sqrt{2e_{\rm bg}}\simeq v_{\rm
                  bg}$. This is confirmed in the simulations of our
                simulation paper. Figure
                \ref{fig:velocity_propagation} shows the time
                evolution of the speeds of the binary and the two
                single stars formed during binary-binary encounters 
                (Model 1 and Model 2 of the simulation paper) at
                $t=$1, 2, 3 and 4$\Myr$.  We differentiate the two
                single stars by their speeds: $S_{\rm fast}$ ($S_{\rm
                  slow}$) refers to the single star with the higher
                (slower) speed, for a given run. At each $t$, the
                three vertically distributed dots correspond to (from
                left to right) the binary (black dots), $S_{\rm fast}$
                (red dots) and $S_{\rm slow}$ (blue dots),
                respectively.  The horizontal dotted line indicates
                the escape velocity. To avoid overcrowding, we offset
                the distributions for $S_{\rm fast}$ and $S_{\rm
                  slow}$ by $+0.1\Myr$ and $+0.2\Myr$ for a given $t$
                (i.e., shift to the right). In the simulations, more
                than 95\% of all outcomes escape from the background
                potential before $t=1\Myr$. Figure
                \ref{fig:velocity_propagation} shows that the speeds
                of rapidly-moving stars ($v/v_{\rm esc}\geq 5-7$ for
                both single stars) barely change once they have
                escaped from the potential. We also refer to the
                cumulative velocity distributions for the two single
                stars and the binary shown in Appendix A of the
                simulation paper.

		However, this approximation may lose its validity for
                stars whose speeds at $r=r_{\rm bg}$ are not
                sufficiently large compared to $v_{\rm esc}$ (i.e.,
                typically more massive escapers). In our numerical
                experiments (simulation paper), we find that for most
                final binaries, $v_{\rm bg}/v_{\rm esc}\simeq
                1-3$. Therefore, in the following (Section
                \ref{sec:results}), we will use Equation
                \ref{eq:v_out} to estimate the final speeds of the
                binaries along with a constant $v(t)$ for ejected
                single stars moving at high speeds.  Using the
                following equations of motions for stars at $t$: for
                $r(t)<r_{\rm bg}$,
		\begin{align}
		\label{eq:EOM_in}
		v(t)&=v_{\rm ej}\cos w(t-t_{\rm ej})\hfill \hspace{1.46in} \text{(Equation \ref{eq:r_harmonic})},\nonumber\\
		r(t)&=\frac{v_{\rm ej}}{w}\sin w(t-t_{\rm ej})+r_{\rm ej}\hfill \hspace{1.2in} \text{(Equation \ref{eq:v_harmonic})}\nonumber,
		\end{align}
		and for $r(t)\geq r_{\rm bg}$ (Equation \ref{eq:r_out})
			\begin{numcases}{v(t)=}
			v_{\rm bg}\hfill \hspace{1.85in}  \text{for single stars};\nonumber\\
			\text{equation \ref{eq:v_out}} \hspace{1.44in}  \text{for binaries}\nonumber
			\end{numcases}
			where $v_{\rm ej}\cos w~t_{\rm bg}=v_{\rm bg}$, we can calculate $v_{\rm ej}$ of the two ejected stars 
		given their observed values.
		After getting $v_{\rm b}(t=t_{\rm ej})$ from Equation \ref{eq:solution_vbrest_overall}, we can calculate what we would 
		observe for the binary at time $t=t_{\rm obs}$, i.e., 
		$v_{\rm b}(t=t_{\rm obs})=v_{\rm b,obs}$. In order to do this, we first 
		use Equation \ref{eq:v_ej_v_bg} to find $v_{\rm bg}$, then Equation 
		\ref{eq:v_out} to find $v_{\rm b, obs}$.
		
		Finally, we summarize the above procedure as follows. For a background potential with an escape 
		velocity not too high compared to the typical ejection velocities of the stars 
		(a valid assumption for runaway stars), 
		when we observe two runaway stars with speeds $v_{\rm s,1,obs}$ and 
		$v_{\rm s,2,obs}$ with a relative angle $\xi$ at $t=t_{\rm obs}$
		\footnote{If their radial distances $r_{\rm s,1,obs}$ and $r_{\rm s,2,obs}$ 
			(in the CM of four stars) are known, $t_{\rm obs} (\gg t_{\rm bg})$ 
			can be approximated for the two fast runaway stars as,
			\begin{align} 
			t_{\rm obs}\simeq r_{\rm s,1,obs}/v_{\rm s,1,obs}\simeq r_{\rm s,2,obs}/v_{\rm s,2,obs}.
			\end{align}
			}, the velocity of the binary at $r=r_{\rm bg}$ and $t=t_{\rm bg}~(<t_{\rm obs})$ left behind by the two runaway stars can be estimated as, 
					\begin{multline}
					\label{eq:solution_vbrest_overall_bg}
					v_{\rm b,bg}(v_{\rm s,1,obs},~ v_{\rm s,2,obs},~m_{\rm s,1},~m_{\rm s,2},~m_{\rm b})\\
					=\sqrt{\frac{1}{m_{\rm b}^{2}}\Big[\mathcal{P}_{\rm s,1}^{2}+\mathcal{P}_{\rm s,2}^{2}+2\mathcal{P}_{\rm s,1}^{}\mathcal{P}_{\rm s,2}^{}\cos\xi\Big]-\frac{1}{2}v_{\rm esc}^{2}\Big[1-\frac{\sqrt{2}r_{\rm ej,2}w}{v_{\rm esc}}\Big]^{2}}
					\end{multline}
					where 
					\begin{align}
					\mathcal{P}_{\rm s,1}=m_{\rm s,1}\sqrt{\frac{1}{2}v_{\rm esc}^{2}\Big[1-\frac{\sqrt{2}r_{\rm ej,1}w}{v_{\rm esc}}\Big]^{2}+(v_{\rm s,1,obs})^{2}},\label{eq:vs1_ej_bg}\\
					\mathcal{P}_{\rm s,2}=m_{\rm s,2}\sqrt{\frac{1}{2}v_{\rm esc}^{2}\Big[1-\frac{\sqrt{2}r_{\rm ej,2}w}{v_{\rm esc}}\Big]^{2}+(v_{\rm s,2,obs})^{2}}.\label{eq:vs2_ej_bg}
					\end{align}
				where $r_{\rm
                                            ej,1}$ ($r_{\rm ej,2}$)
                                          represents the radial
                                          distance between the CM of
                                          the background potential and
                                          the CM of the system made of the
                                          ejected star and the
                                          remaining $N=3$ ($N=2$)
                                          stars at the first (second)
                                          ejection event.

		Finally, the speed and radial distance of the final binary 
		outside the potential ($r\geq r_{\rm bg}$) at any $t~(\geq t_{\rm bg})$ are described as,

		\begin{align}
			v_{\rm b}(t)=\frac{2\sqrt{\frac{2GM_{\rm bg}}{\mathfrak{R}}}\Big[\sqrt{\frac{r_{\rm bg}}{\mathfrak{R}}+\Big(\frac{r_{\rm bg}}{\mathfrak{R}}\Big)^{2}}+\sqrt{\frac{2GM_{\rm bg}}{\mathfrak{R}^{3}}}(t-t_{\rm bg})\Big]}
			{\sqrt{
					1+4\Big[\sqrt{\frac{r_{\rm bg}}{\mathfrak{R}}+\Big(\frac{r_{\rm bg}}{\mathfrak{R}}\Big)^{2}}+\sqrt{\frac{2GM_{\rm bg}}{\mathfrak{R}^{3}}}(t-t_{\rm bg})   \Big]^{2}
				}}\label{eq:vbinaryrest_bg}
			\end{align}
			\begin{multline}
				r_{\rm b}(t)\\
				=\frac{1}{2}\mathfrak{R}\Bigg[-1+\sqrt{
					1+4\Big[\sqrt{\frac{r_{\rm bg}}{\mathfrak{R}}+\Big(\frac{r_{\rm bg}}{\mathfrak{R}}\Big)^{2}}+\sqrt{\frac{2GM_{\rm bg}}{\mathfrak{R}^{3}}}(t-t_{\rm bg}) \Big]^{2}}\Bigg]+r_{\rm bg}\label{eq:rbinaryrest_bg}
			\end{multline}
			where $\mathfrak{R}=r_{\rm bg}/[(v_{\rm b,bg}/v_{\rm esc})^{2}-1]$
			\footnote{As a sanity check, we explore two extreme limits, \ref{sanitycheck1} $t\rightarrow\infty$ and \ref{sanitycheck2} $\rho,~M_{\rm bg}\rightarrow0$ .
				\begin{enumerate}
					\item\label{sanitycheck1} $t\rightarrow\infty$
					\begin{align}
					v_{\rm b}(t>t_{\rm bg})\sim\sqrt{\frac{2GM_{\rm bg}}{\mathfrak{R}}}\sim \sqrt{v_{\rm b,bg}^{2}-v_{\rm esc}^{2}}\sim\begin{cases}
					v_{\rm bg} \hspace{0.1in} \text{for} \,\,\,\ v_{\rm b,bg}\gg v_{\rm esc};\\
					0 \hspace{0.2in} \text{for} \,\,\,\ v_{\rm b,bg}\simeq ~v_{\rm esc}.
					\end{cases}
					\end{align}
Both of the two limiting values are consistent with the speeds at spatial infinity for an $\sim(1/r)$ gravitational potential. 
				\item \label{sanitycheck2}$\rho,~M_{\rm bg}\rightarrow0$
				
				In this limit,
				\begin{align}
\mathfrak{R}&=\frac{r_{\rm bg}v_{\rm esc}}{v_{\rm b,bg}^{2}-v_{\rm esc}^{2}}\sim 0,\\
\sqrt{\frac{2GM_{\rm bg}}{\mathfrak{R}}}&=\sqrt{\frac{2GM_{\rm bg}}{r_{\rm bg}}\Big[\frac{v_{\rm b,bg}}{v_{\rm esc}}-1\Big]^{2}}=\sqrt{v_{\rm b}^{2}-v_{\rm esc}^{2}}\sim v_{\rm b}.
				\end{align}
Now, as the terms with $\mathfrak{R}^{k}$ ($k>0$) converge toward zero, 
				\begin{align}
				r_{\rm b}(t>t_{\rm ej})\sim\sqrt{\frac{2GM_{\rm bg}}{\mathfrak{R}}}t\sim v_{\rm b}t
				\end{align}
which is expected for ejected stars after being completely isolated from their parent stellar group.									
				\end{enumerate}
}.
			
			Given the equations above, we find that, when the following relation is satisfied,
			\begin{align}
				\label{eq:esc_condition}
				\sqrt{\mathcal{P}_{\rm s,1}^{2}+\mathcal{P}_{\rm s,2}^{2}+2\mathcal{P}_{\rm s,1}^{}\mathcal{P}_{\rm s,2}^{}\cos\xi}<m_{\rm b}v_{\rm esc}\sqrt{1+\frac{1}{2}\Big[1-\frac{\sqrt{2}r_{\rm ej,2}w}{v_{\rm esc}}\Big]^{2}}\,,
			\end{align}
			it is more likely that the final binary will 
			have remained bound to the potential. Using Equation \ref{eq:esc_condition}, 
			for two runaway 
			stars and a binary which are believed to form from a 2+2 encounter, 
			we can estimate a rough lower limit for $v_{\rm esc}$ 
			of the parent cluster at the moment of the last ejection event. We will apply the equation above for the Trapezium cluster in Section \ref{sec:Discussionandsummary}.
			
\begin{figure*}
	\centering
	\includegraphics[width=8.7cm]{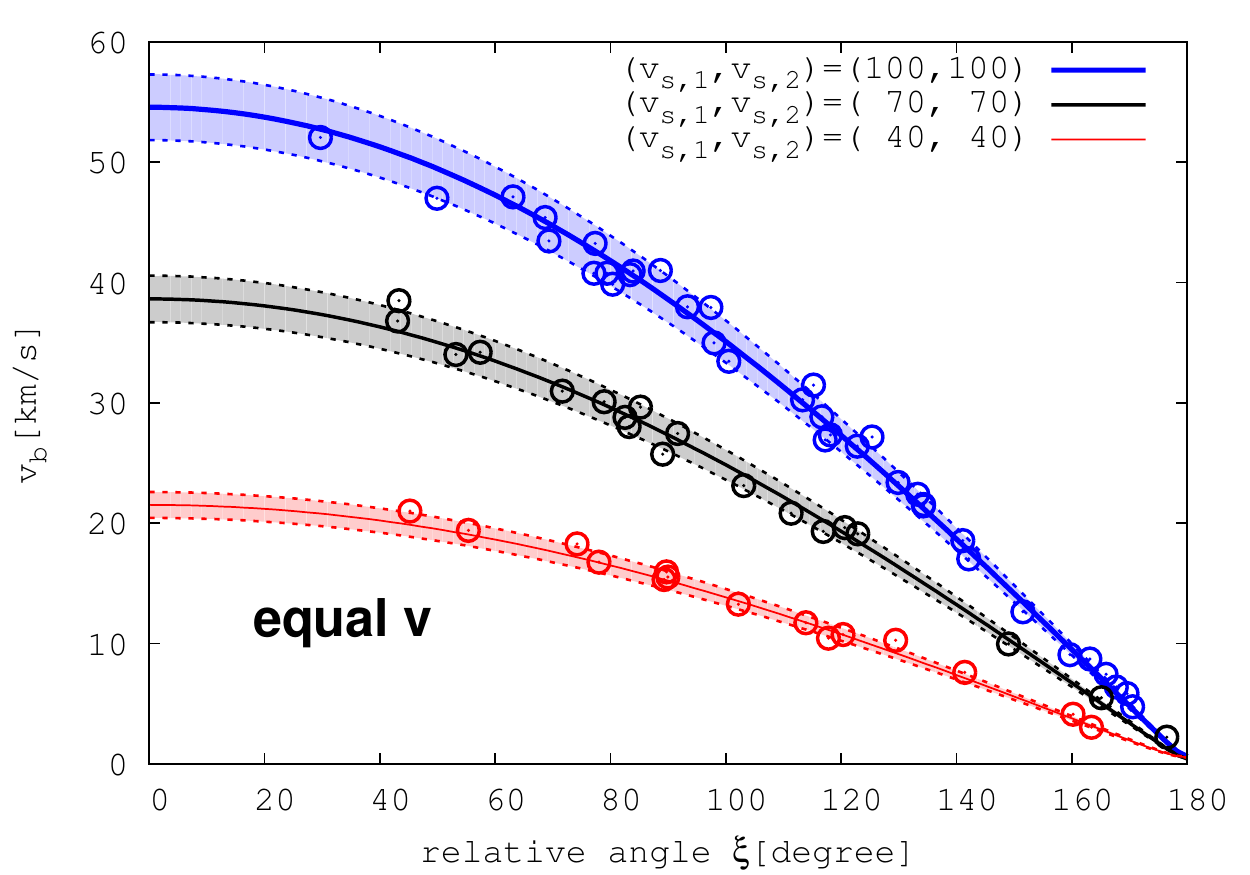}
	\includegraphics[width=8.7cm]{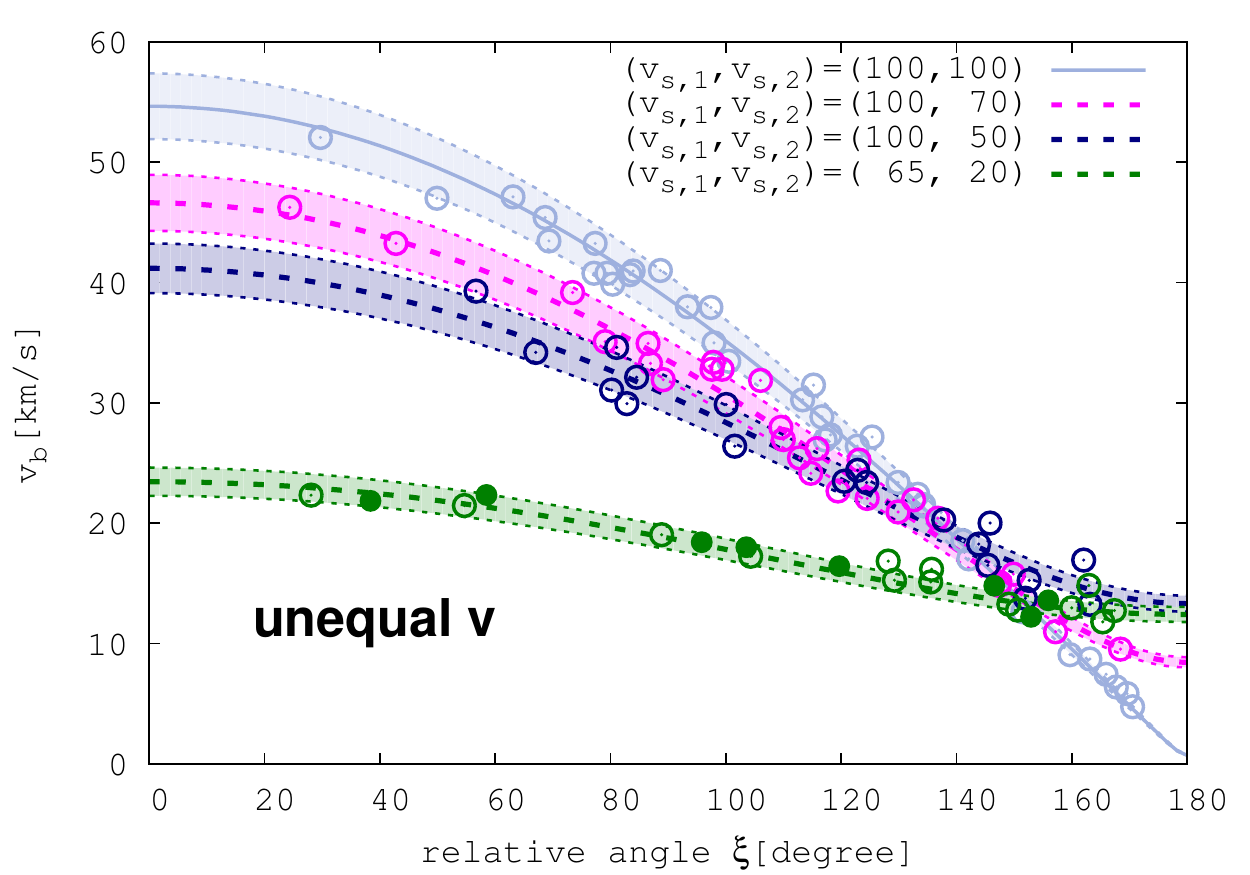}
	\includegraphics[width=8.7cm]{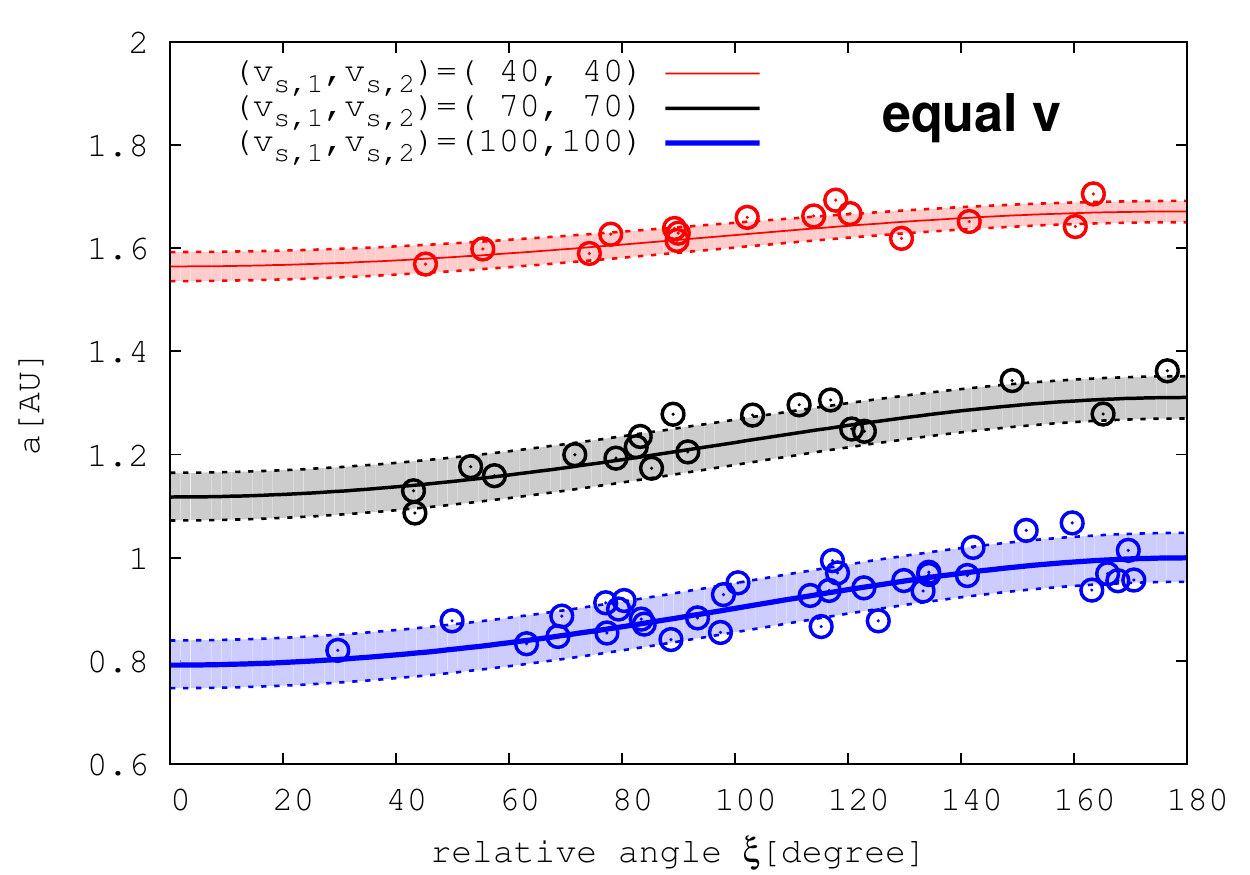}
	\includegraphics[width=8.7cm]{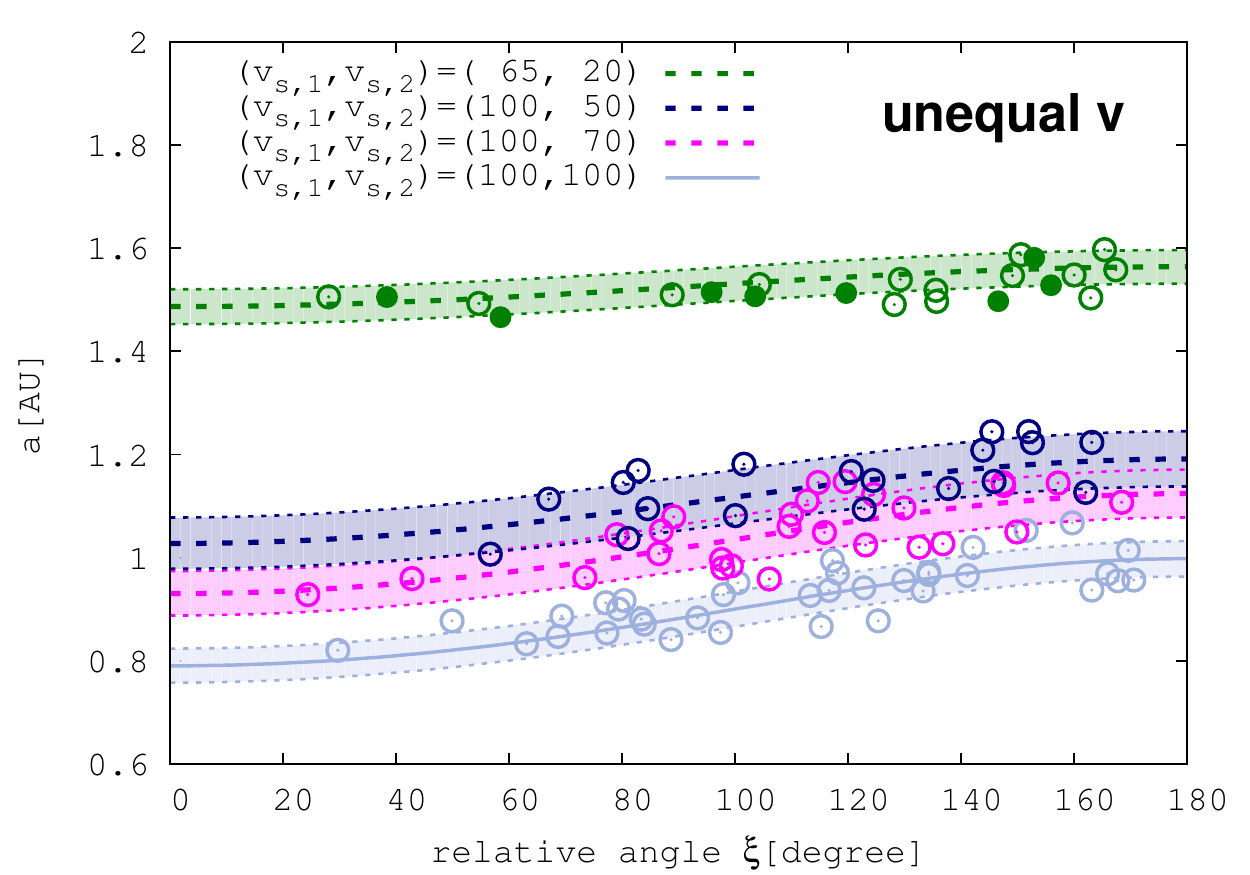}
	\includegraphics[width=8.7cm]{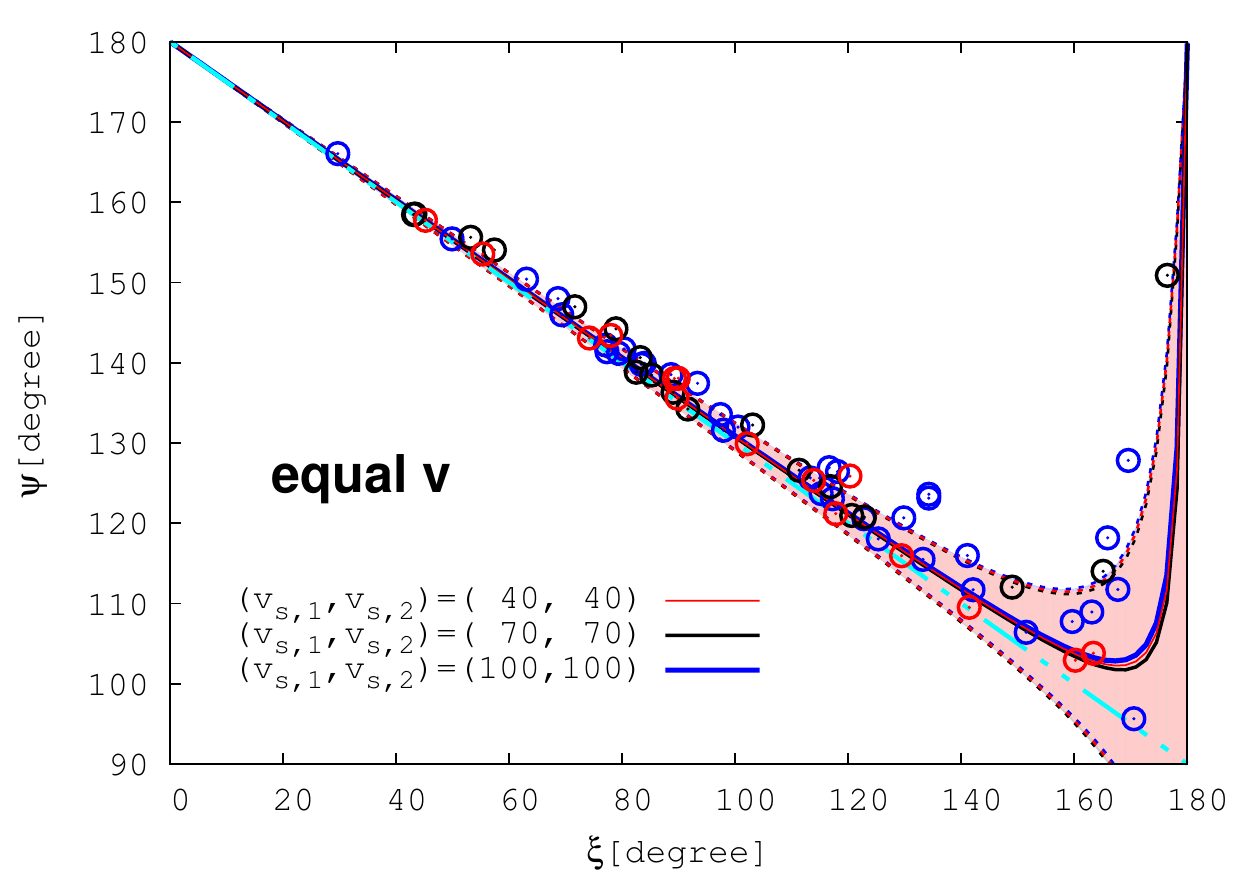}
	\includegraphics[width=8.7cm]{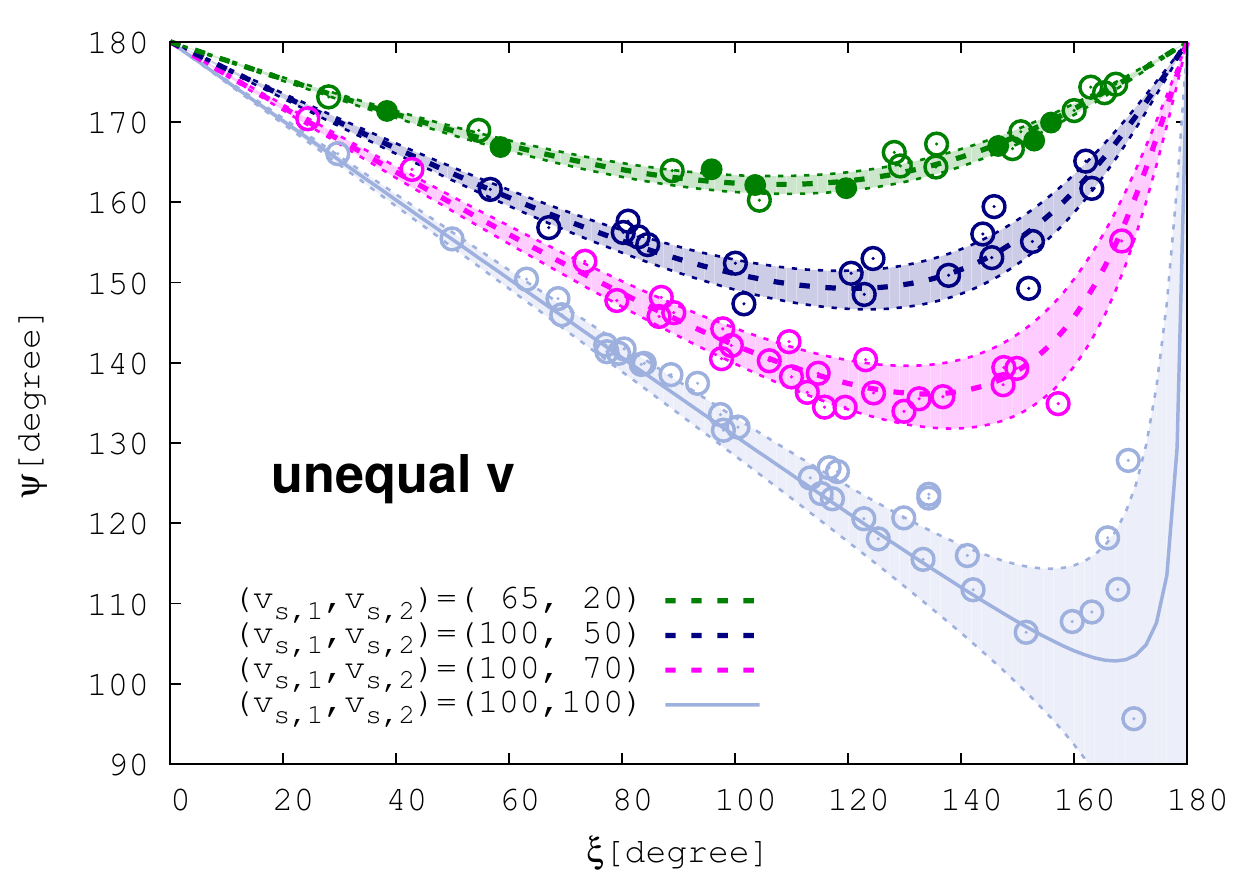}
	\caption{ The speeds of the final binaries $v_{\rm b}$
		(\textit{upper} panel), their semimajor axes $a$
		(\textit{middle} panel) and the relative angle $\Psi$
		(\textit{bottom} panel) expected from our analytic formulae
		\ref{eq:solution_vbrest_overall} and \ref{eq:Psi_2}
		(solid/thick dotted lines) for each combination of the initial
		velocities ($v_{\rm s,1}$, $v_{\rm s,2}$) in unit of$\km\s^{-1}$: equal single
		star velocities in the \textit{left} panels and unequal
		velocities in the \textit{right} panels.  The data points
		(dots) show the results of our numerical binary-binary
		scattering simulations taken from the simulation paper (Model 0
		without a background potential).  For the simulated data
		points, the final speeds of the ejected stars are
		constrained to be within $\sim5\%$ of the speeds shown in
		the plots. The solid lines for $v_{\rm b}$, $a$ and $\Psi$ (also
		in all plots below) are estimated using the median values of
		the speeds of the ejected stars. For each combination of
			speeds, we show the upper and lower limits (thin dotted lines) around 
			the median values (solid lines and thick dotted lines), adopting 
			the same colors as for the dots. Furthermore, we fill the regions 
			between the two limits with the same color. In the \textit{bottom} panel, 
		we depict the case (exactly same $v_{s,1}$ and $v_{s,2}$) 
		corresponding to Equation \ref{eq:Psi_2_same speeds}
		using the cyan-dot-dashed line, running
		diagonally from $\Psi=180^{\circ}$ (left upper corner) to
		$\xi=180^{\circ}$ (right bottom corner). The green
		solid circles in the $\textit{upper}$ and $\textit{middle}$
		panels correspond to data points with $v_{\rm s,1},v_{\rm
			s,2}=(20,65)$, and illustrate the symmetric property of
		Equation \ref{eq:solution_vbrest_overall}. However, we note
		that the good agreement shown in the \textit{bottom} panel
		is not because of the symmetric property of the equation,
		but because we define $\Psi$ as the angle of the binary with
		respect to the more rapidly-moving single star, in a given
		simulation.  In the \textit{middle} panel, we estimate the
		semimajor axis $a$ of the final binary assuming that the
		total energy is conserved. All of the parameters $v_{\rm
			b}$, $a$ and $\Psi$ calculated from our analytic formulae
		show excellent agreement with the numerical simulations.
	}
	\label{fig:analyticformula}
\end{figure*}

\section{Results}
\label{sec:results}
In this section, we apply our analytic formulae to simulated 2+1+1
outcomes formed during binary-binary encounters in the
presence/absence of a homogeneous background potential. The numerical
data are taken from the simulation paper. There,
we performed suites of numerical scattering
experiments between two binaries (which serve as proxies for O-type
stars), both in isolation and in a homogeneous background
potential. However, since we considered
  binary-binary encounters occurring near the system CM and $r_{\rm
    ej}\leq r_{\rm bg}$, we assume that $r_{\rm ej}$ is negligible in
  the following analysis, i.e., $r_{\rm ej}\simeq0$.
We compare our analytic formulae to the results of these simulations,
and confirm their validity.  We further study here the effects of the
background potential on the formation of runaway stars, in terms of
affecting the properties of the runaway stars and final binary. For
more details regarding the setup of the simulations, see the
  simulation paper.

\subsection{Results for scatterings in isolation} \label{results_isolation}

In Figure \ref{fig:analyticformula}, we show the speeds of the final
binaries $v_{\rm b}$ (\textit{upper} panel), the corresponding
semimajor axes $a$ (\textit{middle} panel) and the relative angle
$\Psi$ between the binary and the faster single star for each
simulation (\textit{bottom} panel), given by our analytic formulae
(Equations \ref{eq:solution_vbrest_overall} and \ref{eq:Psi_2},
solid/dotted lines) for different combinations of the ejected single
star speeds ($v_{\rm s,1}$, $v_{\rm s,2}$) in unit of$\km\s^{-1}$.  We plot equal values for
both $v_{\rm s}$ parameters in the \textit{left} column and unequal
values in the \textit{right} column.  To compare, we mark via the data
points the results of the numerical experiments of binary-binary
scatterings (2+1+1 outcome) performed in the simulation
  paper. For the simulated data points, the final speeds of the
ejected stars are constrained to be within 5\% of the speeds shown in
the plots.  The lines for $v_{\rm b}$, $a$ and $\Psi$ (also in all
plots below) are estimated using the median values of the speeds of
the ejected stars. Hence the actual combinations of the speeds for the
estimates are slightly different from those indicated in the plots.
For the analytic calculations, we assume the same total initial energy
and masses for the stars as adopted in the simulated data.
 For each combination of the single star ejection speeds, the
lines and dots share the same color. For comparison, we over-draw the
lines and dots for ($v_{\rm s,1}$, $v_{\rm s,2}$)=(100, 100) in the
\textit{right} panel in faint blue.

The results of our numerical scattering experiments are in excellent
agreement with our analytic formulae.  As shown in the \textit{upper}
panel of Figure \ref{fig:analyticformula}, the speed of the final
binary is the smallest at $\xi=180^{\circ}$, and rises steadily
towards $\xi=0^{\circ}$.  This is easily understood via conservation
of momentum. Finally, in order to confirm the symmetric nature of
Equation \ref{eq:solution_vbrest_overall}, we mark the data points
with $(v_{\rm s,1},v_{\rm s,2})=(65,20)$ (\textit{right} column) by
green hollow circles, and those with $(20,65)$ by green solid
circles. As expected, both sets of dots follow the theoretical line,
which shows that the order of ejection events is irrelevant in terms
of estimating the final binary speed using Equation
\ref{eq:solution_vbrest_overall}.

\begin{figure*}
	\centering
	\includegraphics[width=8.0cm]{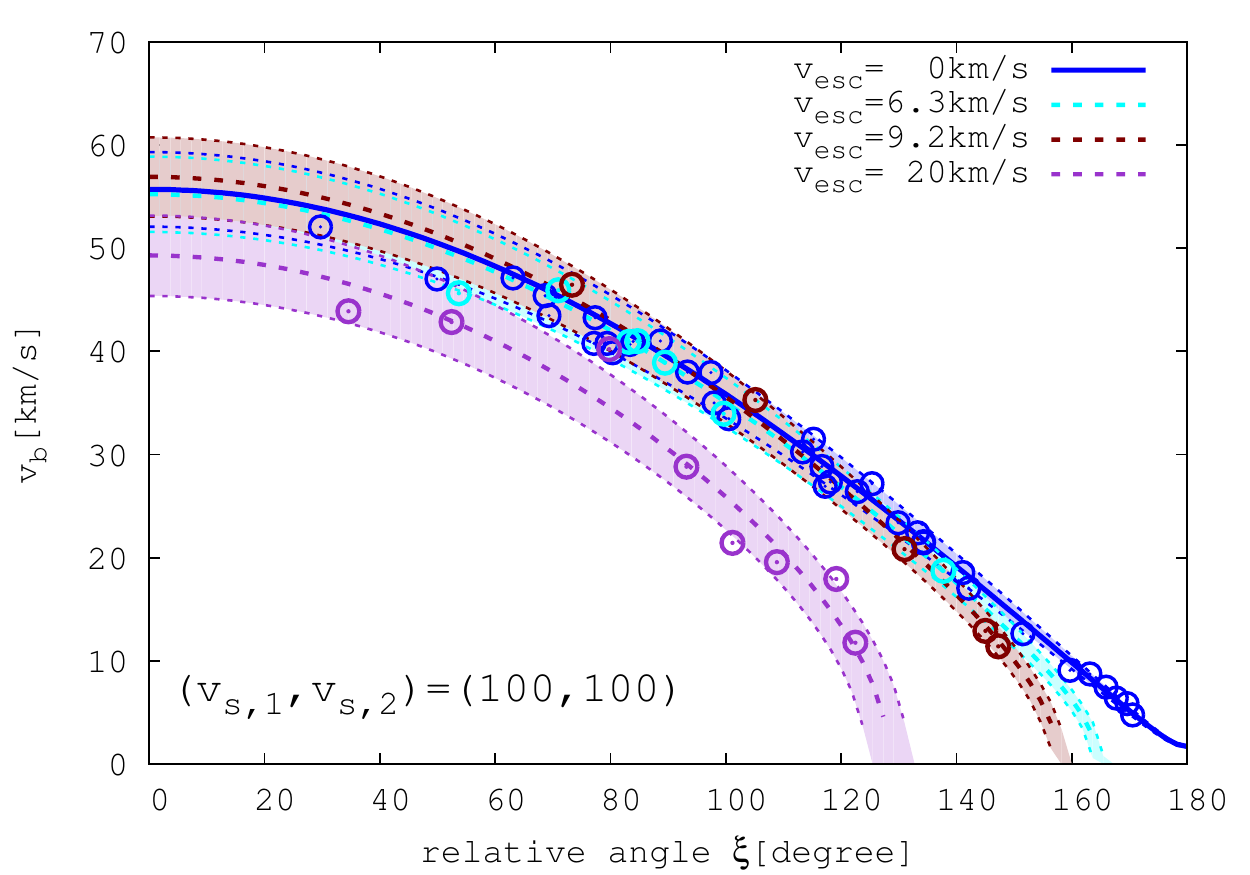}
	\includegraphics[width=8.0cm]{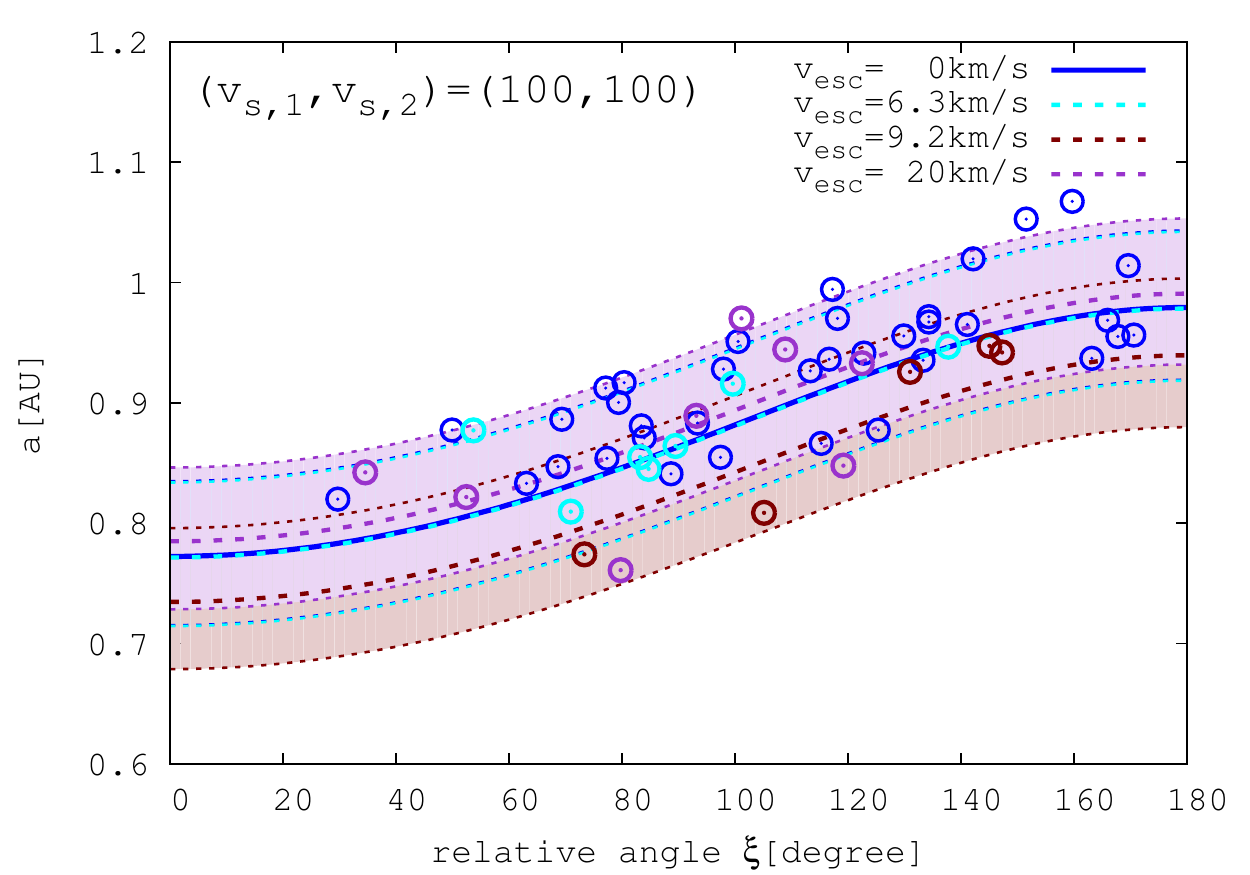}
	\includegraphics[width=8.0cm]{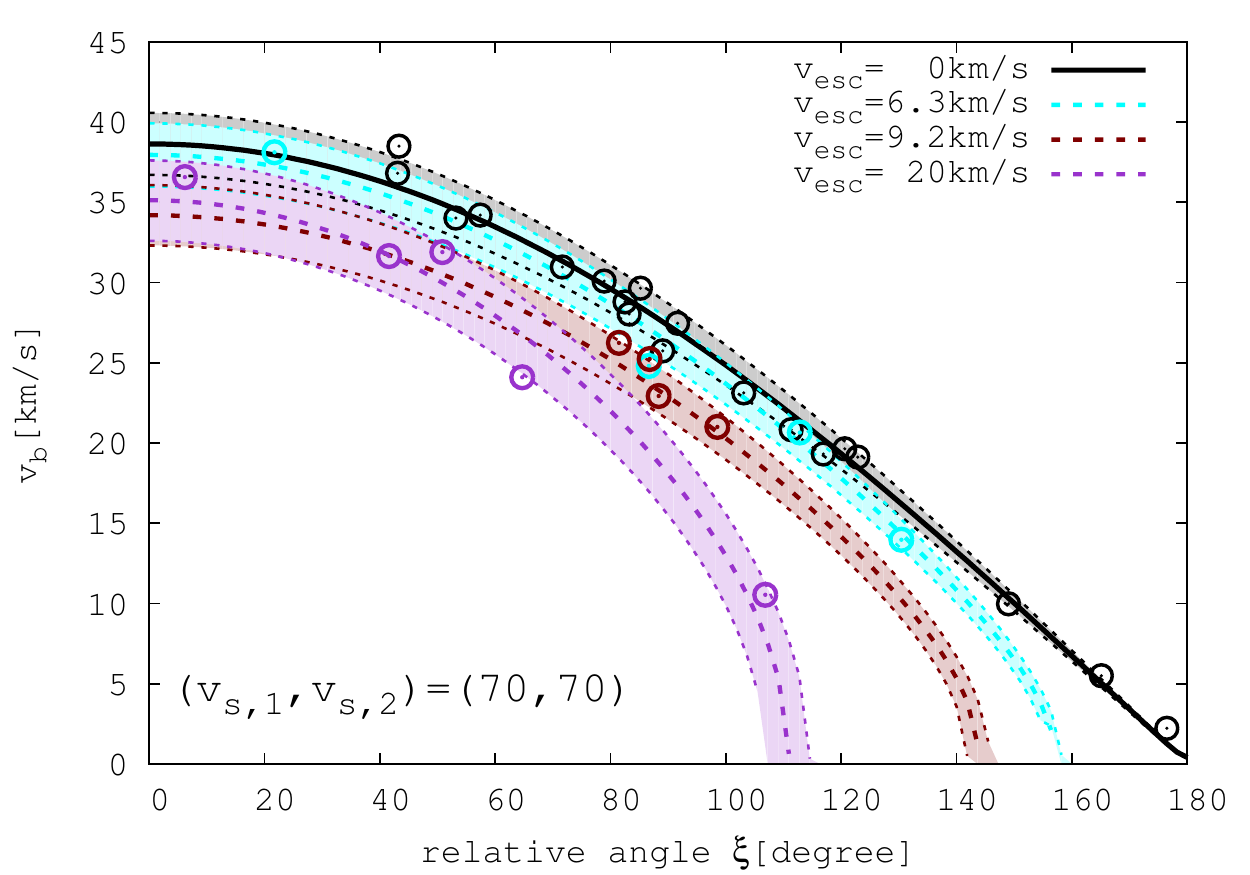}
	\includegraphics[width=8.0cm]{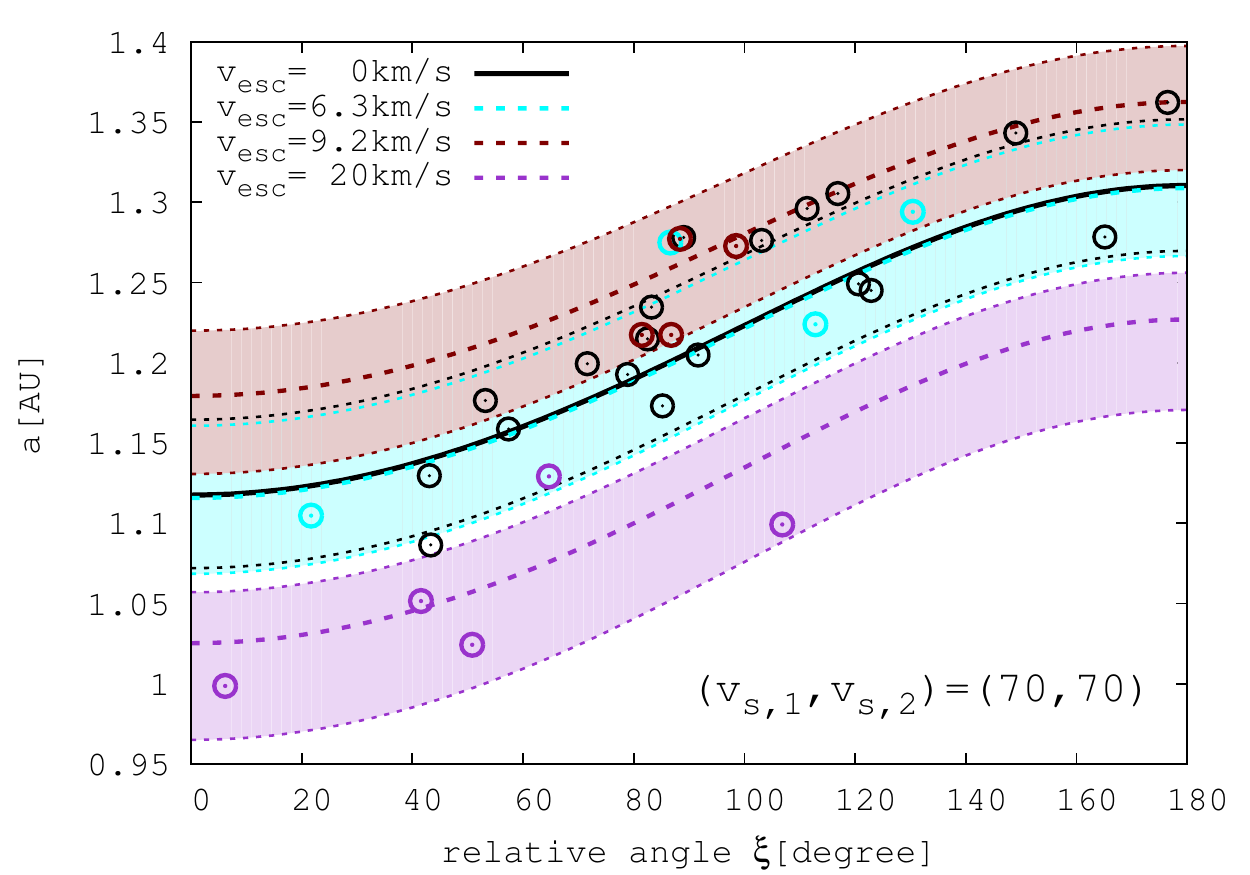}
	\includegraphics[width=8.0cm]{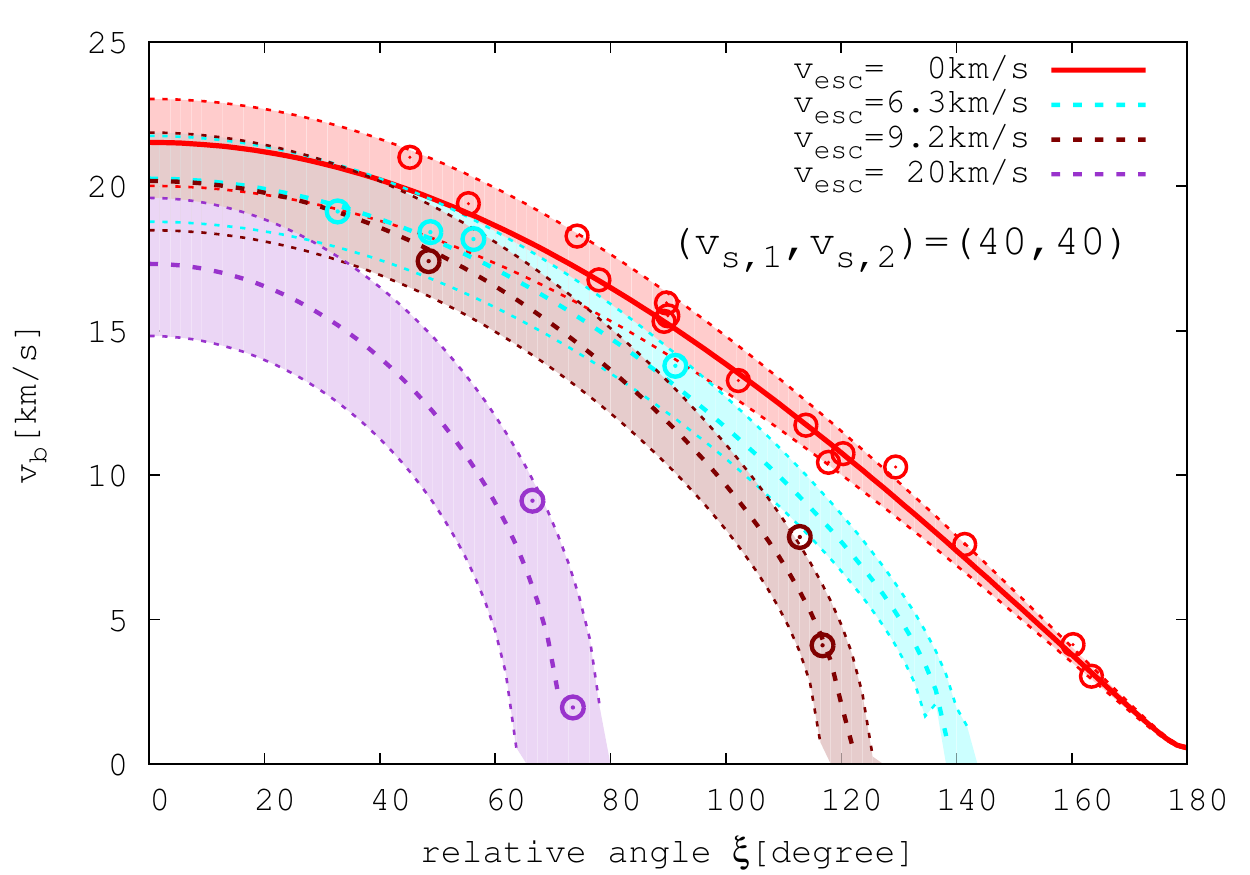}
	\includegraphics[width=8.0cm]{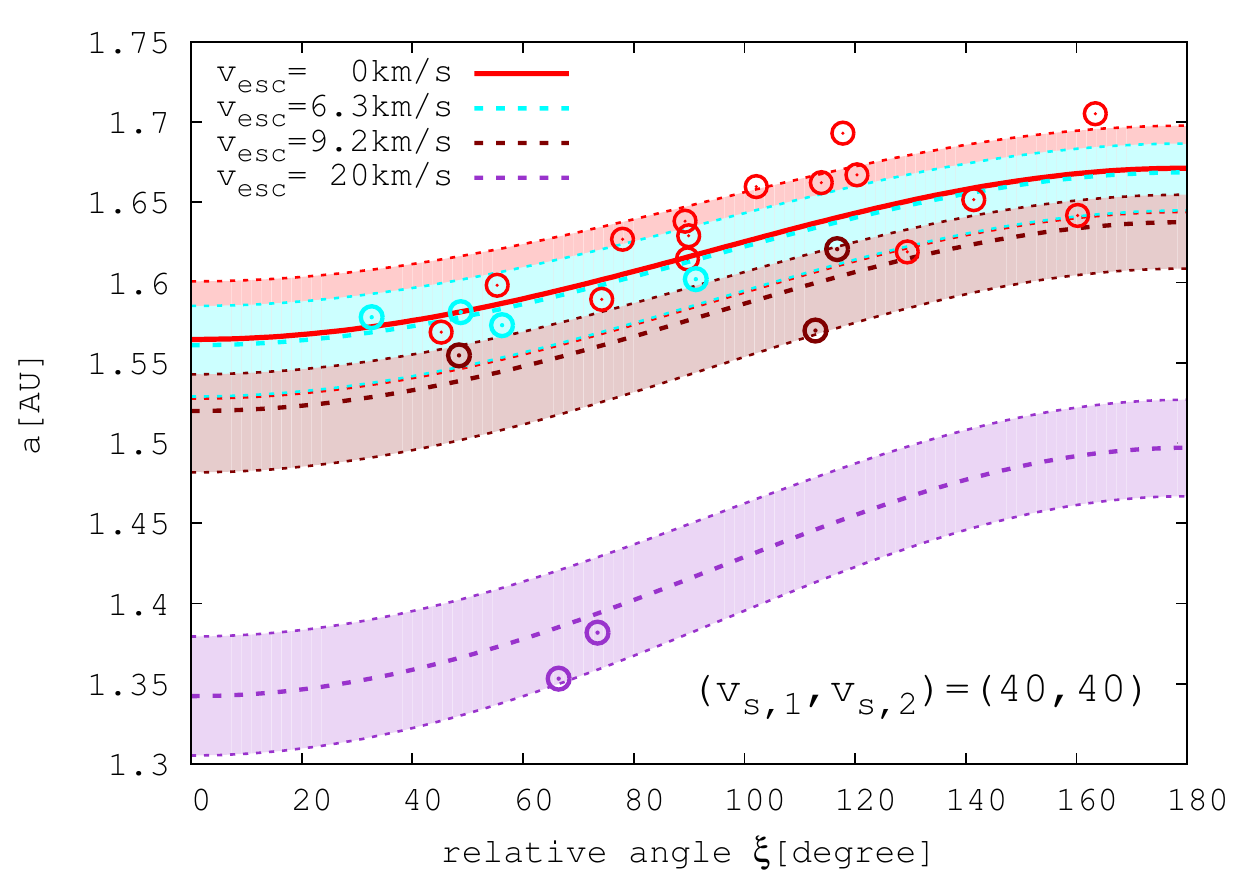}
	\caption{\textit{Equal} $v_{\rm s}$: The speeds
		$v_{\rm b}$ (\textit{left} panels) and the semimajor axes
		$a$ (\textit{right} panels) of the final binaries which have escaped from the potential 
		are shown, as calculated using Equation
		\ref{eq:vbinaryrest_bg} for each combination of the initial
		velocities ($v_{\rm s,1}$, $v_{\rm s,2}$) in unit of$\km\s^{-1}$.
		  The semimajor axes are estimated
		assuming conservation of energy (Equation
		\ref{eq:semimajoraxis}) with the momenta $\mathcal{P}$
		(Equations \ref{eq:vs1_ej_bg} and \ref{eq:vs2_ej_bg}).
		For comparison, we also show the 
		isolated (i.e., no background potential) case (solid lines with the same colors as used in Figure 
		\ref{fig:analyticformula}). Note that the final speeds of the simulated ejected stars are 
		constrained to be within 7\% of the speeds shown in the plot.  
		The upper and lower limits (thin dotted lines) around 
			the median values (solid lines and thick dotted lines) are drawn, 
			with the borders between each limit colored the same as are the dots and lines.
		Even when 
		the background potential is taken into account, our analytic formulae show 
		good agreement with the results of the simulations.}
	\label{fig:analyticformula_bg}
\end{figure*}

In the \textit{middle} panel of Figure \ref{fig:analyticformula}, 
we estimate the semimajor axes $a$ of the final binaries assuming 
conservation of energy. The simulated data from are recorded when 
all stars (single stars and binaries) are sufficiently far from the system CM that 
no further (significant) gravitational interaction occurs between them. At this point, 
the mutual gravitational potentials between the stars are negligible.  Hence, 
the total energy $E_{\rm tot}$ is the sum of the total kinetic energy for all objects 
($KE$) and the binding energy of the binary ($E_{\rm bind}$), namely,
\begin{align}
E_{\rm tot}&=KE_{\rm s,1}+KE_{\rm s,2}+KE_{\rm b}+E_{\rm bind}\nonumber\\
&=\frac{1}{2}m_{\rm s,1}v_{\rm s,1}^{2}+\frac{1}{2}m_{\rm s,2}v_{\rm s,2}^{2}+
\frac{1}{2}m_{\rm b}v_{\rm b}^{2}-\frac{G m_{\rm b,1}m_{\rm b,2}}{2a}
\label{eq:semimajoraxis}
\end{align}
where $m_{\rm b,1}$ and $m_{\rm b,2}$ are the masses of the component stars of the
final binary (i.e., $m_{\rm b}=m_{\rm b,1}+m_{\rm b,2}$). 
Combining the results for $v_{\rm b}$ and $a$ in both panels, for a
given combination of speeds for the two ejected stars (i.e., the same
$KE_{\rm s,1}+KE_{\rm s,2}$ or same $KE_{\rm b}+E_{\rm binding}$),
when two single stars are ejected with a smaller angle ($\xi$) and at
high velocities, conservation of momentum requires that the final
binary moves at a higher velocity (i.e., larger $KE_{\rm b}$ for a
given $KE_{\rm b}+E_{\rm binding}$).  Consequently, a more compact
binary is formed.

In the \textit{bottom} panel, the simulated data and the lines from
our analytic formula (Equation \ref{eq:Psi_2}) show good agreement.
However, we see a discrepancy between the results for equal and
unequal values of the two single star velocities (\textit{left} and
right panels, respectively). For equal speeds, independently of the
values for the speeds, the angles $\Psi$ and $\xi$ have an unique
relation. This is because, as the speeds of the two single stars of
same mass become equal, $\Psi$ loses the dependence on the speeds
themselves. For encounters with the exactly same values of the single
star masses $m_{\s, 1}$ and $m_{\rm s,2}$, Equation \ref{eq:Psi_2}
reduces to the following equation,
\begin{align}
\label{eq:Psi_2_same speeds}
\cos\Psi=-\frac{\sqrt{1+\cos\xi}}{2}.
\end{align}
We depict this case in the \textit{bottom left} panel of Figure
\ref{fig:analyticformula} using the cyan-dotted line, running
diagonally from $\Psi=180^{\circ}$ (left upper corner) to
$\xi=180^{\circ}$ (right bottom corner).  However, the solid lines in
the figure for equal single star velocities (\textit{left} panel)
increase again up to $\Psi=180^{\circ}$ as $\xi$ increases to
$\xi=180^{\circ}$, similar to the lines for unequal velocities
(\textit{right} panel). This is because the single star velocities
actually used for these estimates are not exactly identical (we noted
above that the lines are drawn using the median values of the ejected
stars' speeds).
We further note that the simulated data for both $(v_{\rm s,1},v_{\rm s,2})=(65,~20)$ (green hollow dots) and (20, 60) (green solid dots) nicely satisfy Equation \ref{eq:Psi_2}. This is  because we define $\Psi$ as the angle between the velocity vector of the binary and that of the more rapidly-moving single star.

\subsection{Results for scatterings in a background potential} \label{results_backgroundpotential}

\begin{figure*}
	\centering
	\includegraphics[width=8.0cm]{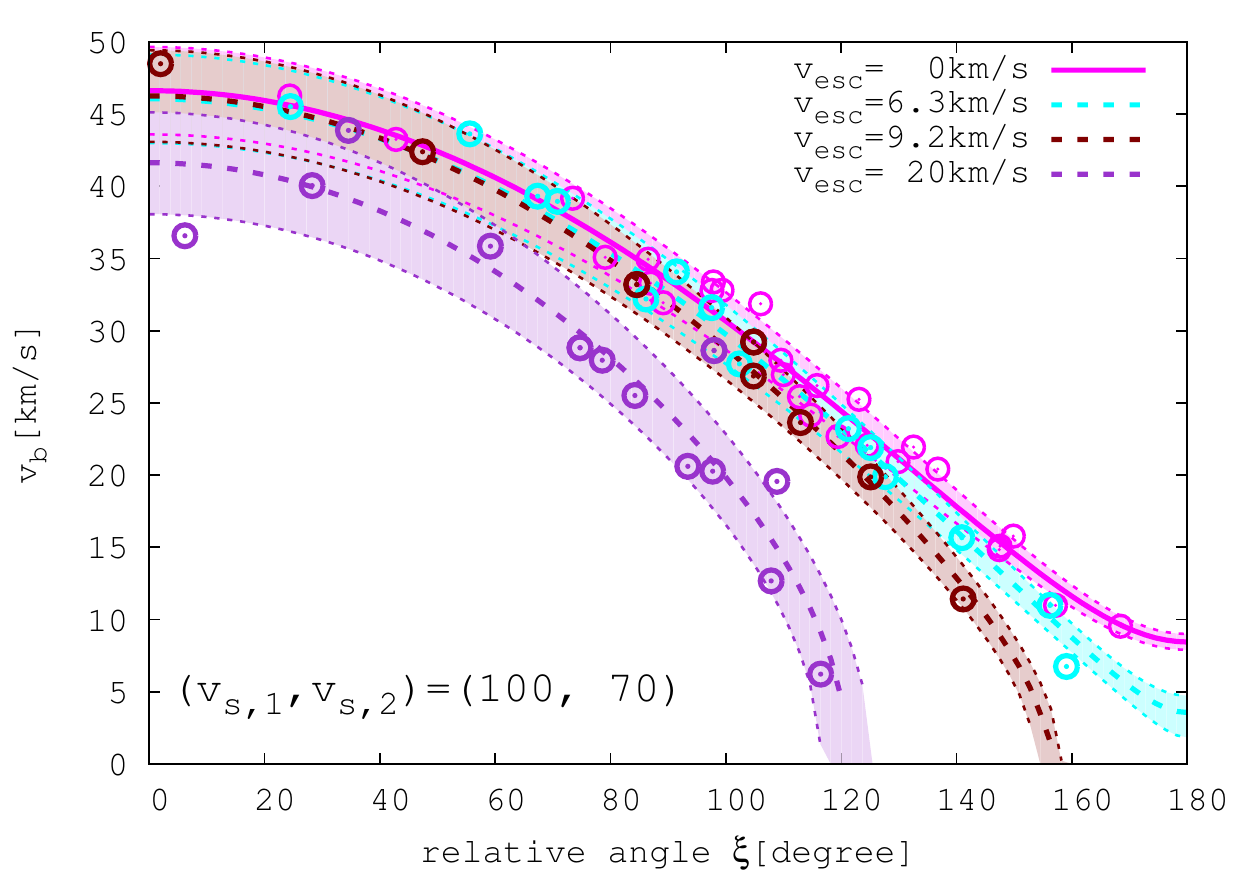}
	\includegraphics[width=8.0cm]{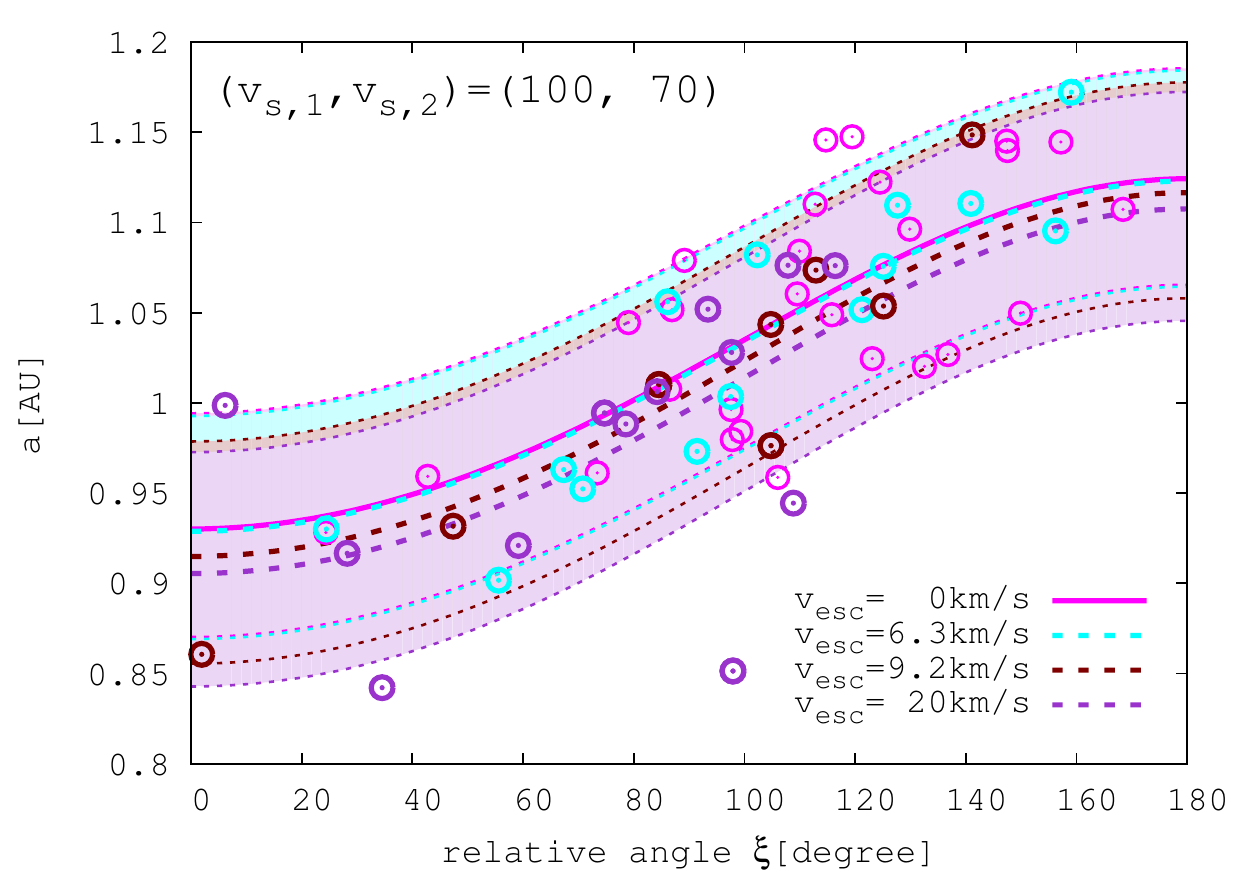}
	\includegraphics[width=8.0cm]{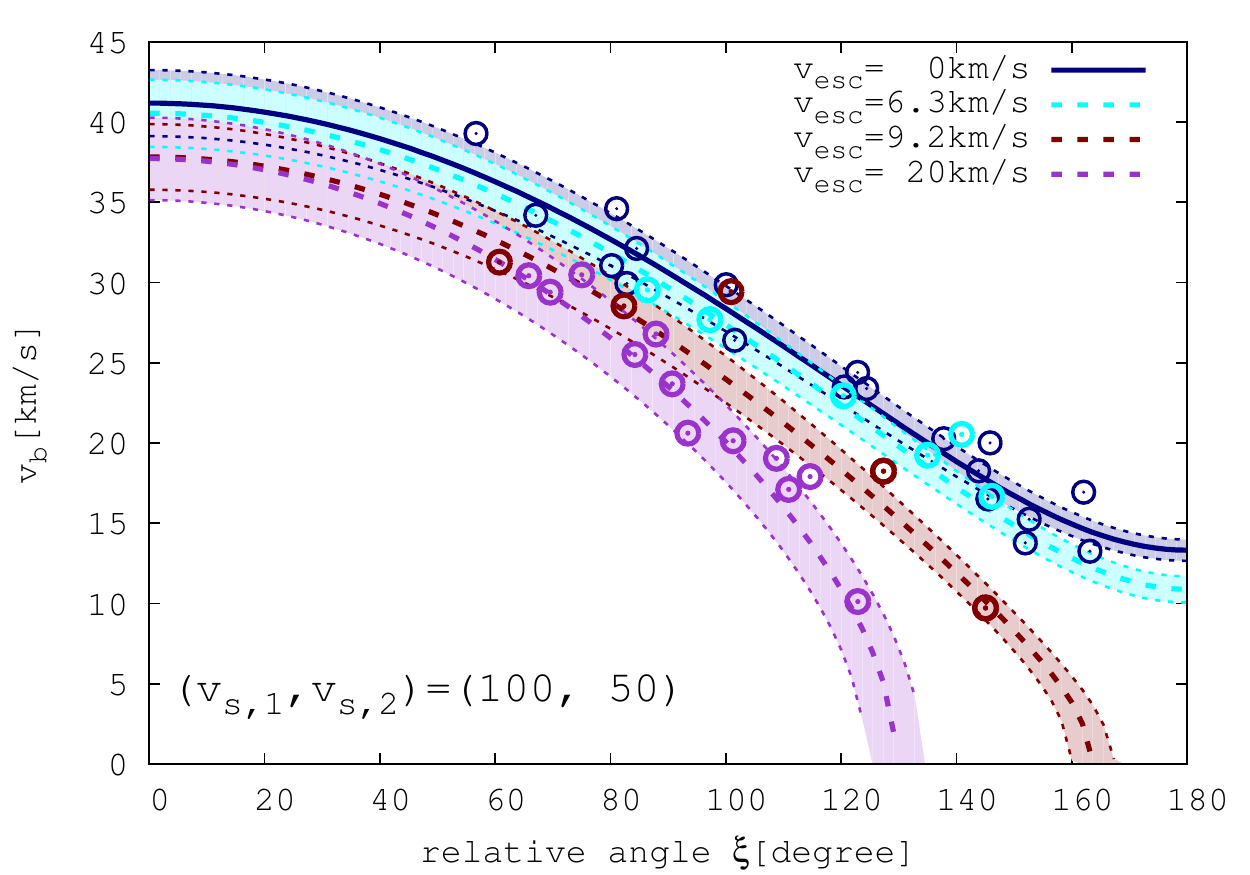}
	\includegraphics[width=8.0cm]{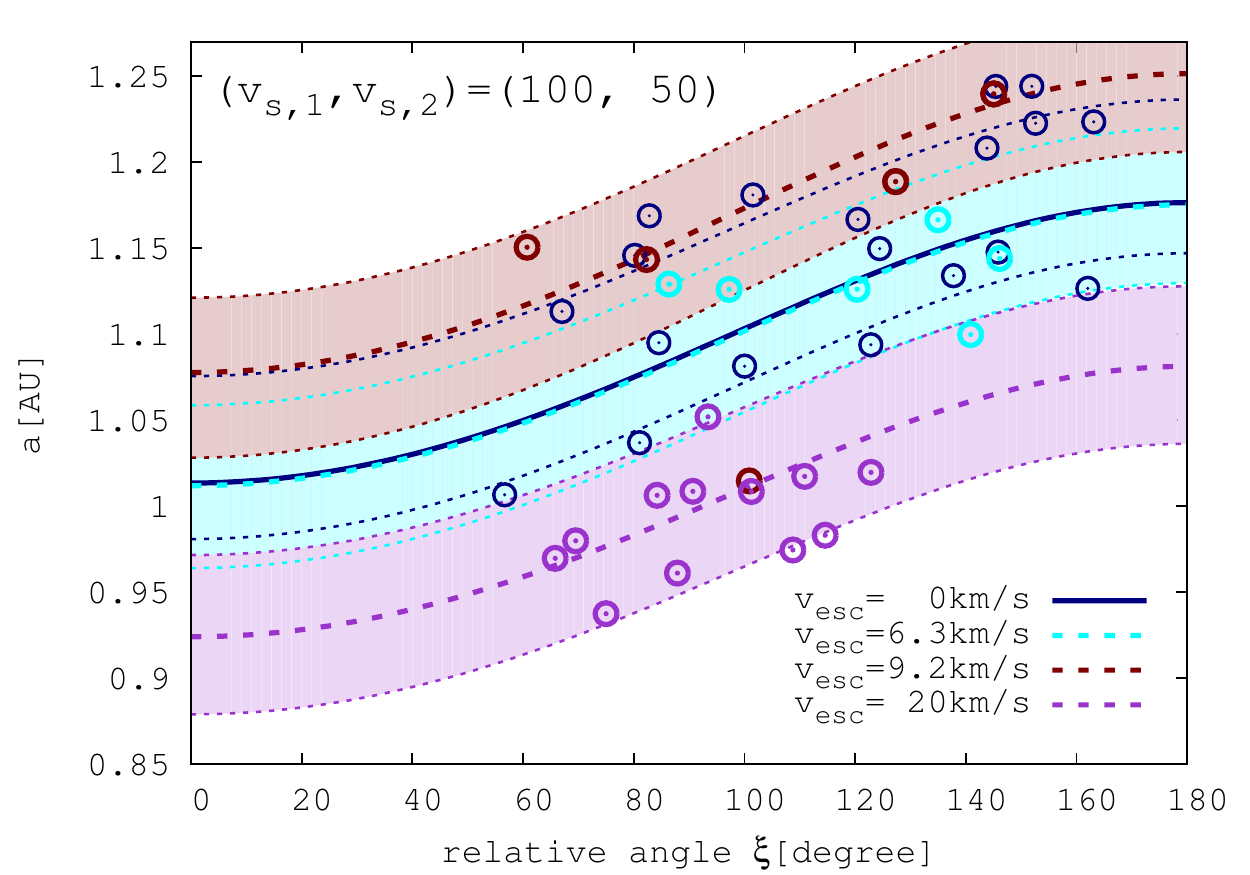}
	\includegraphics[width=8.0cm]{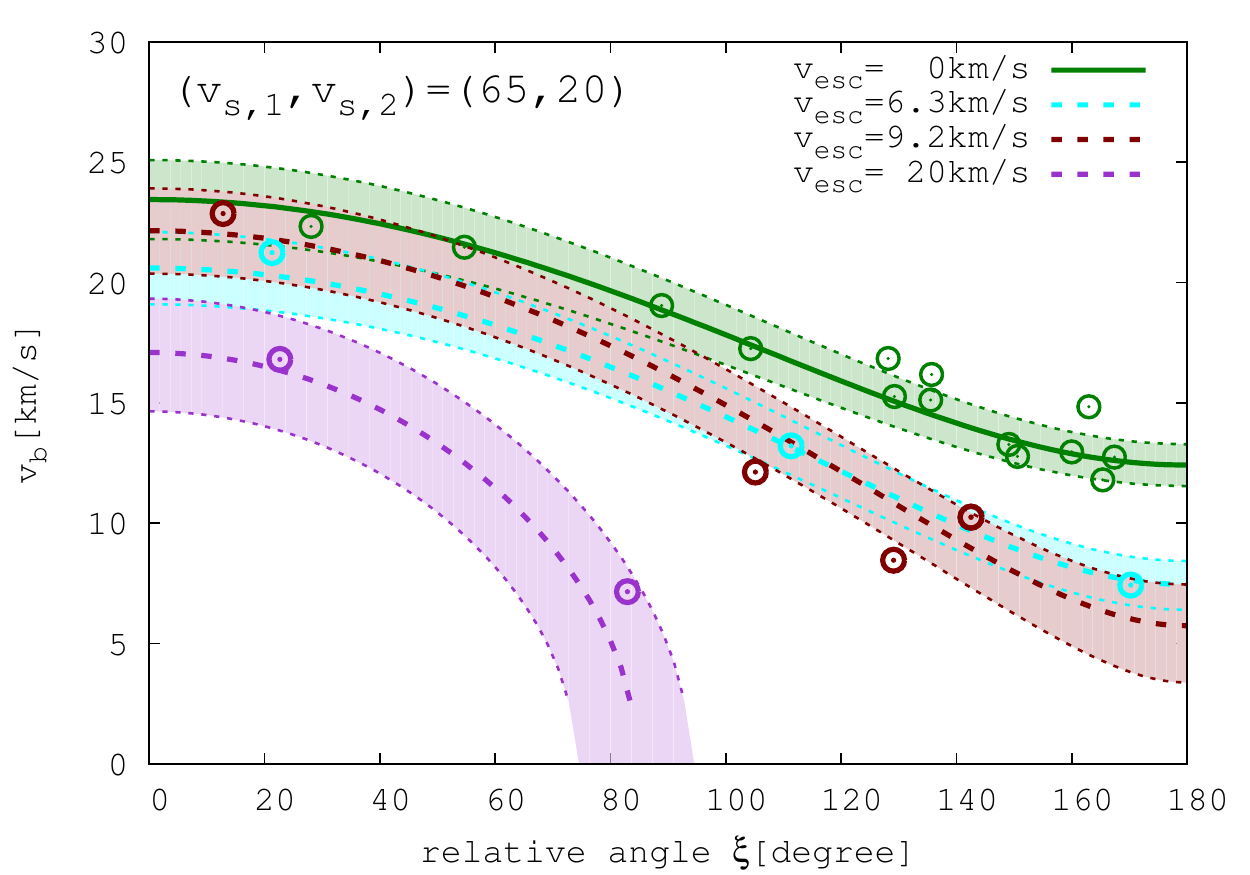}
	\includegraphics[width=8.0cm]{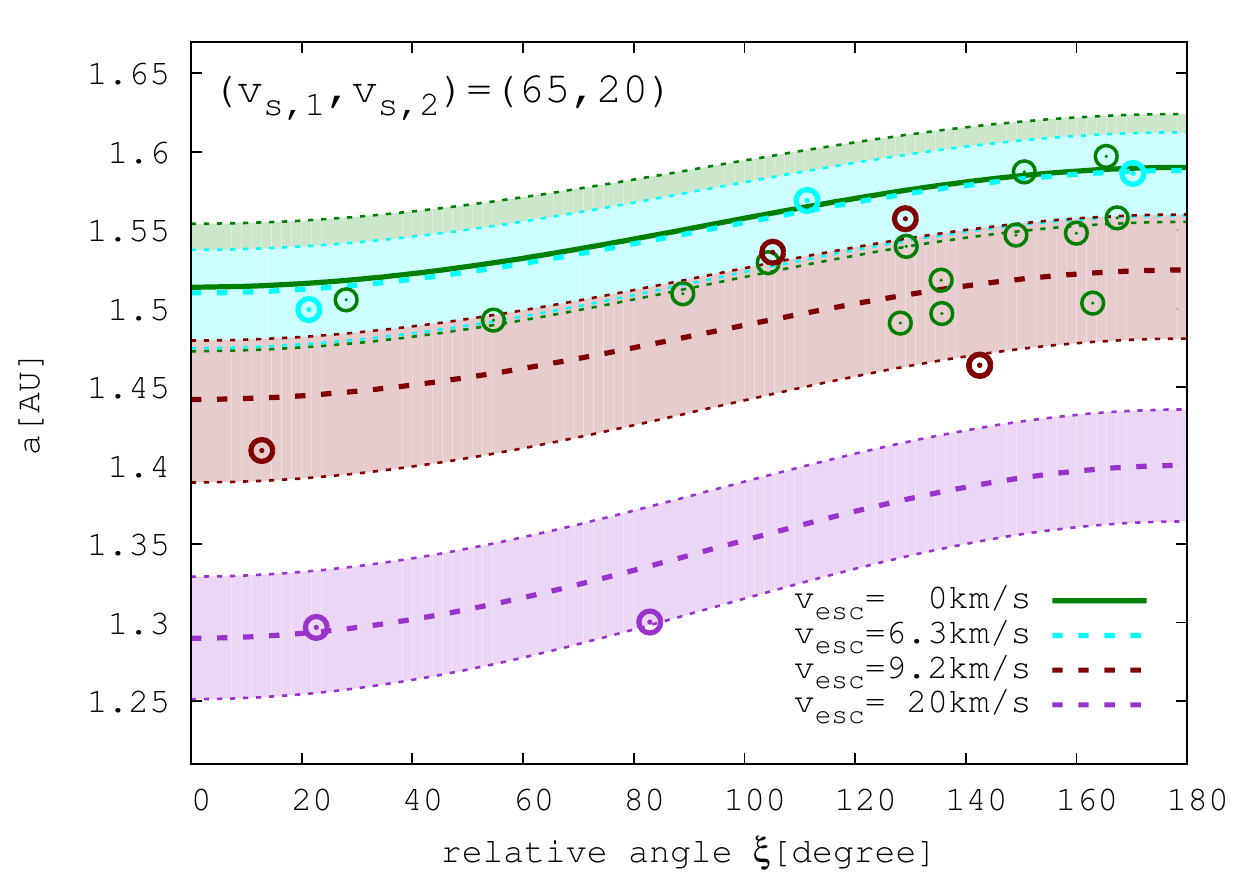}		
	\caption{\textit{Unequal} $v_{\rm s}$ : the speeds $v_{\rm b}$
		(\textit{left} panels) and the semimajor axes $a$
		(\textit{right} panels) of the binaries same as in Figure
		\ref{fig:analyticformula_bg}. The final speeds of the
		simulated ejected stars are constrained to be within 7\% of
		the analytic formulae. The upper and lower limits 
			(thin dotted lines) around the median values (solid lines and 
			thick dotted lines) are drawn. The regions between the limits 
			are filled with the same colors as the dots and lines.}
	\label{fig:unequal_analyticformula_bg}
\end{figure*}

In Figure \ref{fig:analyticformula_bg} and
\ref{fig:unequal_analyticformula_bg}, we show $v_{\rm b}$
(\textit{left} panels) and $a$ (\textit{right} panels) for the final
binaries \textit{which have escaped from the background potential}
using Equation \ref{eq:vbinaryrest_bg} for equal and unequal single
star velocities, respectively. We calculate the semimajor axes as
before, assuming conservation of energy (Equation
\ref{eq:semimajoraxis})\footnote{Note that for such high speeds of
  the single stars and large distances from the potential, $V_{\rm bg}$ has insignificant contribution to the
  total energy.}. However, we have used $\mathcal{P}_{\rm s,2}/m_{\rm
  s,2}$ (Equation \ref{eq:vs1_ej_bg}) and $\mathcal{P}_{\rm
  s,2}/m_{\rm s,2}$ (Equation \ref{eq:vs2_ej_bg}) for the speeds of
the two single stars and $1/m_{\rm b}\sqrt{\mathcal{P}_{\rm
    s,1}^{2}+\mathcal{P}_{\rm s,2}^{2}+2\mathcal{P}_{\rm
    s,1}^{}\mathcal{P}_{\rm s,2}^{}\cos\xi}$ (Equation
\ref{eq:solution_vbrest_overall_bg}) for that of the binary. For
comparison, we also show the isolated (i.e., no background potential)
case (solid lines with the same colors as used in Figure
\ref{fig:analyticformula}). Note that the final speeds of the ejected
stars are constrained to be within 7\% of the speeds shown in the
plots. Even when the background potential is taken into account, our
analytic formulae show good agreement with the results of the
simulations\footnote{Note the increased scatter about the analytic 
predictions in the right panels of Figures \ref{fig:analyticformula_bg} and \ref{fig:unequal_analyticformula_bg}.  We emphasize that 
this is due to the simplifying assumptions made for the velocity in 
deriving the equations of motion. Therefore, a more accurate calculation 
can easily be done for the binary semi-major axis, provided 
this assumption is not made and the higher-order terms are included.}. 
As the background potential gets deeper (i.e., higher
escape velocities), the binary speed is reduced more. Moreover, given
the adopted potentials and corresponding escape velocities, $\xi$ has
an upper limit. For example, for $(v_{s,1},v_{s,2})=(70,~70)$, the
line corresponding to $v_{\rm esc}=20\km \s^{-1}$ (purple line) drops
quickly to zero at $\xi\simeq110^{\circ}$. This maximum angle refers
to the angle required to give a sufficiently high recoil kick to the
binary that it can escape to infinity, as imposed by Equation
\ref{eq:esc_condition}.  The maximum angle decreases as $v_{\rm esc}$
increases (see also Section 3.4 in the simulation paper).

\begin{figure*}
	\centering
	\includegraphics[width=8.7cm]{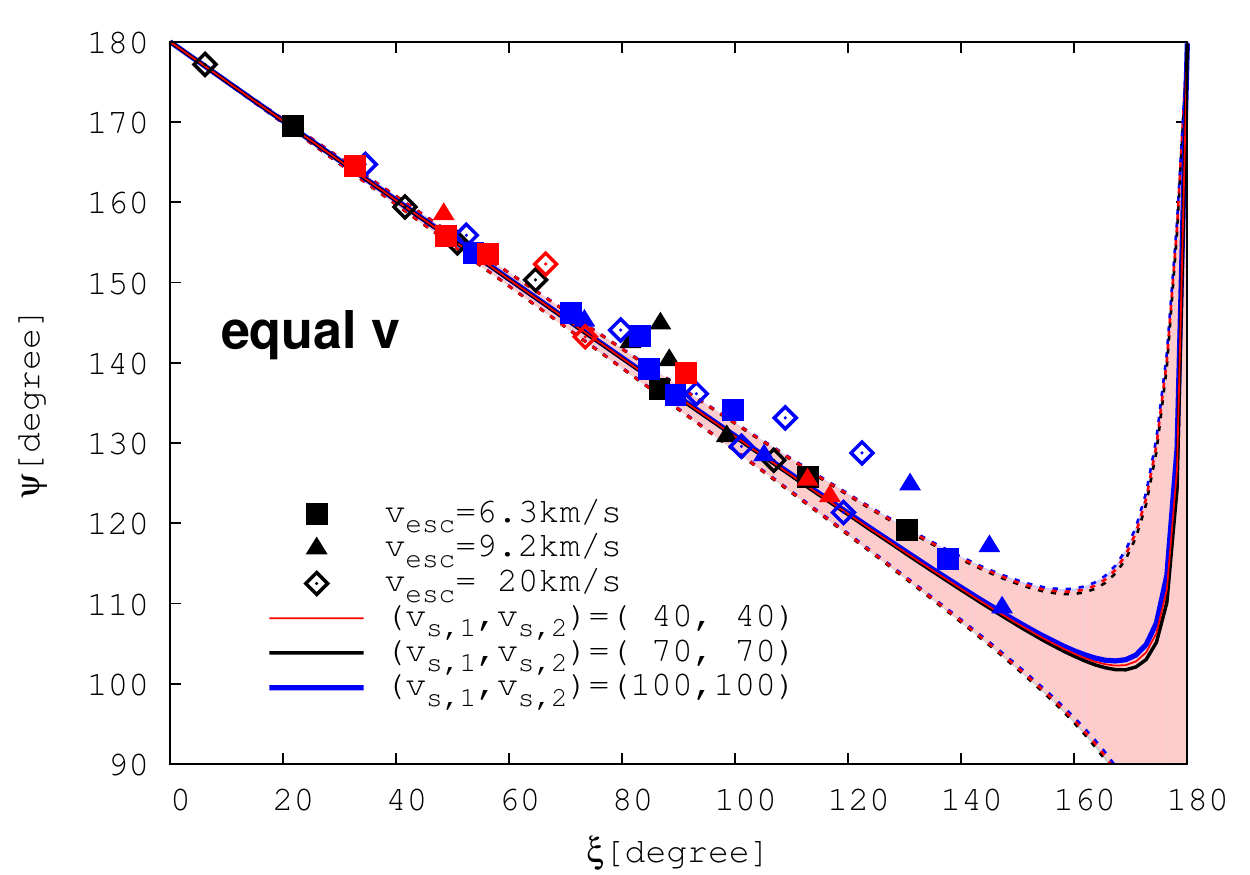}
	\includegraphics[width=8.7cm]{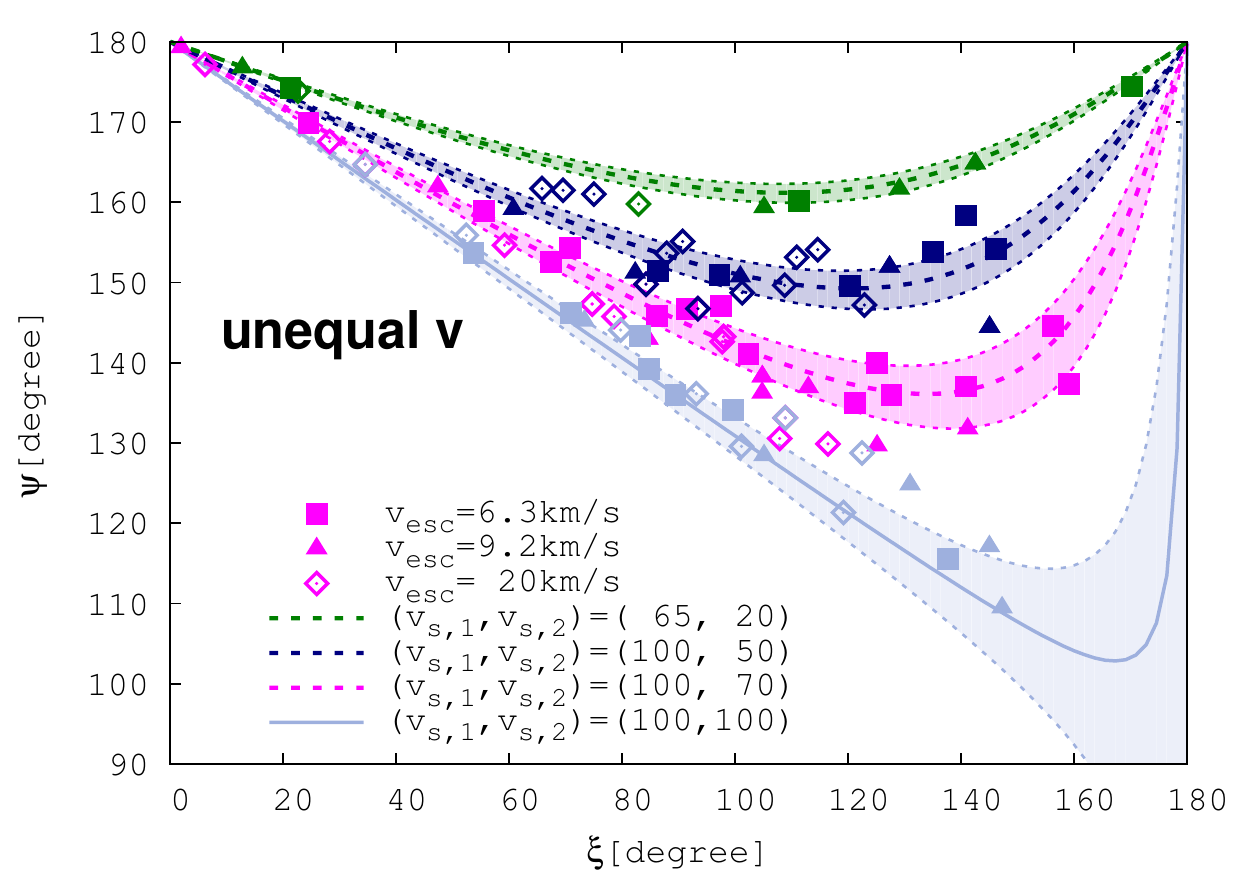}
	\caption{ The relative angles $\Psi$ between the escaped
          binaries and the faster single stars formed in the presence
          of a background potential. The angles still nicely satisfy
          Equation \ref{eq:Psi_2} for both equal (\textit{left}
          column) and unequal (\textit{right} column) single star
          velocities. In order to avoid overcrowding, we do not show
          the simulated data for $v_{\rm esc}=0$ (see the
          \textit{bottom} panels in Figure \ref{fig:analyticformula}
          for the case without a background potential). Different dot
          types are adopted to make a distinction between the
          potential models with different depths : $v_{\rm
            esc}=6.3\km\s^{-1}$ (square), $v_{\rm esc}=9.2\km\s^{-1}$
          (triangle) and $v_{\rm esc}=20\km\s^{-1}$ (diamond). 	The final 
          speeds of the simulated ejected stars are chosen within
          5-7\% of the speeds shown in the figure. The shaded 
          	regions around the median values (solid lines and thick dotted lines) 
          	demarcate the ranges between the upper and lower limits 
          	(thin dotted lines), filled with the same colors as the dots and lines.
           For example, the blue
          square dots refer to the simulated data for
          $(v_{s,1},v_{s,2})=(100,~100)$ with the potential of $v_{\rm
            esc}=6.3\km\s^{-1}$. }
	\label{fig:angle}
\end{figure*}

\begin{figure*}
	\centering
	\includegraphics[width=8.7cm]{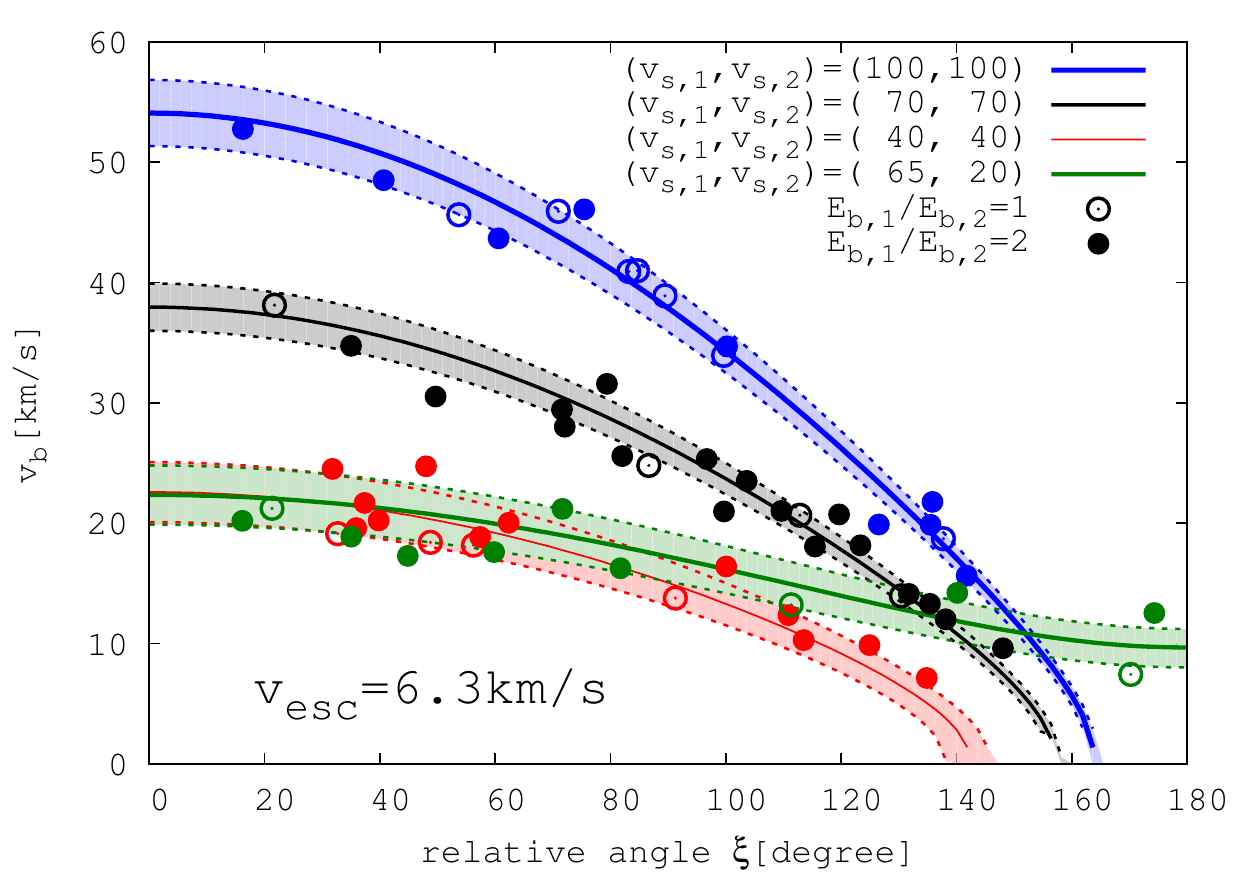}
	\includegraphics[width=8.7cm]{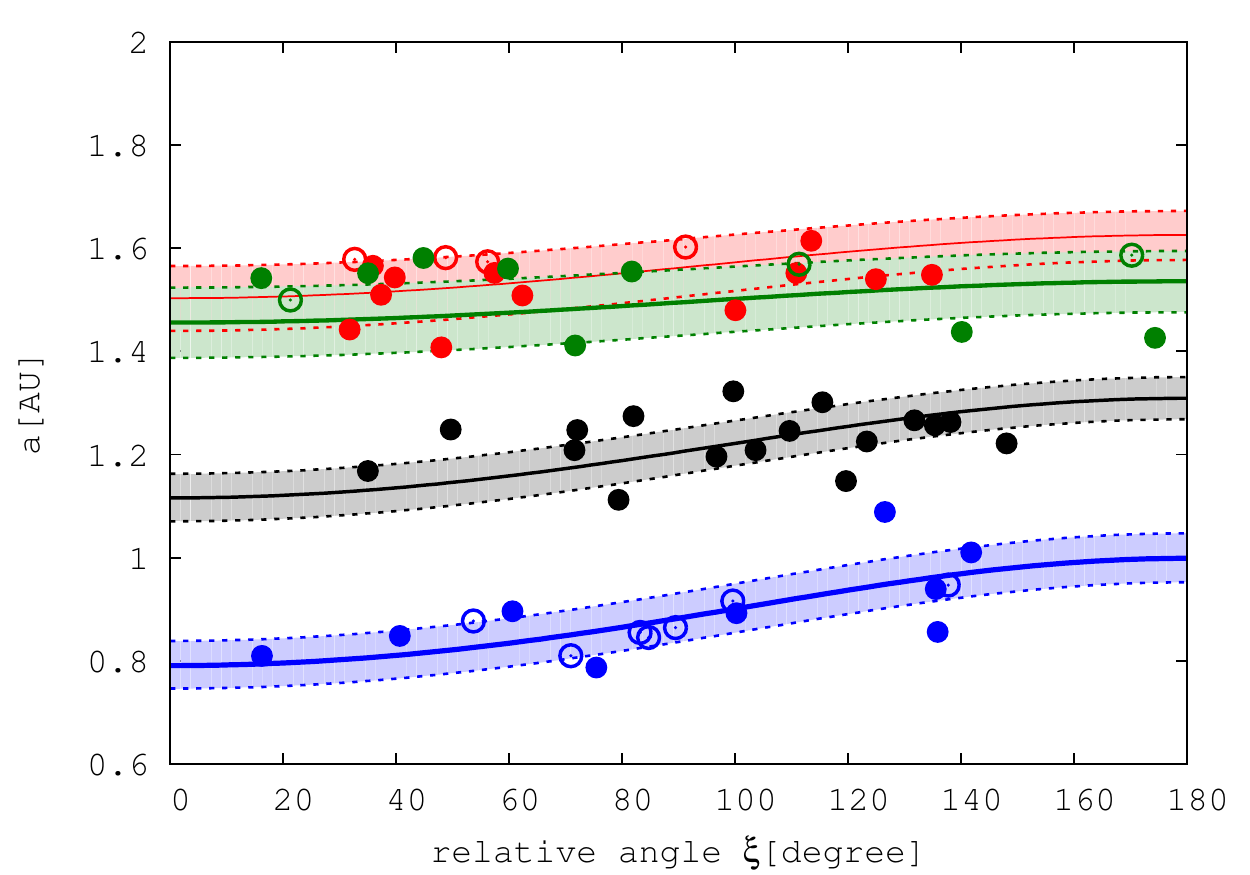}
	\caption{
		The speeds and the semimajor axes of the final binaries, for encounters between 
		two binaries with both equal and unequal energies. We take the results of 
		Model 1 and Model 1-1 from the simulation paper. 
	The final speeds of the simulated ejected stars are chosen within
	5-7\% of the speeds shown in the figure. The shaded regions 
		around the median values (solid lines and thick dotted lines) correspond 
		to the ranges between the upper and lower limits (thin dotted lines), 
		sharing the same colors as the dots and lines.
		In both models, 
		we consider the same homogenous potential, with $M_{\rm bg}=3\times 10^{3}\Msol$ and 
		$v_{\rm esc}=6.3\km \s^{-1}$. In Model 1, the ratio of the binary binding 
		energies is unity whereas in Model 1-1 the more massive binary is more tightly 
		bound than the less massive binary by a factor of two. We have used Equation 
		\ref{eq:rbinaryrest_bg} to estimate $v_{\rm b}$ (solid lines). The hollow circles 
		correspond to the case with equal binding energies ($E_{\rm b,1}/E_{\rm b,2}=1$), and the 
		solid circles correspond to the case with unequal binding energies ($E_{\rm b,1}/E_{\rm b,2}=2$). 
		Regardless of the orbital properties of the initial binaries, our 
		formulae still agree very well with the simulations. This is the case even when the background potential is taken into account.}
	\label{fig:Model1_1}
\end{figure*}

In Figure \ref{fig:angle}, we show the relative angles
  $\Psi$ between the escaped binaries and the faster single stars
  formed in a background potential. The angles still nicely fulfill
  the relation with $\xi$ expected from Equation \ref{eq:Psi_2} for
  both equal and unequal single star velocities.
	In order to avoid overcrowding, we do not mark the simulated
        data for $v_{\rm esc}=0$ (see the \textit{bottom} panels in
        Figure \ref{fig:analyticformula} for the case without a
        background potential). We use different dot types to
distinguish the potential models with different depths : $v_{\rm
  esc}=6.3\km\s^{-1}$ (square), $v_{\rm esc}=9.2\km\s^{-1}$ (triangle)
and $v_{\rm esc}=20\km\s^{-1}$ (diamond). The dots and the lines share
the same color. For example, the blue square dots refer to the
simulated data for $(v_{s,1},v_{s,2})=(100,~100)$ for the background
potential model with $v_{\rm esc}=6.3\km\s^{-1}$.

As an example that our formulae apply to any 2+2 encounters, we
compare in Figure \ref{fig:Model1_1} the final speeds and semimajor
axes of the binaries formed from two different 2+2 scenarios, namely
between binaries with equal and unequal orbital energies. We show the
simulation results for Model 1 and Model 1-1 from the simulation
paper. The final speeds of the simulated ejected stars are within
7-10\% of the analytic predictions. In both models, we assume the same
homogenous potential, with $M_{\rm bg}=3\times 10^{3}$ and $v_{\rm
  esc}=6.3\km \s^{-1}$. In Model 1, the two initial binaries have the
same binding energy whereas in Model 1-1 the more massive binary is
more tightly bound than the less massive binary by a factor of
two. For the given combinations of the final speeds for the two
ejected single stars, we estimate $v_{\rm b}$ (solid lines) using
Equation \ref{eq:rbinaryrest_bg}. The hollow circles correspond to the
case with equal binding energies ($E_{\rm b,1}/E_{\rm b,2}=1$) and the
solid circles correspond to the case with unequal binding energies
($E_{\rm b,1}/E_{\rm b,2}=2$). Regardless of the orbital properties of
the initial binaries, our formulae reproduce very well the results of
the simulations.

In Figure \ref{fig:r_xi} in Appendix \ref{appendix:r_xi}, we also show the radial distance $r_{\rm b}$ from the core at $t=4\Myr$ for both \textit{equal} $v$ (\textit{left} column) and \textit{unequal} $v$ (\textit{right column}) with/without the background potential using Equation \ref{eq:rbinaryrest_bg}.

\section{Discussion:  Astrophysical applications}
\label{sec:Discussionandsummary}

Given the ubiquity of a background potential in the astrophysical sites of dynamical interactions, 
proper theoretical tools that account for its presence are needed.  
Our analytical method derived above offers one such tool, and is ideal for application to 
observations of runaway stars. 

Given an observed pair of runaway stars with a common origin, our method can be used to identify the 
left-over binary.  First, using Equation \ref{eq:esc_condition}, 
we can estimate the probability of observing the left-over binary in its host cluster potential (i.e., to evaluate 
whether or not it should have escaped). Next, we can infer the history of the 
background potential of the host cluster. For example, consider the two runaway stars AE Aur 
and $\mu$ Col, in conjunction with the $\iota$ Orionis binary (all thought to have formed during the same 
binary-binary interaction; \citealt{Blaauw1954,Gies1986,Hoogerwerf+2001}).
 Given their observed speeds 
($\sim 100\km\s^{-1}$), their locations on the sky ($\simeq250\pc$ 
from the CM of all four stars) and the relative angle between the velocity vectors 
of the two ejected single stars ($\xi\simeq 140-150^{\circ}$) 
\citep{Gualandris+2004}, Equation \ref{eq:esc_condition} (assuming $r_{\rm ej}\simeq0$) gives 
$v_{\rm esc}>15.5\km\s^{-1}$ at the time of the last ejection event.
Assuming the background potential has a total mass $M_{\rm bg}=10^{4}\Msol$ \citep{Tielens1985,Hillenbrand1998} and $\rho=1800-3000\Msol/\pc^{3}$ 
\citep{Herbig86} for the Trapezium cluster, the escape velocity of 
the cluster at present (assuming a homogeneous medium) is,

	\begin{align}
		\Big(\frac{v_{\rm esc}}{\km \s^{-1}}\Big)=9\Big(\frac{M_{\rm bg}}{10^{4}\Msol}\Big)^{1/3}\Big(\frac{\rho}{2500\Msol/\pc^{3}}\Big)^{ 1/6}\,.
	\end{align}

Since we know that the $\iota$ Orionis binary resides in the 
Trapezium cluster, this suggests that $2.4\Myr$ ago (i.e., when the two 
runaway stars formed) the Trapezium cluster 			
was denser and/or more massive than it is now. However, we note that this prediction is based on a number of simplified assumptions, such as a homogeneous potential in the core region.  For a more precise estimation, we have to take into account a more realistic (non-homogeneous) potential in the core in addition to the subsequent dynamical evolution of the cluster, including the gas-depletion process \citep{Boily+2003}, evolution in an external tidal field \citep{Baumgardt+2003} and mass segregation \citep{Bonnell+1998,Zwart+2010}.

Second, using the temporal information extracted from the observed positions and 
velocities of the runaway stars, our method can be used to constrain the lifetimes of massive stars.
What's more, in the event that one of the components of the left-over binary has ended its life to 
form a black hole (BH) via direct collapse, then our method can be used to constrain the initial-final 
mass relation for massive stellar remnants.  In this case, our method provides a relation 
between the initial mass of the BH progenitor and the initial binary semi-major axis.  The most probable 
combination of these two parameters can then be constrained using further 
numerical scattering simulations, by evaluating the probability of forming the observed single star velocities 
for every allowed combination of the initial progenitor mass and binary semi-major axis.  

Going one step further, if both binary components form BHs and merge
(or will BHs in the first place), their inspiral could be detectable
via the associated GW emission.  Our method could then be applied to
identify an associated pair of runaway stars, and further constrain
the properties of the BH-BH binary before merger.

In the derivations above, we consider a particular outcome consisting
of two single stars and a binary. However, our analytical treatment is
applicable to any kind of three-body outcome 
(two ejected systems and one left-over system), such as triple-binary
encounters that produce two binaries and one single star
\citep[e.g.][]{Leigh2013}.

We also note that, in reality, it may be more likely that two objects
(single, binary or higher-order hierarchical system) encounter each
other with non-zero centre of mass velocities. However, when it comes
to runaway stars, it is often still valid to treat the ejected objects
in their CM frame and to use the analytic formula derived above.  For
encounters in clusters, the center of mass velocities of the objects
(with respect to the CM of the cluster) should be roughly comparable
to the stellar velocity dispersion. Hence, for clusters with low
velocity dispersions relative to the observed speeds of the runaway
stars, the centre of mass velocities can be ignored.

\section{Summary}
\label{sec:summary}

In this paper, we have studied binary-binary interactions, focusing on
the formation of one binary and two single stars (i.e., the 2+1+1
outcome). Assuming only conservation of momentum in the inertial frame
of each ejection event, we have derived an analytic relation between
the orbital properties of the final binary (speed and semimajor axis)
and the relative angles between the two ejected stars.  This is first
done assuming that the interactions occur in isolation (see Section
\ref{sec:analyticformula:puerlystellardynamics}).  Going one step
further, we have also found an analytic formula to account for the
effects of a homogeneous background potential on the subsequent
motions of the stars (see Section \ref{sec:analyticformula_bg}).  Then
we apply our formula to compare with the results of numerical
simulations taken from a companion simulation paper. We have found an
overall good agreement with the numerical simulations.

\vspace{0.3in}
We summarize the results of this analytic study as follows:

\begin{enumerate}
	\item For \textit{purely stellar interactions}, when a binary and two single stars form,
	 the speeds of the binary $v_{\rm b}$ and the relative angle $\xi$ of two single 
	stars are described as follows,
	\begin{align}
	v_{\rm b}&=\frac{1}{m_{\rm b}}\sqrt{p_{\rm s,1}^{2}+p_{\rm s,2}^{2}+2p_{\rm s,1}^{}p_{\rm s,2}^{}\cos\xi}\,,
	\end{align}
	or, even more simply,
	\begin{align}
	p_{\rm b}&=\sqrt{p_{\rm s,1}^{2}+p_{\rm s,2}^{2}+2p_{\rm s,1}^{}p_{\rm s,2}^{}\cos\xi}\,,
	\end{align}
	where $p_{\rm s,1}$, $p_{\rm s,2}$ and $p_{\rm b}$ are the
        momenta of the first, second ejected stars and the binary,
        respectively, i.e., $p_{i}=m_{i}v_{i}$. The relative angle
        $\Psi$ between the binary and the first ejected star can be
        written as,
	\begin{align}
	\cos\Psi=-\frac{p_{\rm s,1}^{}+p_{\rm s,2}^{}\cos\xi}{\sqrt{p_{\rm s,1}^{2}+p_{\rm s,2}^{2}+2p_{\rm s,1}^{}p_{\rm s,2}^{}\cos\xi}}.
	\end{align}

	\item Since we consider the velocities at the time of the last ejection event, 
	these formulae can be applied independently of the details of the initial 
	binary-binary encounter. In Section \ref{results_backgroundpotential}, we compared
	$v_{\rm b}$ as estimated from the analytic formula to the results of simulations 
	of binary-binary scatterings assuming equal/unequal binding 
	energies (Figure \ref{fig:Model1_1}). We showed that our formula matches the simulated data in all cases.
	
	\item The formula is symmetric upon exchanging the momenta of the two ejected stars. 
 	In our derivations, we 
	differentiate the two single stars in terms of the order of ejection events 
	(i.e., first/second ejected star). But the order of ejection events is irrelevant for  
	estimating the final binary velocity. We have shown this in Figure 
	\ref{fig:analyticformula} (green line). This symmetric property of the 2+1+1 outcome 
	makes comparisons to observations straightforward, since the order of ejection events does not
       need to be known.
	\item For \textit{a background potential} of total mass $M_{\rm bg}$ 
	with outer boundary $r_{\rm bg}$ and escape velocity $v_{\rm esc}$, given two observed
	 runaway stars at $t=t_{\rm obs}$ with velocities measured in the CM frame of all four stars 
	 (i.e., their speeds are $v_{\rm obs,1}$, $v_{\rm obs,2}$ and the relative angle is $\xi$), the speed of the final binary at the outer boundary of the potential (i.e., $r=r_{\rm bg}$ and $t=t_{\rm bg}$) can be expressed as follows, 
							
			\begin{multline}
v_{\rm b,bg}(v_{\rm s,1,obs},~ v_{\rm s,2,obs},~m_{\rm s,1},~m_{\rm s,2},~m_{\rm b})\\
=\sqrt{\frac{1}{m_{\rm b}^{2}}\Big[\mathcal{P}_{\rm s,1}^{2}+\mathcal{P}_{\rm s,2}^{2}+2\mathcal{P}_{\rm s,1}^{}\mathcal{P}_{\rm s,2}^{}\cos\xi\Big]-\frac{1}{2}v_{\rm esc}^{2}\Big[1-\frac{\sqrt{2}r_{\rm ej,2}w}{v_{\rm esc}}\Big]^{2}},\
			\end{multline}
			where 
			\begin{align}
							\mathcal{P}_{\rm s,1}=m_{\rm s,1}\sqrt{\frac{1}{2}v_{\rm esc}^{2}\Big[1-\frac{\sqrt{2}r_{\rm ej,1}w}{v_{\rm esc}}\Big]^{2}+(v_{\rm s,1,obs})^{2}},\\
		\mathcal{P}_{\rm s,2}=m_{\rm s,2}\sqrt{\frac{1}{2}v_{\rm esc}^{2}\Big[1-\frac{\sqrt{2}r_{\rm ej,2}w}{v_{\rm esc}}\Big]^{2}+(v_{\rm s,2,obs})^{2}}.\
			\end{align}	
			Here $m_{\rm s,1}, m_{\rm s,2}$ and $m_{\rm b}$ are, respectively, the masses of the two 
			single stars and the binary. Finally, the speed and the location of a binary which has escaped from the potential at $t~(>t_{\rm bg}$) can be estimated as,
			
			\begin{align}
							v_{\rm b}(t)=\frac{2\sqrt{\frac{2GM_{\rm bg}}{\mathfrak{R}}}\Big[\sqrt{\frac{r_{\rm bg}}{\mathfrak{R}}+\Big(\frac{r_{\rm bg}}{\mathfrak{R}}\Big)^{2}}+\sqrt{\frac{2GM_{\rm bg}}{\mathfrak{R}^{3}}}(t-t_{\rm bg}) \Big]}
							{\sqrt{
									1+4\Big[\sqrt{\frac{r_{\rm bg}}{\mathfrak{R}}+\Big(\frac{r_{\rm bg}}{\mathfrak{R}}\Big)^{2}}+\sqrt{\frac{2GM_{\rm bg}}{\mathfrak{R}^{3}}}(t-t_{\rm bg})   \Big]^{2}
								}}
								\end{align}
								
								\begin{multline}
								r_{\rm b}(t)\\
								=\frac{1}{2}\mathfrak{R}\Bigg[-1+\sqrt{
									1+4\Big[\sqrt{\frac{r_{\rm bg}}{\mathfrak{R}}+\Big(\frac{r_{\rm bg}}{\mathfrak{R}}\Big)^{2}}+\sqrt{\frac{2GM_{\rm bg}}{\mathfrak{R}^{3}}}(t-t_{\rm bg})   \Big]^{2}}\Bigg]+r_{\rm bg}
								\end{multline}
								
								where $\mathfrak{R}=r_{\rm bg}/[(v_{\rm b,bg}/v_{\rm esc})^{2}-1]$. We have also derived the equations of motion 
			for the stars while they are still moving outward in the potential (see Equation \ref{eq:EOM_in}).

	\item Given the derived formulae, we have found a condition for the complete escape
	of a binary from the potential, 
				\begin{align}
		\sqrt{\mathcal{P}_{\rm s,1}^{2}+\mathcal{P}_{\rm s,2}^{2}+2\mathcal{P}_{\rm s,1}^{}\mathcal{P}_{\rm s,2}^{}\cos\xi}\geq m_{\rm b}v_{\rm esc}\sqrt{1+\frac{1}{2}\Big[1-\frac{\sqrt{2}r_{\rm ej,2}w}{v_{\rm esc}}\Big]^{2}}
				\end{align}
				
	\item In summary, we have derived analytic formulae which can
          be applied to three-body outcomes (two ejected systems and one left-over system)
           with/without a background
          potential. We expect that these formulae can be applied
          directly to observations of runway stars, and hence to be
          particularly useful with the expected detection of hundreds
          of runaway stars with the GAIA satellite.

\end{enumerate}

\vspace{0.5cm}

\section*{Acknowledgements}
The authors are grateful to the referee, Joseph Antognini, for his very 
meaningful comments and suggestions which helped 
improve the presentation of our paper. 
Results in this paper were obtained using the
high-performance LIred computing system at the Institute for Advanced
Computational Science at Stony Brook University, which was obtained
through the Empire State Development grant NYS \#28451.








\appendix
\section{Derivation for $v_{\rm b}$ with vector components-purely stellar dynamics}
\label{appendix:derivation1}
\begin{figure}
	\centering
	\includegraphics[width=8.7cm]{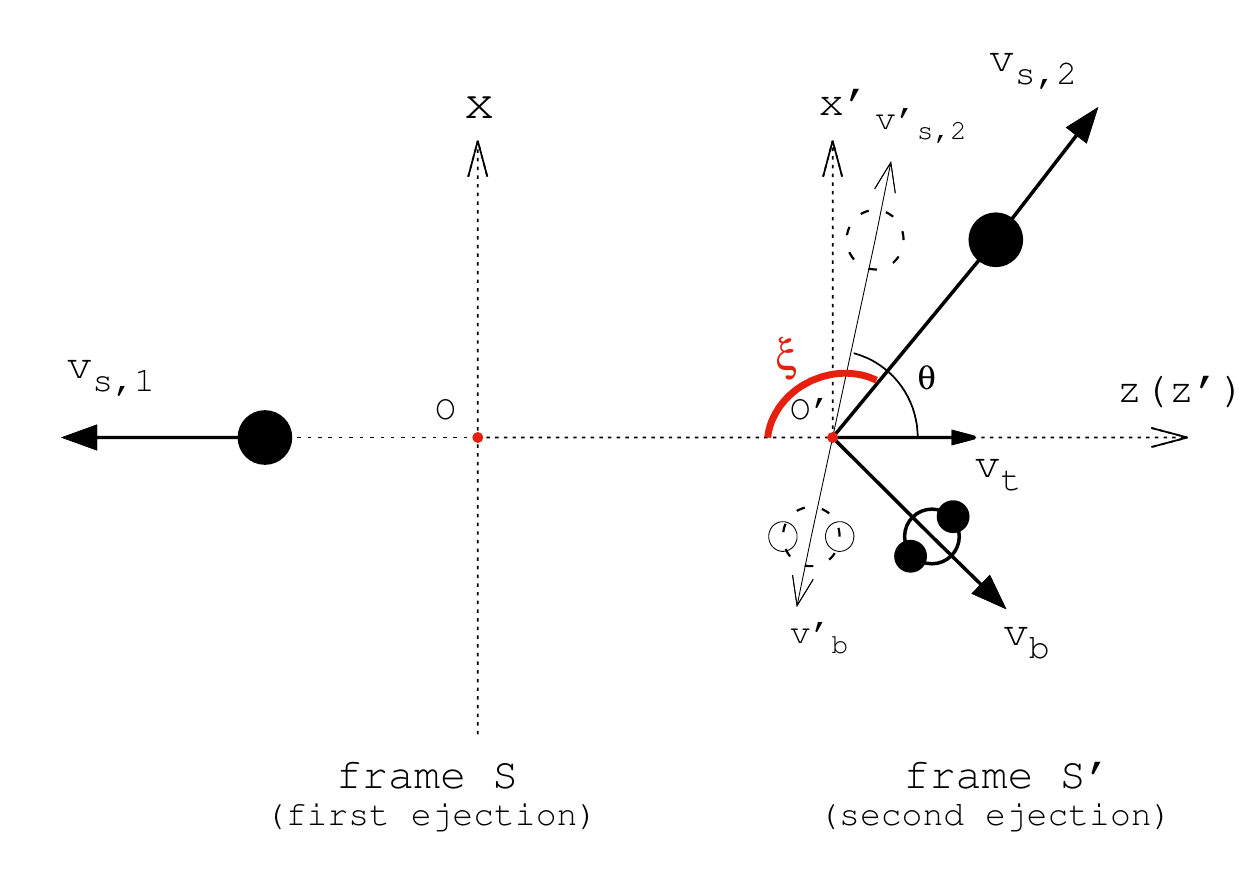}
	\caption{A schematic diagram showing the two ejection
          events. At the first ejection event, a star of mass
          $m=m_{s,1}$ is ejected in -z direction with velocity
          $\textbf{\textit{v}}_{\rm s,1}$ in frame S (with origin O),
          leaving behind a $N=$ 3 system of mass $m=m_{\rm t}$ moving
          at $\textbf{\textit{v}}_{\rm t}=(m_{\rm s,1}/m_{\rm
            t})\textbf{\textit{v}}_{\rm s,1}$. At the second (last)
          ejection event, another star of mass $m=m_{\rm s,2}$ is
          ejected with velocity $\textbf{\textit{v}}'_{\rm s,2}$ and
          the final binary of mass $m=m_{\rm b}$ is recoiled with
          velocity $(m_{\rm s,2}/m_{\rm b})\textbf{\textit{v}}'_{\rm
            s,2}$ in frame S' (with origin O'). The velocities defined
          in frame S' will be denoted with a prime symbol ('). The
          angle $\theta$ refers to the angle between the z axis
          in frame S' and the velocity of the second ejected star
		and $\xi$ is the relative angle between the two single stars in frame S.}
	\label{fig:appdix_coordinate}
\end{figure}
\label{appendix:appdix_analyticformula}
When two binaries collide, a chaotic interaction ensues until two single stars 
are ejected sequentially, leaving behind a binary pair.  For each ejection event, we adopt 
the CM frame of the ejected star and the 
remaining system (i.e., excluding any previously ejected stars). 
This derivation could prove useful for identifying the first ejected 
single star formed during a binary-binary interaction, when applying 
our method to observations of runaways where only one single star and 
an associated binary have been identified.

After the first ejection 
event, a single star of mass $m_{\rm s,1}$ and the remaining system of 
mass $m_{\rm t}$ recede in opposite directions (in the CM frame of all four stars).  We 
assume a velocity of $\textbf{\textit{v}}_{\rm s,1}$ for the first ejected single star, directed 
along the $(-z)$ axis from the CM of all four stars (denoted frame $S$). 
The remaining $N=3$ 
system recedes with velocity $\textbf{\textit{v}}_{\rm t}=(m_{\rm s,1}/m_{\rm t})\textbf{\textit{v}}_{\rm s,1}$ 
along the $z$-axis. 
At the time of the second ejection event, another single star is ejected with velocity 
$\textbf{\textit{v}}'_{\rm s,2}$ in the $x-z$ plane in the CM frame of all three 
remaining stars (denoted 
frame $S'$, with its $z'-$axis oriented such that it coincides with the $z-$axis in frame $S$).  
The final binary recedes in the opposite direction with velocity 
$\textbf{\textit{v}}'_{\rm b}=-(m_{\rm s,2}/m_{\rm b})\textbf{\textit{v}}'_{\rm s,2}$ 
relative to the second ejected single star in frame S. Hereafter, the velocities defined 
in frame S' will be denoted with a prime symbol ('). We present a schematic diagram in 
Figure \ref{fig:appdix_coordinate} showing the two ejection 
events in the two different reference frames. The velocities of both single stars and the 
final binary can be expressed in frame $S$ using polar coordinates 
as follows,
\begin{align}
\label{eq:appdix_velocity}
\textbf{\textit{v}}_{\rm s,1} &= \begin{bmatrix}
0\\
0\\
-v_{\rm s,1}
\end{bmatrix}\\
\textbf{\textit{v}}_{\rm s,2} &= 
\begin{bmatrix}
v'_{\rm s,2}~\sin\theta\\
0\\
v'_{\rm s,2}~\cos\theta+\Big(\frac{m_{\rm s,1}}{m_{\rm t}}\Big)v_{\rm s,1}
\end{bmatrix}\\
\textbf{\textit{v}}_{\rm b} &= \begin{bmatrix}
-v_{\rm b}'~\sin\theta\\
0\\
-v_{\rm b}'~\cos\theta+\Big(\frac{m_{\rm s,1}}{m_{\rm t}}\Big)v_{\rm s,1}
\end{bmatrix}=\begin{bmatrix}
-\Big(\frac{m_{\rm s,2}}{m_{\rm b}}\Big)v'_{\rm s,2}~\sin\theta\\
0\\
-\Big(\frac{m_{\rm s,2}}{m_{\rm b}}\Big)v'_{\rm s,2}~\cos\theta+\Big(\frac{m_{\rm s,1}}{m_{\rm t}}\Big)v_{\rm s,1}
\end{bmatrix}
\end{align}
where $\theta$ is the polar angle relative to the $z'-$axis.
Accordingly, their speeds in frame $S$ are,
\begin{subequations}
	\label{eq:speed}
	\begin{align}
	|\textbf{\textit{v}}_{\rm s,1}|
	&=v_{\rm s,1}\label{eq:appdix_speeds1}\\
	|\textbf{\textit{v}}_{\rm s,2}|
	&=v_{\rm s,2}=\sqrt{v{'}_{\rm s,2}^{2}+2\alpha~v{'}_{\rm s,2}v_{\rm t}+v_{\rm t}^{2}}\label{eq:appdix_speeds2}\\
	|\textbf{\textit{v}}_{\rm b}|
	&=v_{\rm b}=\sqrt{v{'}_{\rm b}^{2}-2\alpha~v'_{\rm b}v_{\rm t}+v_{\rm t}^{2}}\label{eq:appdix_speedb}\
	\end{align}
\end{subequations}
where $\alpha=\cos\theta$.
Given the preceding velocities, we can estimate the relative angle $\xi$ between the 
two ejected single stars, 
\begin{align}
\label{eq:appdix_relativeangle}
\cos\xi
&=\frac{\textbf{\textit{v}}_{\rm s,1}\cdot\textbf{\textit{v}}_{\rm s,2}}{v_{\rm s,1}v_{\rm s,2}}
=-\frac{v'_{\rm s,2}\alpha+v_{\rm t}}{\sqrt{v{'}_{\rm s,2}^{2}+2\alpha~v'_{\rm s,2}v_{\rm t}+v_{\rm t}^{2}}}\nonumber\\ 
&=\frac{\alpha+\frac{v_{\rm t}}{v'_{\rm s,2}}}{\sqrt{(\frac{v_{\rm t}}{v{'}_{\rm s,2}})^{2}+2\alpha (\frac{v_{\rm t}}{v'_{\rm s,2}})+1}}=-\frac{\alpha+{A}}{\sqrt{{A}^{2}+2\alpha{A}+1}}
\end{align}
where ${A}=v_{\rm t}/v'_{\rm s,2}$. 

Next, we motivate our definitions for the angles 
$\xi$ and $\alpha$. We have introduced two inertial frames of reference ($S$ and $S'$). 
In frame $S$ (i.e., the CM frame of all four stars), $\xi$ corresponds to the relative angle 
between the velocity vectors of the two ejected single stars.  
The variable $\alpha$, on the other hand, is defined in the $S'$ frame (i.e., the CM frame of 
the remaining system of three stars left-over after the first ejection event).  It measures the 
relative angle between the velocity vector of 
the second ejected star and the CM velocity vector of the remaining (temporary) $N=$ 3 system 
(i.e., frame $S'$ itself, or the direction opposite to the direction of motion of 
the first ejected single star). In what follows, we will identify every allowed combination 
of these two angles.

In order to find an expression for $v_{\rm b}$ as a function of 
$v_{\rm s,1},~v_{\rm s,2}$, $\xi$ and the masses $m$ of the stars, 
i.e., $v_{\rm b}=v_{\rm b}(v_{\rm s,1},~v_{\rm s,2},~\xi,~m)$, 
we combine Equations \ref{eq:appdix_speeds2} and \ref{eq:appdix_relativeangle} and 
solve for $\alpha$ as a function of $\xi$. Hereafter, we will use the vector $m$ to refer to the masses of two ejected single stars and one binary collectively, or $(m_{\rm s,1},~m_{\rm s,2},~m_{\rm b})$ unless it is necessary to denote each of them separately. Plugging $\alpha(\xi)$ 
back into equation \ref{eq:appdix_relativeangle}, we obtain ${A} =$ ${A}(\xi,v_{\rm s,2})$. Finally, 
using the relation $v'_{\rm b}=(m_{\rm s,2}/m_{\rm b})v'_{\rm s,2}$ in conjunction with 
Equation \ref{eq:appdix_speedb}, we obtain $v_{\rm b}=v_{\rm b}(v_{\rm s,1},~v_{\rm s,2},~\xi,~m)$. 
For tractability of the newly defined parameters, we use Greek letters (e.g., $\alpha$ and $\xi$) for angle-related parameters and capital Roman letters (e.g., ${A}$ and ${B}$) for velocity- (momentum-) or distance-related parameters.

Now, we start with calculating an expression for ${A}$ from Equation 
\ref{eq:appdix_relativeangle} as follows,
\begin{align}
\label{eq:appdix_beta}
{A}=-\alpha\pm \cot\xi \sqrt{1-\alpha^{2}}.
\end{align}
We can get another expression for ${A}$ from Equation \ref{eq:appdix_speeds2}.  That is, 
solving for $v'_{\rm s,2}$ in Equation \ref{eq:appdix_speeds2},
\begin{align}
\label{eq:appdix_v2CM}
v'_{\rm s,2}&=-\alpha v_{\rm t}\pm \sqrt{\alpha v_{\rm t}^{2}-v_{\rm t}^{2}+v_{\rm s,2}^{2}}\nonumber\\
&=v_{\rm t}(-\alpha\pm\sqrt{\alpha^{2}-1+{B}^{2}})
\end{align}
where we define ${B}=v_{\rm s,2}/v_{\rm t}$. Using the relation ${A}=v_{\rm t}/v'_{\rm s,2}$, we find,
\begin{align}
\label{eq:appdix_beta2}
{A}&=\frac{1}{-\alpha\pm\sqrt{\alpha^{2}-1+{B}^{2}}}.
\end{align}
Equating Equations \ref{eq:appdix_beta} and \ref{eq:appdix_beta2}, 
\begin{align}
\label{eq:appdix_equation}
\Big[-\alpha\pm \cot\xi \sqrt{1-\alpha^{2}}\Big]\Big[-\alpha\pm\sqrt{\alpha^{2}-1+{B}^{2}}\Big]=1
\end{align}
Given the equation above, we can find an expression for $\alpha$ in terms of $\xi$ and ${B}$.
There are four possible solutions for $\alpha(\xi,{B})$ (or two, adopting a different range for $\xi$ and ${B}$):
\begin{align}
\label{eq:appdix_alpha}
\alpha(\xi,{B}) = \begin{cases}
\vspace{0.1in}
&-\frac{|1-{B}~\cos\xi|}{\sqrt{1+{B}^{2}-2{B}\cos\xi}}\\
\vspace{0.1in} 
&\frac{|1-{B}~\cos\xi|}{\sqrt{1+{B}^{2}-2{B}\cos\xi}}\\
\vspace{0.1in}
&-\frac{|1+{B}~\cos\xi|}{\sqrt{1+{B}^{2}+2{B}\cos\xi}}\\
\vspace{0.1in}
&\frac{|1+{B}~\cos\xi|}{\sqrt{1+{B}^{2}+2{B}\cos\xi}}\\
\end{cases}
\end{align}

The final relative orientation between the two ejected single stars, characterized by the parameters $\xi$ and 
$\alpha$, allow or forbid different combinations of the plus and minus signs (in each closed 
bracket) in Equation \ref{eq:appdix_equation}.  This, in turn, specifies the solution for $\alpha(\xi,{B})$, of 
the four possible solutions in Equation \ref{eq:appdix_alpha}. Hereafter, we consider two separate 
ranges for $\xi$, i.e., $0<\xi<\uppi/2$ and $\uppi/2<\xi<\uppi$. For each range of $\xi$, 
we must further distinguish between all possible cases using 
$\alpha$ (which can be positive/negative). We will henceforth use the "same (opposite) direction" to 
refer to cases where the relative angle between the final velocity vectors of the two single stars is 
smaller (larger) than $\uppi/2$.

\subsection{Case 1. $\frac{\uppi}{2}<\xi\leq\uppi$ and positive $\alpha$ ($0\leq\theta\leq\frac{\pi}{2}$)}

In Mode 1, a single star is ejected in the same direction 
	as the CM velocity of frame $S$ ($0\leq\theta\leq\frac{\pi}{2}$) and in the opposite direction with respect 
	to the first ejected star ($\textbf{\textit{v}}_{\rm s,2}\cdot\textbf{\textit{e}}_{\rm z}>0$ for $\frac{\uppi}{2}<\xi\leq\uppi$ where $\textbf{\textit{e}}_{z}$ is the unit vector in the $z$ direction). 
For this case, there are two possible directions of motion for the final binary, depending on the relative 
magnitudes of the speed of the binary and the CM speed of frame $S'$. In other words, 
if the binary is kicked in the $-z'$ direction at a speed smaller (higher) than the CM speed 
of frame $S'$, then in frame $S$ the binary is seen as moving in the $z (-z)$ direction. These 
geometry-based arguments allow for only one sign for both ${A}$ and $v'_{\rm s,2}$. 
If ${A}>0$, $v'_{\rm s,2}>0$, $\cot\xi<0$ and $\alpha\geq0$, then 
the only allowed expressions for ${A}$ and $v'_{\rm s,2}$ are,
\begin{align}
\label{eq:appdix_solution_beta}
{A}&=-\alpha-\cot\xi \sqrt{1-\alpha^{2}}
\end{align}
\begin{align}
\label{eq:appdix_vs2CM}
v'_{\rm s,2}&=v_{\rm t}\Big[-\alpha+\sqrt{\alpha^{2}-1+{B}^{2}}\Big]
\end{align}
Given a real number for $v'_{\rm s,2}$, 
we impose the condition that $\alpha^{2}-1+{B}^{2}>0$.  We will return to 
this condition later (Condition \ref{con:appdix_1_case1} below).  Equation \ref{eq:appdix_equation} now reduces to,
\begin{align}
\label{eq:appdix_equation_case1}
\Big[-\alpha-\cot\xi \sqrt{1-\alpha^{2}}\Big]\Big[-\alpha+\sqrt{\alpha^{2}-1+{B}^{2}}\Big]=1
\end{align}

The corresponding solutions for $\alpha(\xi,{B})$ are: 

For  ${B}\geq1$,
\begin{numcases}{\alpha(\xi,{B})=}
-\frac{|1+{B}\cos\xi|}{\sqrt{1+{B}^{2}+2{B}\cos\xi}}\hspace{0.2in} 1+{B}\cos\xi\geq0\,;\label{eq:appdix_solution_alpha_case1_1}\\
\frac{|1+{B}\cos\xi|}{\sqrt{1+{B}^{2}+2{B}\cos\xi}}\hspace{0.25in}~1+{B}\cos\xi<0\,\,.\label{eq:appdix_solution_alpha_case1_2}
\end{numcases}

For  ${B}<1$,
\begin{numcases}{\alpha(\xi,{B})=}
-\frac{|1+{B}\cos\xi|}{\sqrt{1+{B}^{2}+2{B}\cos\xi}}\hspace{0.25in}~{B}+\cos\xi>0\,\,.\label{eq:appdix_solution_alpha_case1_3}
\end{numcases}

Since $\alpha(\xi,{B})\geq0$, all solutions with a negative sign in front are ruled out. 
However, before we can solve for $v_{\rm b}$, we must consider a couple of additional 
conditions.  These are designed to ensure the validity of our expression for $\alpha$, 
and to find a proper range for $\xi$ that is allowed in this case.  For this and every subsequent case, 
we will refer to the following conditions : 
\begin{enumerate}
	\item  \label{con:appdix_1_case1}$\alpha^{2}-1+{B}^{2}\geq0$. 
	
	This is to ensure that $v'_{\rm s,2}$ is a real number. In case 1, this condition is always satisfied, 
	\begin{align}
	\alpha^{2}-1+{B}^{2}={B}^{2}\Big[\frac{\cos^{2}\xi+{B}^{2}+2{B}\cos\xi}{1+{B}^{2}+2{B}\cos\xi}\Big]\geq0
	\end{align}
	This is satisfied for all solutions (Equation \ref{eq:appdix_alpha})\footnote{In general, for $\alpha\sim\frac{|1\pm{B}~\cos\xi|}{\sqrt{1\pm{B}^{2}+2{B}\cos\xi}}$,
		
		\begin{align}
		\alpha^{2}-1+{B}^{2}={B}^{2}\Big[\frac{\cos^{2}\xi+{B}^{2}\pm2{B}\cos\xi}{1+{B}^{2}\pm2{B}\cos\xi}\Big]>0
		\end{align}

	}. Therefore, we assume $v'_{\rm s,2}$ is a real number for all allowed expressions for $\alpha$.

	\item ${A}>0$.

		For the solution for $\alpha(\xi,{B})$ (Equation \ref{eq:appdix_solution_alpha_case1_2}), it is enough to use either of two different expressions for ${A}$ (Equations \ref{eq:appdix_solution_beta} and \ref{eq:appdix_vs2CM}) to show 
		that this condition is satisfied. Using Equation \ref{eq:appdix_vs2CM}, it is easy to see that $v'_{\rm s,2}$ is 
		positive for ${B}>1$. For ${B}<1$,
	
	\begin{align}
	v'_{\rm s,2}
	&=v_{\rm t}\Big[\frac{1+{B}\cos\xi}{\sqrt{1+{B}^{2}+2{B}\cos\xi}}+
	\frac{|{B}({B}+\cos\xi)|}{\sqrt{1+{B}^{2}+2{B}\cos\xi}}\Big]\nonumber\\
	&=v_{\rm t}\sqrt{1+{B}^{2}+2{B}\cos\xi}>0
	\end{align}

	\item $\textbf{\textit{v}}_{s,2}\cdot \textbf{\textit{e}}_{z}>0$. 
	
	This condition is clearly satisfied for positive $\alpha$.
	
\end{enumerate}
\vspace{0.3in}
Finally, with Equations \ref{eq:appdix_vs2CM}, \ref{eq:appdix_speedb} and $v_{\rm b}=(m_{\rm s,2}/m_{\rm b})v_{\rm s,2}$, we are able to find a solution for $v'_{\rm b}$ for \textbf{Case 1},
\begin{align}
\label{eq:appdix_solution_vbrest}
v_{\rm b}&=v_{\rm t}\sqrt{\Big(\frac{m_{\rm s,2}}{m_{\rm b}}\Big)^{2}({B}^{2}+2{B}\cos\xi+1)+2\Big(\frac{m_{\rm s,2}}{m_{\rm b}}\Big)(1+{B}\cos\xi)+1}
\end{align}
for $\cos\xi\leq\frac{-1}{{B}}$ and ${B}>1$.

\subsection{Case 2. $\frac{\uppi}{2}\leq\xi\leq\uppi$ and negative $\alpha$ ($\frac{\pi}{2}<\theta\leq\pi$)}
In this case, both the final binary ($-1\leq\alpha<0$) and the second
ejected star ($\frac{\uppi}{2}\leq\xi\leq\uppi$) are moving in the
opposite direction (or $\textbf{\textit{v}}_{s,2}\cdot
\textbf{\textit{e}}_{z}>0$) to the first ejected star, in frame
$S$. This geometric property allows for both signs for both ${A}$ and
$v'_{\rm s,2}$. Hence, it is necessary to consider each combination of
the signs individually. We refer to each combination as (+,+), (+,-),
(-,+) and (-,-), respectively.

\subsubsection{ \textbf{(+,+) case}}
\label{case:appdix_2_++}

For negative $\alpha$, we know that both ${A}$ and $v'_{\rm s,2}$ are positive, 
\begin{numcases}{}
{A}=-\alpha+\cot\xi \sqrt{1-\alpha^{2}}	\label{eq:appdix_solution_beta_case2++},\,\\
v'_{\rm s,2}=v_{\rm t}\Big[-\alpha+\sqrt{\alpha^{2}-1+{B}^{2}}\Big].	\label{eq:appdix_solution_vs2cm_case2++}
\end{numcases}
Accordingly, Equation \ref{eq:appdix_equation} becomes
\begin{align}
\Big[-\alpha+\cot\xi \sqrt{1-\alpha^{2}}\Big]\Big[-\alpha+\sqrt{\alpha^{2}-1+{B}^{2}}\Big]=1.
\end{align}
There is only one solution for the equation above, namely
\begin{align}
\label{eq:appdix_alpha_case2++}
\alpha(\xi,{B})&=
-\frac{|1-{B}~\cos\xi|}{\sqrt{1+{B}^{2}-2{B}\cos\xi}},\nonumber\\
=&-\frac{1-{B}~\cos\xi}{\sqrt{1+{B}^{2}-2{B}\cos\xi}},
\end{align}
where we have used $\cos\xi\leq0$ in the second equality. 
As mentioned in \textbf{Case 1}, there are a couple of conditions for $\xi$ which must be satisfied in order to ensure physically meaningful solutions.
Condition \ref{con:appdix_case2_++1} below needs to be confirmed since in this case the solutions for ${A}$ and $v_{\rm s,2}$ are not manually chosen to be positive. The conditions are:

\begin{enumerate}
	
	\item\label{con:appdix_case2_++1}	${A}>0$. 
	
	As mentioned before, we can use either of the two expressions for ${A}$ to show that this condition is satisfied. 
	Clearly, ${A}>0$ from Equation  \ref{eq:appdix_solution_vs2cm_case2++} for $\alpha<0$.
	
	\item $\textbf{\textit{v}}_{s,2}\cdot \textbf{\textit{e}}_{z}\geq0$. 
	
	We write this relation as follows,
	\begin{align}
	\label{eq:appdix_case2++}
	\textbf{\textit{v}}_{s,2}\cdot \textbf{\textit{e}}_{x}&=v'_{\rm s,2}\alpha+v_{\rm t}\nonumber\\
	&=v_{\rm t}\Big(-\alpha+\sqrt{\alpha^{2}-1+{B}^{2}}\Big)\alpha+v_{\rm t}\geq0.
	\end{align}
	We find from the relation above that 
	\begin{numcases}{}
	-\frac{1}{\sqrt{1+{B}^{2}}}\leq\alpha<-\sqrt{1-{B}^{2}} \hspace{0.25in} \text{for} \,\,\,\,\ 0\leq{B}<1,\label{eq:appdix_case2++_1}\\
	-\frac{1}{\sqrt{1+{B}^{2}}}\leq\alpha<0 \hspace{0.7in}\text{for}\,\,\,\,\  {B}\geq1.\label{eq:appdix_case2++_2}
	\end{numcases}
	
	However, for any ${B}$ (and given an expression for $\alpha$; Equation \ref{eq:appdix_alpha_case2++}), $\alpha$ is always $< -\frac{1}{\sqrt{1+{B}^{2}}}$
	\footnote{For $\cos\xi<0$, \begin{align}
		\alpha^{2}-\frac{1}{1+{B}^{2}}&\sim(1-{B})^{2}(1+{B}^{2})-(1+{B}^{2}-2{B}\cos\xi)\nonumber\\
		&\sim{B}^{2}\cos\xi({B}^{2}\cos\xi+\cos\xi-2{B})\geq0
		\end{align}
	}. Therefore, we can not find solutions for $\alpha$ which satisfy all the required conditions in this case. 
\end{enumerate}

\subsubsection{\textbf{(+,-) case}}

Given these signs for ${A}$ and $v'_{\rm s, 2}$, we have
\begin{numcases}{}
{A}=-\alpha+\cot\xi \sqrt{1-\alpha^{2}}	\label{eq:solution_beta_case2+-}\,,\\
v'_{\rm s,2}=v_{\rm t}\Big[-\alpha-\sqrt{\alpha^{2}-1+{B}^{2}}\Big],	\label{eq:appdix_solution_vs2cm_case2+-}
\end{numcases}
and Equation \ref{eq:appdix_equation} becomes
\begin{align}
\Big[-\alpha+\cot\xi \sqrt{1-\alpha^{2}}\Big]\Big[-\alpha-\sqrt{\alpha^{2}-1+{B}^{2}}\Big]=1.
\end{align}
We have different solutions for $\alpha(\xi,{B})$ for ${B}>1$ and ${B}<1$:

For  ${B}>1$,
\begin{numcases}{\alpha(\xi,{B})=}
-\frac{|1+{B}\cos\xi|}{\sqrt{1+{B}^{2}+2{B}\cos\xi}}\hspace{0.2in} 1+{B}\cos\xi<0\,;\label{eq:appdix_solution_alpha_case2+-1}\\
\frac{|1+{B}\cos\xi|}{\sqrt{1+{B}^{2}+2{B}\cos\xi}}\hspace{0.25in}~1+{B}\cos\xi>0\,\,.\label{eq:appdix_solution_alpha_case2+-2}
\end{numcases}

For  ${B}<1$,
\begin{numcases}{\alpha(\xi,{B})=}
-\frac{|1+{B}\cos\xi|}{\sqrt{1+{B}^{2}+2{B}\cos\xi}}\hspace{0.25in}~{B}+\cos\xi<0\,\,.\label{eq:appdix_solution_alpha_case2+-3}\\
\frac{|1+{B}\cos\xi|}{\sqrt{1+{B}^{2}+2{B}\cos\xi}}\hspace{0.3in}~{B}+\cos\xi>0\,\,.\label{eq:appdix_solution_alpha_case2+-4}
\end{numcases}

Since the solutions \ref{eq:appdix_solution_alpha_case2+-2} and \ref{eq:appdix_solution_alpha_case2+-4} are positive definite, they can be ruled out due to the fact that $\alpha<0$. In addition, for ${B}>1$ (i.e., where the solution \ref{eq:appdix_solution_alpha_case2+-1} is valid), we exclude solution \ref{eq:appdix_solution_alpha_case2+-1} because $v'_{\rm s,2}$ becomes negative. Only solution \ref{eq:appdix_solution_alpha_case2+-4} remains. However, we need to check the following conditions for the remaining solution to ensure it is physically allowed:

\begin{enumerate}
	\item  	${A}>0$. 
	
	Starting with Equation \ref{eq:appdix_solution_vs2cm_case2+-},
	\begin{align}
	v'_{\rm s,2}&=\frac{|1+{B}\cos\xi|}{\sqrt{1+{B}^{2}+2{B}\cos\xi}}-\frac{{B}|{B}+\cos\xi|}{\sqrt{1+{B}^{2}+2{B}\cos\xi}},\nonumber\\
	&=\frac{1+{B}\cos\xi}{\sqrt{1+{B}^{2}+2{B}\cos\xi}}-\frac{{B}({B}+\cos\xi)}{\sqrt{1+{B}^{2}+2{B}\cos\xi}|},\nonumber\\
	&=\sqrt{1+{B}^{2}+2{B}\cos\xi}>0
	\end{align}
	where in the second equality, we have used the conditions for ${B}$ that ${B}<1$ and ${B}+\cos\xi>0$. For these conditions, $1+{B}\cos\xi>0$.
	
	\item $\textbf{\textit{v}}_{s,2}\cdot \textbf{\textit{e}}_{z}>0$. 
	
	We write this relation as follows,
	\begin{align}
	\textbf{\textit{v}}_{s,2}\cdot \textbf{\textit{e}}_{z}&=v'_{\rm s,2}\alpha+v_{\rm t}\nonumber\\
	&=v_{\rm t}\Big(-\alpha-\sqrt{\alpha^{2}-1+{B}^{2}}\Big)\alpha+v_{\rm t}>0.
	\end{align}
	Re-arranging terms, we find that
	\begin{align}
	-1\leq\alpha<-\sqrt{1-{B}^{2}} \hspace{0.25in} \text{for} \,\,\,\,\ 0<{B}<1,\label{eq:appdix_case2+-_1}
	\end{align}
	Inserting Equation \ref{eq:appdix_solution_alpha_case2+-3} into Equation \ref{eq:appdix_case2+-_1}, it turns out that this condition is always satisfied since
	\begin{align}
	\alpha^{2}-(1-{B})^{2}=\frac{{B}^{2}({B}+\cos\xi)^{2}}{1+{B}^{2}+2{B}\cos\xi}>0
	\end{align}
	
\end{enumerate}
\vspace{0.3in}
Hence, we have only one solution for this case,
\begin{align}
\label{eq:appdix_solution_vs2CM_case2+-}
v'_{\rm s,2}&=v_{\rm t}\sqrt{{B}^{2}+2{B}\cos\xi+1}
\end{align}
\begin{align}
\label{eq:appdix_solution_vbrest_case2+-}
v_{\rm b}&=v_{\rm t}\sqrt{\Big(\frac{m_{\rm s,2}}{m_{\rm b}}\Big)^{2}({B}^{2}+2{B}\cos\xi+1)+2\Big(\frac{m_{\rm s,2}}{m_{\rm b}}\Big)(1+{B}\cos\xi)+1}
\end{align}
for $0<{B}<1$ and $-1<\cos\xi<-{B}$.

\subsubsection{\textbf{(-,+) case}}

Given these signs for ${A}$ and $v_{\rm s, 2}$,
\begin{numcases}{}
{A}=-\alpha-\cot\xi \sqrt{1-\alpha^{2}}	\label{eq:appdix_solution_beta_case2-+}\,,\\
v'_{\rm s,2}=v_{\rm t}\Big[-\alpha+\sqrt{\alpha^{2}-1+{B}^{2}}\Big],	\label{eq:appdix_solution_vs2cm_case2-+}
\end{numcases}
Equation \ref{eq:appdix_equation} becomes
\begin{align}
\Big[-\alpha-\cot\xi \sqrt{1-\alpha^{2}}\Big]\Big[-\alpha+\sqrt{\alpha^{2}-1+{B}^{2}}\Big]=1.
\end{align}
We have obtained the same solutions as for the \textbf{\textit{(+,-) case}}, but with different 
ranges for $\xi$:

For  ${B}\geq1$,
\begin{numcases}{\alpha(\xi,{B})=}
-\frac{|1+{B}\cos\xi|}{\sqrt{1+{B}^{2}+2{B}\cos\xi}}\hspace{0.2in} 1+{B}\cos\xi\geq0\,;\label{eq:appdix_solution_alpha_case2-+1}\\
\frac{|1+{B}\cos\xi|}{\sqrt{1+{B}^{2}+2{B}\cos\xi}}\hspace{0.25in}~1+{B}\cos\xi<0\,\,.\label{eq:appdix_solution_alpha_case2-+2}
\end{numcases}
For  ${B}<1$,
\begin{numcases}{\alpha(\xi,{B})=}
-\frac{|1+{B}\cos\xi|}{\sqrt{1+{B}^{2}+2{B}\cos\xi}}\hspace{0.25in}~{B}+\cos\xi\geq0\,\,.\label{eq:appdix_solution_alpha_case2-+3}\\
\frac{|1+{B}\cos\xi|}{\sqrt{1+{B}^{2}+2{B}\cos\xi}}\hspace{0.25in}~{B}+\cos\xi>0\,\,.\label{eq:appdix_solution_alpha_case2-+4}
\end{numcases}

Solutions \ref{eq:appdix_solution_alpha_case2-+2} and \ref{eq:appdix_solution_alpha_case2-+4} (positive definite) can be ruled out if $\alpha<0$. Note that, for ${B}<1$, $1+{B}\cos\xi>0$ (the numerator of solution \ref{eq:appdix_solution_alpha_case2-+3}). As before, the conditions that must be satisfied to ensure the validity of the solutions are:
\begin{enumerate}
	\item  	${A}>0$. 
	
	This condition is already fulfilled given the choice of signs in Equations \ref{eq:appdix_solution_beta_case2-+} and \ref{eq:appdix_solution_vs2cm_case2-+}.
	
	\item $\textbf{\textit{v}}_{s,2}\cdot \textbf{\textit{e}}_{x}\geq0$. 
	
	Given the same expressions as for Equations \ref{eq:appdix_case2++}, \ref{eq:appdix_case2++_1} and \ref{eq:appdix_case2++_2},
	
	the necessary conditions are described in terms of $\xi$,\
	\begin{align}
	-\frac{2{B}}{1+{B}^{2}}\leq\cos\xi<0 \hspace{0.65in} \text{for all} \,\ {B},\label{eq:appdix_case2-+_3}
	\end{align}
	
\end{enumerate}

Combining the allowed ranges for $\xi$ (see Equations \ref{eq:appdix_solution_alpha_case2-+1} and \ref{eq:appdix_solution_alpha_case2-+3}), we have different conditions for $\xi$ depending on whether or not ${B}$ is larger than 1, i.e.,
\begin{numcases}{}
-{B}\leq\cos\xi<0 \hspace{0.7in} \text{for} \,\,\,\,\,\ 0\leq{B}<1 \label{eq:appdix_case2-+_5}\\ 
-\frac{1}{{B}}\leq\cos\xi<0 \hspace{0.7in}\text{for}\,\,\,\,\  {B}\geq1.\label{eq:appdix_case2-+_6}
\end{numcases}

Note that provided $-(2{B})/(1+{B}^{2})<-{B}$ for $0\leq{B}<1$, the allowed range for $\cos\xi$ is stringently determined by ${-B}$, i.e., $-B\leq\cos\xi$\footnote{\label{appdix_relativemagnitude}.  A hierarchical comparison of the magnitudes of $\frac{2{B}}{1+{B}^{2}}$, $\frac{1}{{B}}$ and ${B}$ is as follows,
	\begin{align}
	\begin{cases}
	\frac{1}{{B}}>\frac{2{B}}{1+{B}^{2}}>{B} \hspace{0.25in} \text{for}\,\,\,\ 0<{B}<1,\\
	{B}\geq\frac{2{B}}{1+{B}^{2}}\geq\frac{1}{{B}} \hspace{0.25in} \text{for}\,\,\,\, {B}\geq1,\\
	\end{cases}
	\end{align}
}.
Finally, the solutions can be written, 
\begin{align}
\label{eq:appdix_solution_vs2CM_case2-+}
v'_{\rm s,2}&=v_{\rm t}\sqrt{{B}^{2}+2{B}\cos\xi+1}
\end{align}
\begin{align}
\label{eq:appdix_solution_vbrest_case2-+}
v_{\rm b}&=v_{\rm t}\sqrt{\Big(\frac{m_{\rm s,2}}{m_{\rm b}}\Big)^{2}({B}^{2}+2{B}\cos\xi+1)+2\Big(\frac{m_{\rm s,2}}{m_{\rm b}}\Big)(1+{B}\cos\xi)+1}
\end{align}
for 
\begin{numcases}{}
-{B}\leq\cos\xi<0 \hspace{0.7in} \text{for} \,\,\,\ 0\leq{B}<1 \nonumber\\ 
-\frac{1}{{B}}\leq\cos\xi<0 \hspace{0.7in}\text{for}\,\,\,\,\  {B}\geq1.\nonumber
\end{numcases}

\subsubsection{\textbf{(-,-) case}}

In this case, there is no solution for negative values of $\alpha$ that satisfy Equation \ref{eq:appdix_equation}.

\subsection{Case 3. $0\leq\xi<\frac{\uppi}{2}$ and positive $\alpha$ ($0\leq\theta<\frac{\pi}{2}$)}
In this case, the ranges for $\xi$ and $\alpha$ are incompatible. For positive $\alpha$, 
the second ejected star is moving in the same direction as frame $S$ itself (or, the positive 
$z$ component of $\textbf{\textit{v}}_{\rm s ,2}$). Therefore, 
$\textbf{\textit{v}}_{s,2}\cdot \textbf{\textit{e}}_{z}$ should be positive, which
forbids the range $0\leq\xi<\frac{\uppi}{2}$. Therefore, this case is not physically allowed. 

\subsection{Case 4. $0\leq\xi\leq\frac{\uppi}{2}$ and negative $\alpha$ ($\frac{\pi}{2}\leq\theta\leq\pi$)}
In this case, the second ejected star is moving (in frame $S'$) in the direction opposite 
to the CM motion of frame $S'$ ($\frac{\pi}{2}\leq\theta\leq\pi$). 
In frame S, the two ejected stars recede in the same direction (-z direction, $0\leq\xi<\frac{\uppi}{2}$). The binary, however, is recoiled in the opposite direction (z direction, $\textbf{\textit{v}}_{s,2}\cdot \textbf{\textit{e}}_{z}>0$). In this case, it is more likely for the binary to move at high speeds relative to the other cases. 
As for case 2, since it is possible for ${A}$ and $r_{\rm s,2}$ to have either sign, we have to consider each case individually.

\subsubsection{ \textbf{(+,+) case}}

We take the positive signs in both Equations \ref{eq:appdix_beta} and \ref{eq:appdix_v2CM},
\begin{align}
\label{eq:appdix_solution_case4++beta}
{A}=-\alpha+\cot\xi \sqrt{1-\alpha^{2}}
\end{align}
\begin{align}
\label{eq:appdix_solution_case4++vs2CM}
v'_{\rm s,2}=v_{\rm t}\Big[-\alpha+\sqrt{\alpha^{2}-1+{B}^{2}}\Big]
\end{align}
We have two sets of solutions for different ranges of ${B}$,

For ${B}\geq1$,
\begin{numcases}{\alpha(\xi,{B})=}
-\frac{|1-{B}\cos\xi|}{\sqrt{1+{B}^{2}-2{B}\cos\xi}}\hspace{0.2in} 1-{B}\cos\xi\geq0\,;\label{eq:appdix_solution_alpha_case4++1}\\
\frac{|1-{B}\cos\xi|}{\sqrt{1+{B}^{2}-2{B}\cos\xi}}\hspace{0.25in}~1-{B}\cos\xi<0\,\,.\label{eq:appdix_solution_alpha_case4++2}
\end{numcases}
and for ${B}<1$,
\begin{numcases}{\alpha(\xi,{B})=}
-\frac{|1-{B}\cos\xi|}{\sqrt{1+{B}^{2}-2{B}\cos\xi}}\hspace{0.2in} {B}-\cos\xi\geq0\,;\label{eq:appdix_solution_alpha_case4++3}\\
\frac{|1-{B}\cos\xi|}{\sqrt{1+{B}^{2}-2{B}\cos\xi}}\hspace{0.25in}~{B}-\cos\xi<0\,\,.\label{eq:appdix_solution_alpha_case4++4}
\end{numcases}
The condition $-1\leq\alpha<0$ excludes the solutions \ref{eq:appdix_solution_alpha_case4++2} and \ref{eq:appdix_solution_alpha_case4++4}. As before, we must check the following conditions:

\begin{table*}
	\centering
	\setlength\extrarowheight{4pt}
	\begin{tabulary}{0.7\linewidth}{ c| c c}
		\hline
		case&$v'_{\rm b}$ & $\xi$ \\
		\hline
		case 1 & $\frac{1}{m_{\rm b}}\sqrt{p_{\rm s,1}^{2}+p_{\rm s,2}^{2}+2p_{\rm s,1}p_{\rm s,2}\cos\xi}$& $\cos^{-1}(\frac{-1}{{B}})\leq\xi\leq\uppi$ for ${B}>1$.\\
		case 2 (++) & No solutions	&-\\
		case 2 (+-) &  $\frac{1}{m_{\rm b}}\sqrt{p_{\rm s,1}^{2}+p_{\rm s,2}^{2}+2p_{\rm s,1}p_{\rm s,2}\cos\xi}$ & $\cos^{-1}(-{B})<\xi<\uppi$ for ${B}<1$.\\
		case 2 (-+) &  $\frac{1}{m_{\rm b}}\sqrt{p_{\rm s,1}^{2}+p_{\rm s,2}^{2}+2p_{\rm s,1}p_{\rm s,2}\cos\xi}$ & $\frac{\uppi}{2}<\xi\leq \cos^{-1}(\frac{-1}{{B}})$ for ${B}\geq1$ and $\frac{\uppi}{2}<\xi\leq \cos^{-1}(-{B})$ for ${B}<1$.\\
		case 2 (--) & No solutions & - \\
		case 3  & No solutions	&-\\
		case 4 (++) & No solutions & - \\
		case 4 (+-) & No solutions & - \\
		case 4 (-+) &  $\frac{1}{m_{\rm b}}\sqrt{p_{\rm s,1}^{2}+p_{\rm s,2}^{2}+2p_{\rm s,1}p_{\rm s,2}\cos\xi}$ & $0\leq\xi\leq \frac{\uppi}{2}$ for any ${B}$ \\
		case 4 (--) & No solutions & - \\
		\hline
	\end{tabulary}
	\caption{$p_{\rm s,1}$ and $p_{\rm s,2}$ are the momenta of the first and second ejected star, respectively, i.e., $p_{{\rm s},i}=m_{\rm i}v_{{\rm s},i}$ ($i=1,~2$). ${B}=(m_{\rm t}v_{\rm s,2})/(m_{\rm s,1}v
		_{\rm s,1})$ where $m_{\rm t}$ is the mass of the substellar system which the first ejected star left behind, i.e., $m_{\rm t}=m_{\rm s,2}+m_{\rm b}$. }
	\label{tab:appdix_analyticsolution}
\end{table*}
\begin{enumerate}
	
	\item  	${A}>0$. 
	
	This condition is already satisfied given the choice of signs in Equations \ref{eq:appdix_solution_case4++beta} and \ref{eq:appdix_solution_case4++vs2CM}.

	\item $\textbf{\textit{v}}_{s,2}\cdot \textbf{\textit{e}}_{z}\leq0$. 
	
	We already derived an expression for $\textbf{\textit{v}}_{s,2}\cdot \textbf{\textit{e}}_{z}$ (see Equations \ref{eq:appdix_case2++}, \ref{eq:appdix_case2++_1} and \ref{eq:appdix_case2++_2}). The corresponding condition can be written as,
	\begin{align}
	\label{eq:appdix_case4++_xcomp}
	\textbf{\textit{v}}_{s,2}\cdot \textbf{\textit{e}}_{z}&=v_{\rm t}\Big(-\alpha+\sqrt{\alpha^{2}-1+{B}^{2}}\Big)\alpha+v_{\rm t}\leq0,
	\end{align}
	We find that 
	\begin{align}{}
	-1\leq\alpha<-\frac{1}{\sqrt{1+{B}^{2}}}\hspace{0.25in} \text{for all} \,\,\,\,\ {B},\label{eq:appdix_case4++_1}
	\end{align}
	Here, for ${B}<1$, Equation \ref{eq:appdix_case4++_xcomp} imposes one other constraint such that $\alpha<-\sqrt{1-{B}^{2}}$.  This condition is less strict compared to the above since $-\frac{1}{\sqrt{1+{B}^{2}}}<-\sqrt{1-{B}^{2}}$.
	The conditions above are described in terms of $\xi$,
	\begin{align}
	\frac{2{B}}{1+{B}^{2}}\leq\cos\xi<1 \hspace{0.65in} \text{for all} \,\ {B},\label{eq:appdix_case4++_3}
	\end{align}
	
	Here, we find that Equation \ref{eq:appdix_case4++_3} is in direct conflict with
	the conditions for $\xi$ identified by \ref{eq:appdix_solution_alpha_case4++1} and \ref{eq:appdix_solution_alpha_case4++3} (see footnote \ref{appdix_relativemagnitude}). Therefore, 
	no solutions exist in this case. 
	
\end{enumerate}

\subsubsection{\textbf{(+,-) case}}
In this case, solutions with negative $\alpha$ do not exist.

\subsubsection{\textbf{(-,+) case}}
Given the choice of the signs in the expressions for ${A}$ and $v'_{\rm s,2}$,
\begin{align}
\label{eq:appdix_solution_case4-+beta}
{A}=-\alpha-\cot\xi \sqrt{1-\alpha^{2}}
\end{align}
\begin{align}
\label{eq:appdix_solution_case4-+vs2CM}
v'_{\rm s,2}=v_{\rm t}\Big[-\alpha+\sqrt{\alpha^{2}-1+{B}^{2}}\Big],
\end{align}
there exists a solution with negative $\alpha$,
\begin{align}
\alpha(\xi,{B})=-\frac{|1+{B}\cos\xi|}{\sqrt{1+{B}^{2}+2{B}\cos\xi}},\label{eq:appdix_solution_alpha_case4-+}
\end{align}
Accordingly, the conditions for ${A}$ and $\xi$ are,

\begin{enumerate}
	\item ${A}>0$
	
	Clearly, $v'_{\rm s,2}$ (Equation \ref{eq:appdix_solution_case4-+vs2CM}) is positive for $\alpha<0$. 
	
	\item $\textbf{\textit{v}}_{s,2}\cdot \textbf{\textit{e}}_{z}\leq0$. 
	
	This is satisfied for any range of $\xi$. This can be understood as follows. 
	When we compare $\alpha(\xi,{B})$ in Equations \ref{eq:appdix_solution_alpha_case4-+} and $-1/\sqrt{1+{B}^{2}}$ in Equation \ref{eq:appdix_case4++_1} (still valid in this case), they are the same at $\cos\xi=0$. Since $\alpha(\xi,{B})$ is a monotonically decreasing function of $\cos\xi$ for fixed ${B}$ (as $\cos\xi$ increases from 0 to 1, $\alpha$ decreases), $\alpha$ is always smaller than $-1/\sqrt{1+{B}^{2}}$, which means Equation \ref{eq:appdix_case4++_1} is fulfilled for any $\xi$.
\end{enumerate}
\vspace{0.3in}
Overall, in this case, we have one solution that is valid for any value of ${B}$, which is,
\begin{align}
\label{eq:appdix_solution_vs2CM_case4-+}
v'_{\rm s,2}&=v_{\rm t}\sqrt{{B}^{2}+2{B}\cos\xi+1}
\end{align}
\begin{align}
\label{eq:appdix_solution_vbrest_case4-+}
v_{\rm b}&=v_{\rm t}\sqrt{\Big(\frac{m_{\rm s,2}}{m_{\rm b}}\Big)^{2}({B}^{2}+2{B}\cos\xi+1)+2\Big(\frac{m_{\rm s,2}}{m_{\rm b}}\Big)(1+{B}\cos\xi)+1}
\end{align}

\subsubsection{\textbf{(-,-) case}}

With negative signs in both of the expressions for ${A}$ and $v'_{\rm s,2}$,
\begin{align}
\label{eq:appdix_solution_case4--beta}
{A}=-\alpha-\cot\xi \sqrt{1-\alpha^{2}}
\end{align}
\begin{align}
\label{eq:appdix_solution_case4--vs2CM}
v'_{\rm s,2}=v_{\rm t}\Big[-\alpha-\sqrt{\alpha^{2}-1+{B}^{2}}\Big],
\end{align}
we obtain the following solutions for $\alpha(\xi,{B})$:

For ${B}\geq1$,
\begin{numcases}{\alpha(\xi,{B})=}
-\frac{|1-{B}\cos\xi|}{\sqrt{1+{B}^{2}-2{B}\cos\xi}}\hspace{0.2in} 1-{B}\cos\xi<0\,;\label{eq:appdix_solution_alpha_case4--1}\\
\frac{|1-{B}\cos\xi|}{\sqrt{1+{B}^{2}-2{B}\cos\xi}}\hspace{0.25in}~1-{B}\cos\xi\geq0\,\,.\label{eq:appdix_solution_alpha_case4--2}
\end{numcases}
and for ${B}<1$,
\begin{numcases}{\alpha(\xi,{B})=}
\frac{|1-{B}\cos\xi|}{\sqrt{1+{B}^{2}-2{B}\cos\xi}}\hspace{0.25in}~{B}-\cos\xi\geq0\,\,.\label{eq:appdix_solution_alpha_case4--4}
\end{numcases}
Equation \ref{eq:appdix_solution_alpha_case4--1} only satisfies the required condition if $\alpha<0$. For the same reason as in Case 2 (+,-), for ${B}>1$, $v'_{\rm s,1}$ becomes negative, such that no solutions exist in this case.

\vspace{0.3in}

After exploring all cases, we have found one simple expression for $v_{\rm b}$ which is valid for all ranges of ${B}$ and $\xi$,
\begin{align}
\label{eq:appdix_solution_vbrest_overall2}
v_{\rm b}&=v_{\rm t}\sqrt{\Big(\frac{m_{\rm s,2}}{m_{\rm b}}\Big)^{2}({B}^{2}+2{B}\cos\xi+1)+2\Big(\frac{m_{\rm s,2}}{m_{\rm b}}\Big)(1+{B}\cos\xi)+1}\nonumber\\
&=\frac{1}{m_{\rm b}}\sqrt{p_{\rm s,1}^{2}+p_{\rm s,2}^{2}+2p_{\rm s,1}^{}p_{\rm s,2}^{}\cos\xi}
\end{align}
Or even more simply,
\begin{align}
\label{eq:appdix_solution_vbrest_overall}
p_{\rm b}&=\sqrt{p_{\rm s,1}^{2}+p_{\rm s,2}^{2}+2p_{\rm s,1}^{}p_{\rm s,2}^{}\cos\xi}
\end{align}
where $p_{\rm s,1}$, $p_{\rm s,2}$ and  $p_{\rm b}$ are the momenta of the first, second ejected stars and the binary, respectively, i.e., $p_{i}=m_{i}v_{i}$. We summarize the physically-allowed solutions for each case in Table \ref{tab:appdix_analyticsolution}.

\section{The radial distance $r_{\rm b}$ as a function of $\xi$}
\label{appendix:r_xi}
\begin{figure*}
	\centering
	\includegraphics[width=8.0cm]{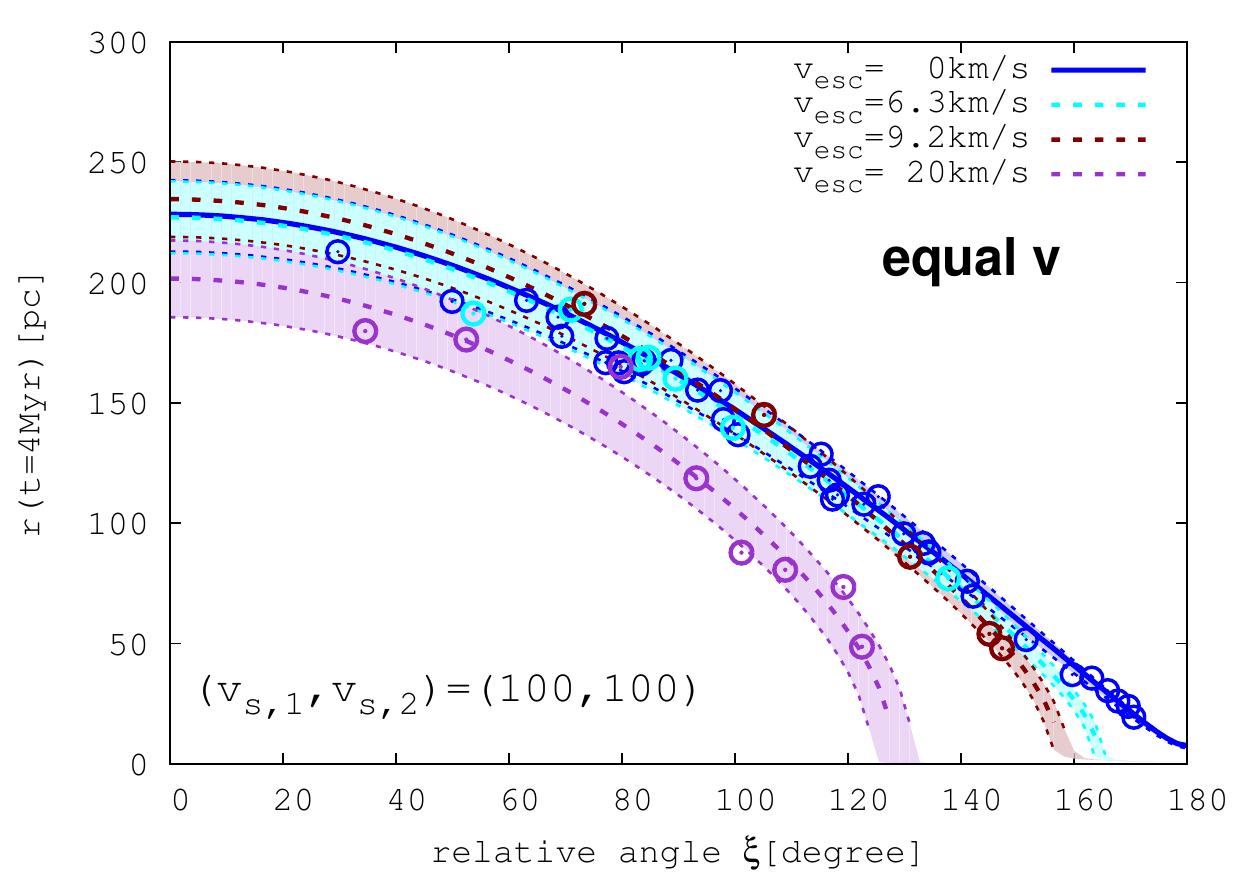}
	\includegraphics[width=8.0cm]{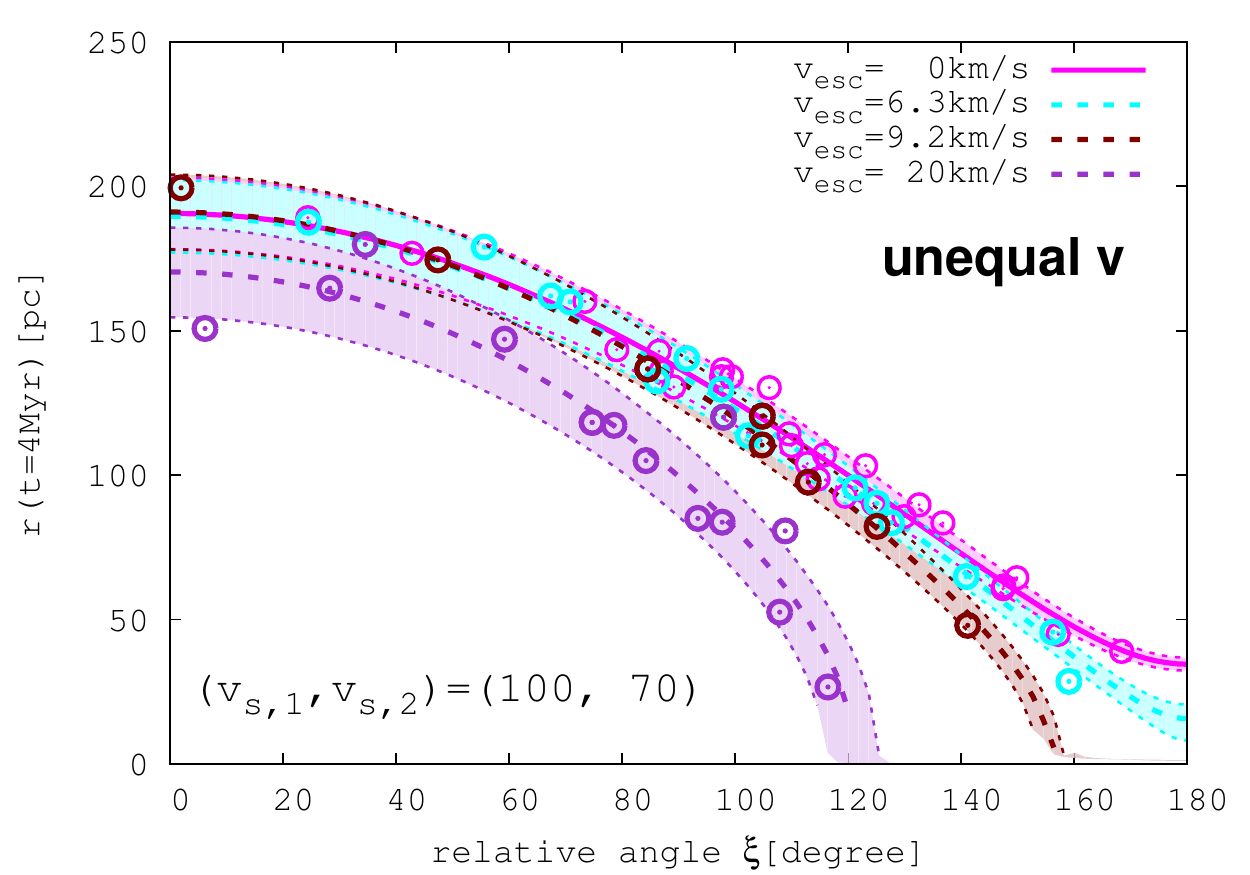}
	\includegraphics[width=8.0cm]{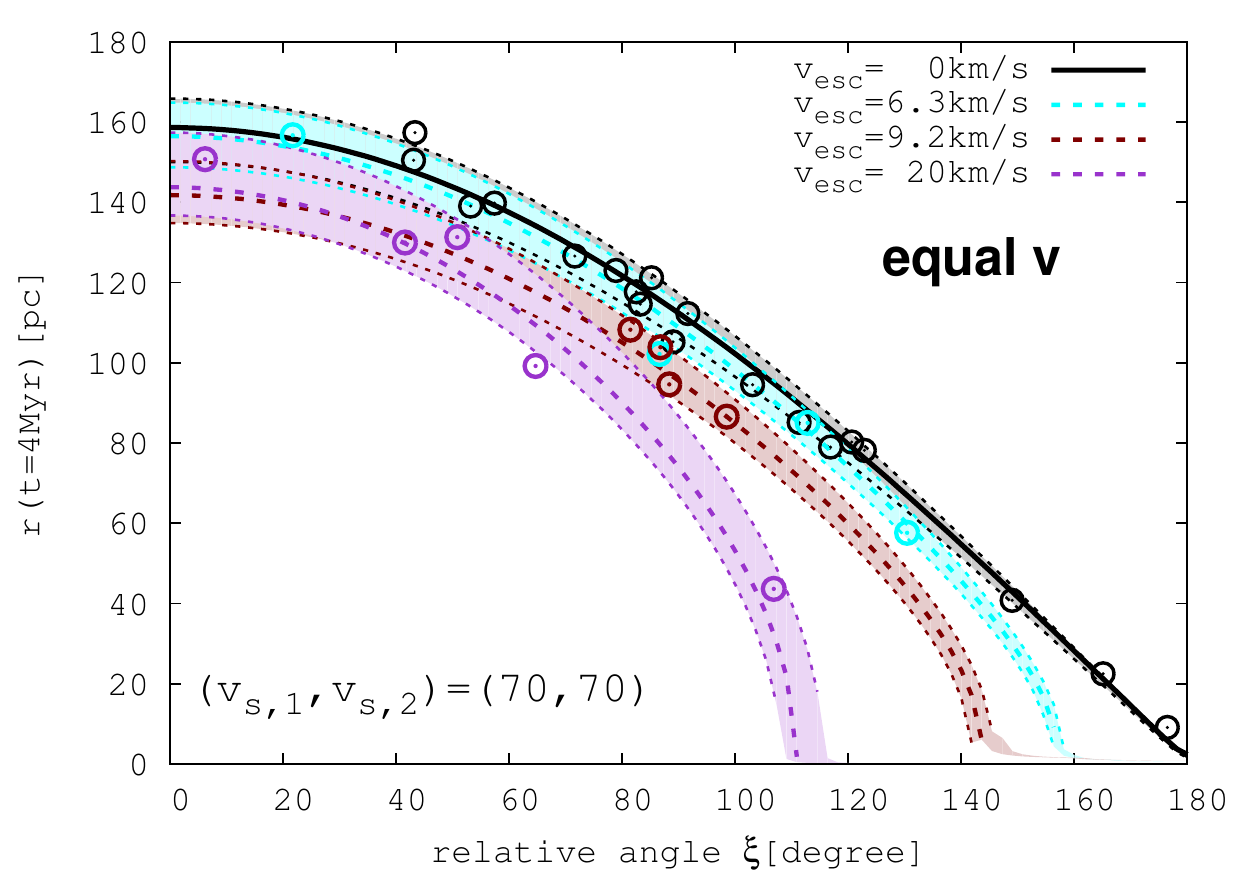}
	\includegraphics[width=8.0cm]{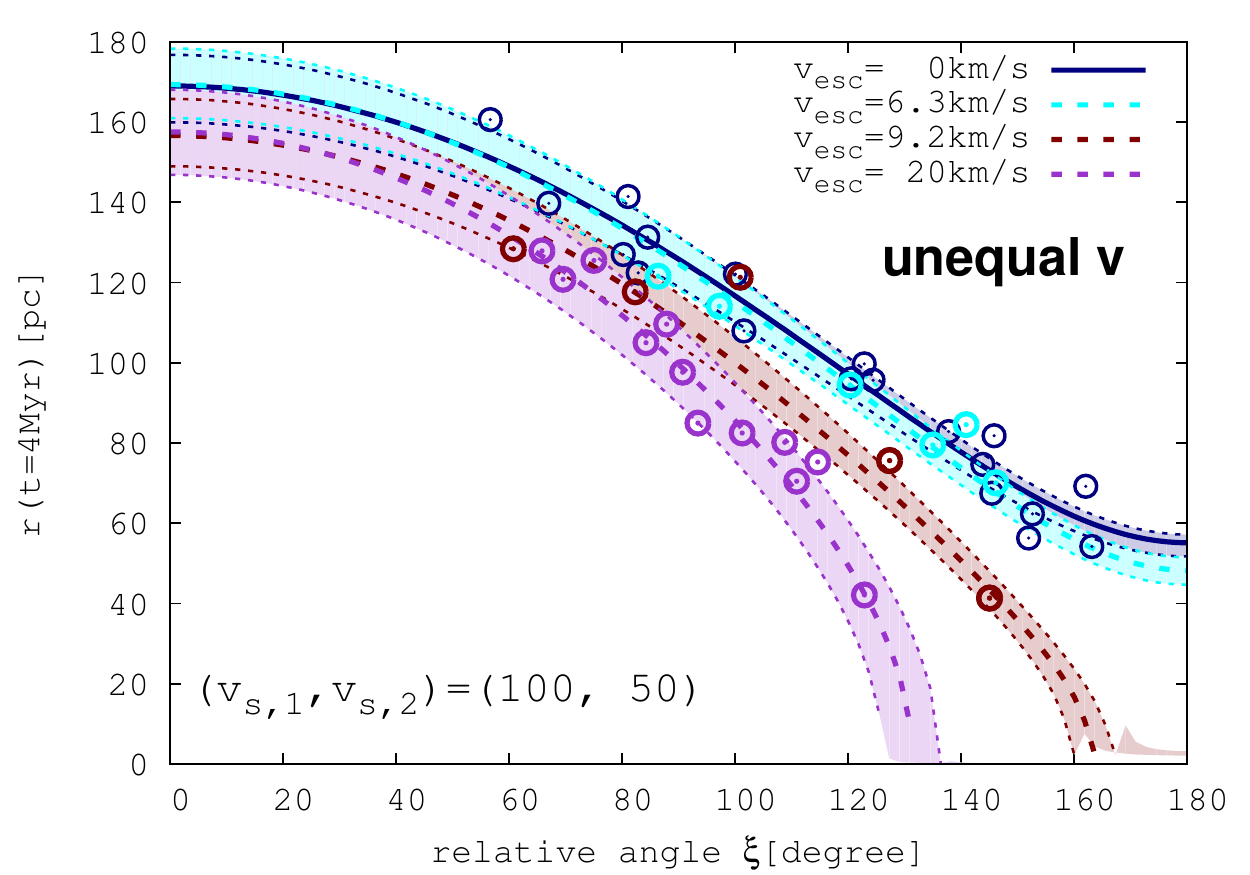}		
	\includegraphics[width=8.0cm]{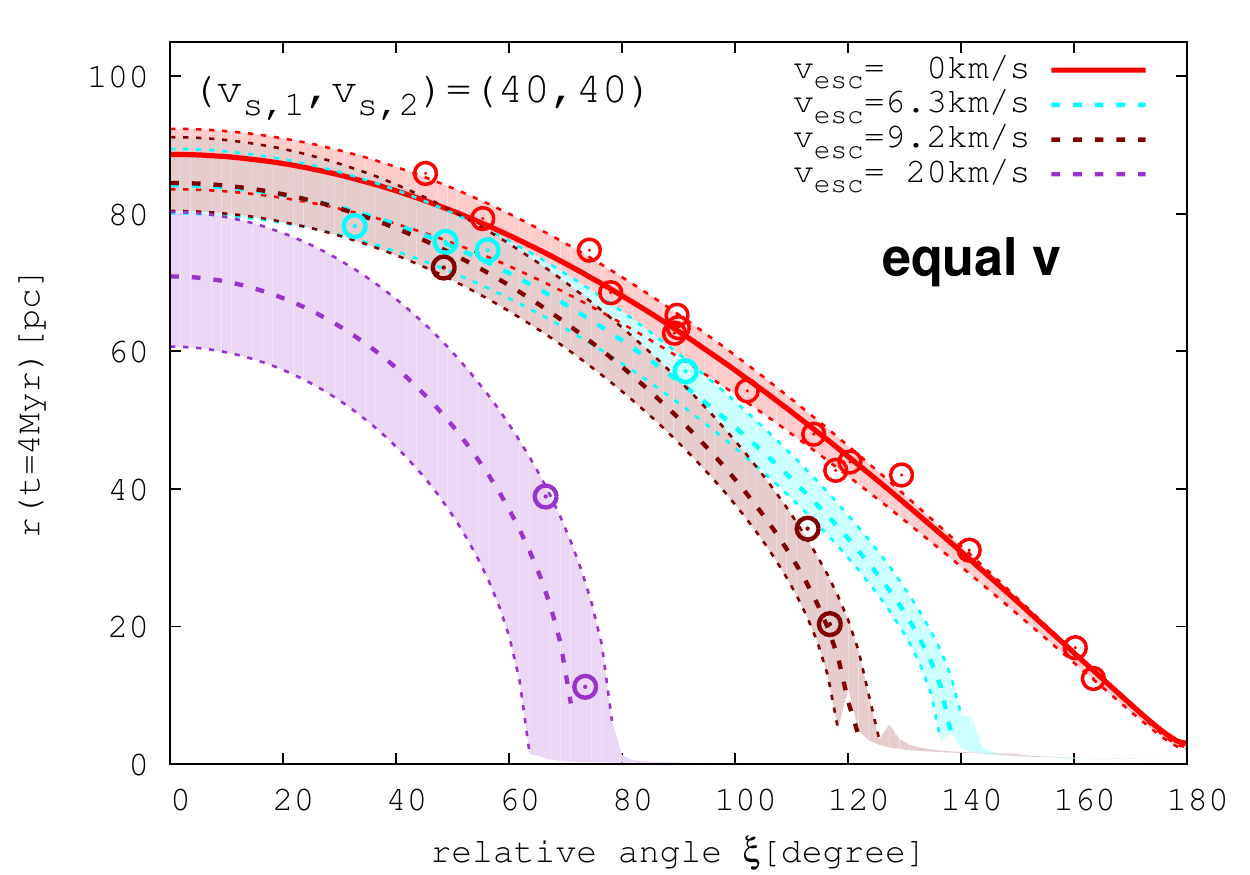}
	\includegraphics[width=8.0cm]{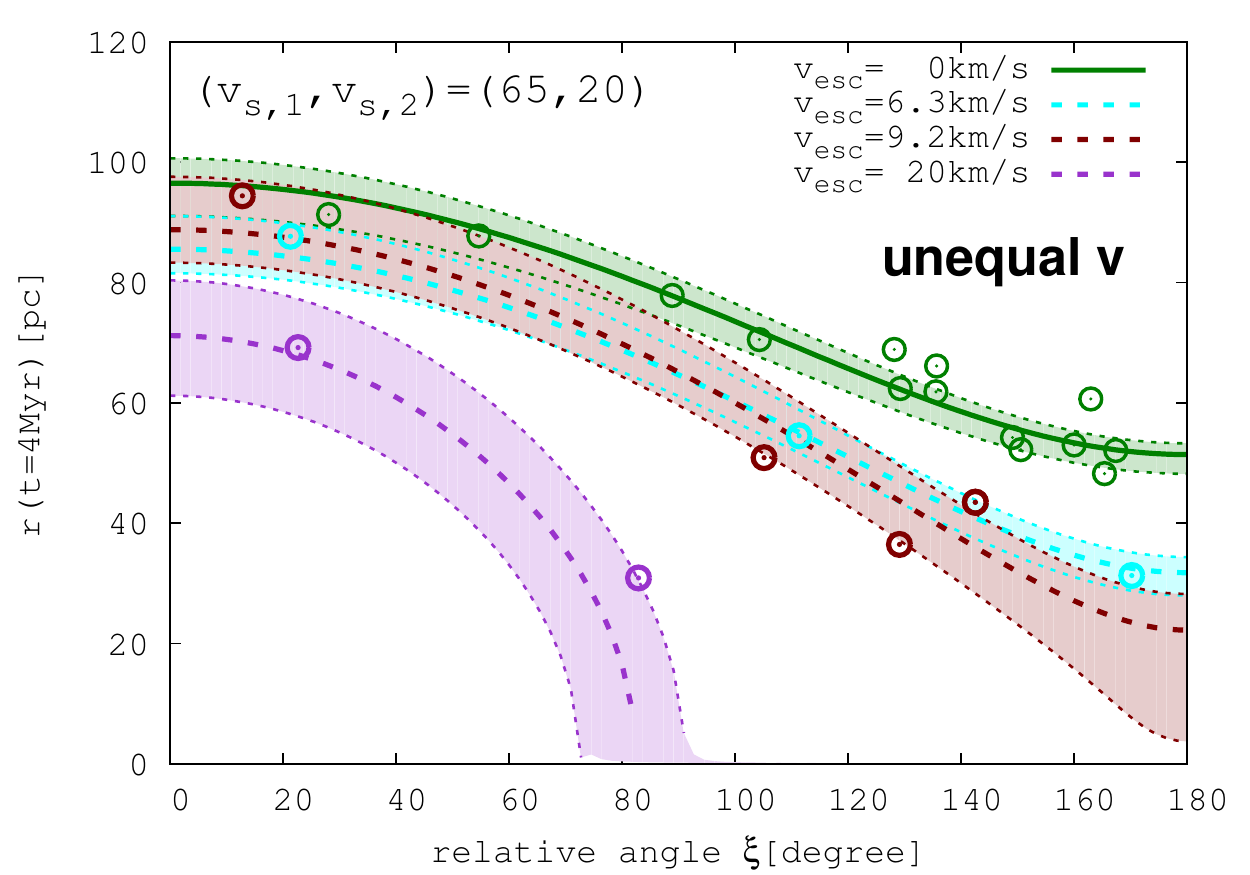}
	\caption{The radial distance $r_{\rm b}$ from the core at $t=4\Myr$ 
		for both \textit{equal} $v$ (\textit{left} column) and \textit{unequal} $v$ 
		(\textit{right column}) with/without the background potential. We use 
		Equation \ref{eq:rbinaryrest_bg} to draw the thick dotted lines ($v_{\rm esc}\neq0$). 
		The final speeds of the
		simulated ejected stars are constrained to be within 7\% of
		the analytic formulae. The upper and lower limits (thin dotted lines) around the median 
		values (solid lines and thick dotted lines) are depicted. The regions between the limits 
		are filled with the same colors as the dots and lines.}
	\label{fig:r_xi}
\end{figure*}

\bsp	
\label{lastpage}
\end{document}